\documentclass[11pt]{article}
\date{\today}

\usepackage{geometry}
\usepackage{amsmath,amsfonts,amssymb}  % AMS math, fonts, and symbols
\usepackage{bbold}

\usepackage{graphicx}       % Required for inserting figures
\usepackage{subcaption}      % More flexible subcaption support
\usepackage{float}           % Controls figure placement

\usepackage[natbibapa]{apacite} % APA style citations with natbib options
\usepackage[hidelinks]{hyperref} % Removes link boxes and enables clickable references

\usepackage{comment}  % Used to comment out blocks of text
\usepackage[normalem]{ulem}  % Supports text strikeout
\usepackage[dvipsnames]{xcolor} % Supports colored text
\usepackage[table,xcdraw]{xcolor} 

\usepackage{authblk} % Supports multiple authors
\usepackage{placeins} % for \FloatBarrier:Want all figures/tables of that section to appear before the next section.
\usepackage{tabularx}
\usepackage{booktabs} 
\usepackage{threeparttable}
\usepackage{microtype}
\title{Portfolio Optimization under Heavy Tails and Asymmetric Volatility: Evidence from Taiwan-Exposed ETFs}
\author[1]{Ting-Jung Lee}
\author[2]{Abootaleb Shirvani}
\author[1]{Farzana Afroz}
\author[1]{Svetlozar T. Rachev}
\author[3]{Frank J. Fabozzi}

\affil[1]{\small Department of Mathematics \& Statistics, Texas Tech University, Lubbock, TX 79409-1042, USA, email: tinglee@ttu.edu, fafroz@ttu.edu, zari.rachev@ttu.edu}
\affil[2]{\small Department of Mathematical Science, Kean University, email: ashirvan@kean.edu}
\affil[3]{\small Carey Business School, Johns Hopkins University, Baltimore, MD 21218, USA, email: fabozzi321@aol.com}
\affil[*]{\small Corresponding author, tinglee@ttu.edu}

\begin{document}
\maketitle

\begin{abstract}
Taiwan's central role in global semiconductor manufacturing exposes Taiwan-related ETFs to technology concentration, geopolitical uncertainty, and supply-chain disruptions, resulting in return distributions characterized by heavy tails, volatility clustering, and asymmetric responses to negative shocks. This paper analyzes thirty U.S.-listed ETFs with Taiwan exposure from February 2015 to February 2025 using tail-risk diagnostics, asymmetric volatility modeling, and portfolio optimization under mean--variance and conditional value-at-risk (CVaR) criteria. Hill tail-index estimates document heavy-tailed behavior across the ETF universe. Although the ETFs exhibit broadly similar asymptotic tail-decay behavior, semiconductor-focused ETFs produce substantially larger VaR and CVaR estimates than diversified benchmarks, indicating that cross-sectional differences in extreme downside risk are driven primarily by differences in return scale rather than tail-index estimates. GJR-GARCH estimates reveal persistent, asymmetric volatility, and the apparent long memory in squared returns is largely attributable to conditional heteroskedasticity rather than genuine fractional integration. CVaR optimization produces substantially more concentrated allocations than mean--variance optimization, with the CVaR tangent portfolio allocating a large weight to SMH during the post-COVID AI-driven expansion. Portfolio rankings depend on the performance measure: the Sharpe ratio and STARR measure favor the equally weighted portfolio, whereas the Rachev ratio favors CVaR-based portfolios. Overall, the results suggest that variance-based frameworks alone provide an incomplete characterization of risk in technology-concentrated investment environments and that variance-based and tail-sensitive performance measures may favor different portfolios over the same sample period.
\end{abstract}


%\noindent\textbf{JEL Classification:} G11; G12; G15; C58; C22

\section{Introduction}
The global semiconductor industry has become increasingly concentrated in Taiwan, where Taiwan Semiconductor Manufacturing Company (TSMC) and a small number of associated firms account for a dominant share of the world's advanced-node semiconductor manufacturing used in artificial intelligence (AI), high-performance computing, and advanced consumer electronics. TSMC currently produces the majority of the world's leading-edge semiconductors and occupies a strategically important position within global technology supply chains \citep{TSMC2025}. Consequently, exchange-traded funds (ETFs) with exposure to Taiwan are influenced not only by broad market conditions but also by semiconductor demand cycles, AI-driven capital expenditure, supply-chain disruptions, and geopolitical developments in East Asia. These forces create a distinctive investment environment characterized by asymmetric return dynamics, heavy tails, and persistent volatility clustering, posing important challenges for traditional portfolio frameworks that rely primarily on variance as a measure of risk.

Modern portfolio theory is founded on the mean--variance framework of \citet{markowitz1952}, in which investors select portfolios by trading off expected returns against return variance. Although analytically tractable and widely used, variance treats positive and negative deviations symmetrically and therefore does not distinguish between favorable and unfavorable outcomes. This limitation is particularly relevant in semiconductor-concentrated ETF markets, where returns frequently exhibit skewness, excess kurtosis, and large downside movements associated with supply-chain disruptions, geopolitical events, and abrupt shifts in technology demand. Such departures from Gaussian assumptions, together with heavy tails and volatility clustering, are well documented in financial return series \citep{mandelbrot1963variation,cont2001empirical,campbell2001have,mcneil2015quantitative}.
These considerations motivate the use of tail-sensitive risk measures and extreme-value methodologies. Value-at-Risk (VaR) summarizes potential losses through quantiles of the return distribution, whereas Conditional Value-at-Risk (CVaR), also known as Expected Shortfall, measures the expected magnitude of losses beyond the VaR threshold. Unlike VaR, CVaR satisfies the axioms of coherent risk measures established by \citet{artzner1999coherent} and provides a more comprehensive measure of extreme downside risk. Furthermore, \citet{rockafellar2000optimization,rockafellar2002conditional} showed that CVaR optimization can be formulated as a tractable optimization problem, making it attractive for portfolio construction under non-normal return distributions. Extreme value theory complements this framework by characterizing tail behavior, while the semi-parametric Hill estimator provides a direct measure of tail heaviness without imposing restrictive parametric assumptions \citep{hill1975simple,embrechts1997,mcneil2015quantitative}.
Tail behavior alone does not fully characterize financial risk because return volatility evolves dynamically over time. A large empirical literature documents volatility clustering, conditional heteroskedasticity, and asymmetric volatility responses, whereby negative return shocks generate larger increases in future volatility than positive shocks of comparable magnitude \citep{bollerslev1986generalized,ding1993long,black1976,glosten1993relation}. These effects may be especially pronounced in technology-intensive and semiconductor-related ETFs, where export restrictions, geopolitical tensions, and fluctuations in semiconductor demand can simultaneously affect multiple firms through shared supply chains and demand cycles.

An important question is whether the strong persistence frequently observed in squared returns reflects genuine long-range dependence or highly persistent conditional heteroskedasticity. Fractionally integrated volatility models permit long-memory behavior in conditional variance dynamics \citep{baillie1996fractionally}, yet apparent long memory may also arise from structural breaks, aggregation effects, and multi-component volatility processes \citep{granger1980,diebold2001long,corsi2009}. Distinguishing between these competing explanations is important for volatility forecasting, risk management, and portfolio construction.

The rapid growth of ETFs has generated substantial interest in their risk characteristics and portfolio implications. ETFs provide investors with diversified exposure to international equity markets, sector-specific themes, and alternative investment strategies \citep{elton2009modern}. As their number and diversity continue to expand, portfolio selection has become increasingly important. In particular, \citet{demiguel2009optimal} showed that simple equally weighted portfolios often performed surprisingly well relative to optimized strategies because estimation error could offset the theoretical benefits of optimization. More recently, \citet{jaffri2025} examined portfolio optimization for Pakistan-exposed ETFs using tail-sensitive risk measures, while \citet{divelgama2026} analyzed the risk characteristics and performance of Asian ETFs, highlighting the importance of market-specific risk factors and concentration effects.

The present paper extends this line of research to Taiwan-exposed ETFs, a market distinguished by its concentration in semiconductor-related industries, AI-driven demand, and significant geopolitical uncertainty. However, neither study examined the relative performance of long-memory and persistent GARCH models, the interaction between semiconductor concentration and CVaR optimization, or the dependence of portfolio rankings on the choice of performance measure. Moreover, relatively little empirical evidence exists on Taiwan-exposed ETFs despite their increasing importance in global semiconductor supply chains and AI-related investment themes.

Taiwan provides a particularly informative setting for studying tail-sensitive portfolio allocation because it combines technology-sector concentration with unique sources of geopolitical and supply-chain risk. Cross-strait tensions introduce low-probability but potentially severe disruption scenarios that are difficult to characterize using variance-based measures alone. More generally, geopolitical risk has been shown to affect asset prices, increase uncertainty, and amplify downside risk in financial markets \citep{caldara2022,pastor2013}. The post-2020 expansion of artificial intelligence has further amplified these dynamics by generating extraordinary demand for advanced semiconductors and substantial appreciation in semiconductor-related equities, consistent with the broader literature describing artificial intelligence as a general-purpose technology whose productivity gains emerge through complementary investments and technological diffusion \citep{brynjolfsson2021productivity}. As shown in Section~\ref{sec:tailrisk_volatility}, semiconductor-focused ETFs exhibit materially greater downside tail risk than diversified benchmarks, consistent with fat-tailed return dynamics associated with technology concentration and exposure to common systematic shocks.

Taken together, these considerations suggest that the period from 2015 to 2025, spanning a relatively stable pre-COVID expansion, a severe market disruption, and an extraordinary AI-driven semiconductor boom, provides an informative setting for examining whether tail-sensitive portfolio methodologies produce conclusions that differ materially from variance-based analysis and for identifying the sources of volatility persistence in technology-concentrated investment environments.

These considerations motivate four research questions that guide the empirical analysis. Do Taiwan-exposed ETFs exhibit tail-risk characteristics that variance-based measures systematically understate? Does the apparent long-memory behavior observed in ETF volatility reflect genuine fractional integration or persistent conditional heteroskedasticity? Does CVaR-based portfolio optimization produce materially different allocations from mean--variance optimization in semiconductor-concentrated markets? And do portfolio performance rankings depend on whether variance-based or tail-sensitive reward-to-risk measures are employed?

This paper makes four contributions corresponding to these research questions. First, it develops a unified empirical framework integrating Hill tail-index estimation, asymmetric GJR--GARCH modeling, rolling FIGARCH estimation, and CVaR-based portfolio optimization for thirty Taiwan-exposed, emerging-market, Asia--Pacific, technology, and semiconductor-sector ETFs over February 2015--February 2025. To the best of our knowledge, this combination of methods has not previously been applied in this setting. Second, it shows that the apparent long memory in ETF volatility is largely attributable to asymmetric conditional heteroskedasticity rather than genuine fractional integration, as long-memory diagnostics collapse toward zero after GJR--GARCH filtering. Third, it documents that CVaR optimization produces materially more concentrated allocations than mean--variance optimization, with semiconductor ETFs frequently emerging as corner solutions during the post-COVID AI-driven expansion. Fourth, it shows that portfolio performance rankings depend on the evaluation criterion: the Sharpe ratio and STARR measure favor the equally weighted portfolio, whereas the Rachev ratio favors CVaR-based allocations.

The remainder of the paper is organized as follows. Section~\ref{sec:data} describes the data and market characteristics. Sections~\ref{sec:risk_tail_behavior} and \ref{sec:tailrisk_volatility} examine ETF-level tail risk, volatility persistence, and asymmetric dynamics. Sections~\ref{sec:historical_optimization} and \ref{sec:portfolio_performance} develop the portfolio optimization framework and evaluate risk-adjusted performance. Section~\ref{sec:conclusion} concludes with the main findings, their implications, and directions for future research.

%%%%%%%%%%%%%%%%%%%%%%%%%%%%%%%%%%%%%%%%%%%%%%%%%%%
%%
\section{Data Description and Market Characteristics}\label{sec:data}
This section describes the ETF dataset, benchmark construction, and market characteristics used throughout the empirical analysis. Because Taiwan plays a central role in global semiconductor manufacturing and advanced technology supply chains, ETFs with exposure to Taiwan provided a natural setting for examining concentration risk, asymmetric return behavior, and tail-sensitive portfolio allocation. The selected ETF universe included Taiwan-focused funds, emerging-market ETFs, global technology ETFs, and semiconductor-sector ETFs, thereby facilitating comparisons across different degrees of geographic and sectoral concentration.

In addition to describing the sample construction, this section examines cumulative return performance and benchmark behavior over the period from February 2015 to February 2025, encompassing major market events such as the COVID-19 crisis and the subsequent AI-driven technology expansion. Particular attention is given to the roles of semiconductor concentration and diversification in shaping long-run return and volatility patterns. These observations provide the empirical context for the tail-risk, volatility, portfolio-optimization, and performance-evaluation analyses presented in subsequent sections.

The COVID-19 pandemic generated one of the most severe episodes of financial market volatility in modern history, producing unprecedented cross-sectional differences in returns across industries and firms \citep{baker2020unprecedented,ramelli2020feverish}. Consequently, the sample period considered in this study provides a valuable opportunity to examine the behavior of Taiwan-related ETFs under conditions of extreme market stress, followed by the post-pandemic expansion in technology and semiconductor-related industries driven by rapid advances in artificial intelligence.

\subsection{Data}

The empirical analysis uses daily return data for a sample of thirty U.S.-listed exchange-traded funds (ETFs) with exposure to Taiwan, the broader Asia--Pacific region, and emerging markets. The sample period extends from February 20, 2015, to February 20, 2025, covering diverse market environments, including periods of stable growth, heightened geopolitical uncertainty, the COVID-19 crisis, and the subsequent AI-driven technology expansion associated with increasing semiconductor demand.

The ETF universe was selected to capture multiple dimensions of market exposure relevant to Taiwan-related investments. It includes Taiwan-focused ETFs, emerging-market ETFs, Asia--Pacific regional ETFs, global technology ETFs, semiconductor-sector ETFs, and globally diversified international equity ETFs. This heterogeneity permits comparisons across varying degrees of sector concentration, geographic exposure, and portfolio diversification.

Daily adjusted closing prices for all ETFs, together with the Dow Jones Industrial Average (DJIA) and the S\&P 500 Index, were obtained from Bloomberg. Risk-free rates were proxied using 3-month and 1-year U.S. Treasury yields obtained from the U.S. Department of the Treasury. Treasury yields were converted to daily rates to ensure consistency with the return frequency used throughout the analysis.

The sample was restricted to ETFs with sufficiently long trading histories to support reliable estimation of volatility, tail risk, and rolling portfolio statistics. ETFs without sufficient return histories over the full sample period were excluded to maintain stable rolling-window estimation.

The ETF universe exhibits substantial variation in sector concentration and market exposure. Broadly diversified international ETFs, such as ACWX, VEU, and VXUS, provide broad global equity exposure, whereas semiconductor-focused ETFs, including SOXX and SMH, provide concentrated exposure to the technology and semiconductor sectors. Taiwan-focused ETFs, particularly EWT, occupy an intermediate position by combining Taiwan-specific market exposure with substantial exposure to semiconductor-related risk factors.

\begin{table}[H]
\centering
\caption{Descriptive statistics for the ETF sample and market benchmarks.}
\label{tab:summary_stats}
\begin{threeparttable}

\footnotesize

\resizebox{.9\textwidth}{!}{
\begin{tabular}{lrrrrrrr}
\hline
\textbf{Ticker} &
\textbf{Mean} &
\textbf{Std Dev} &
\textbf{Skewness} &
\textbf{Ex. Kurtosis} &
\textbf{Min} &
\textbf{Max} &
\textbf{Max DD} \\
& (\%) & (\%) & & & (\%) & (\%) & (\%)\\
\hline

EWT  & 4.977  & 21.529 & $-$1.740 & 17.279 & $-$16.997 & 6.525  & 41.272 \\
IXUS & 2.272  & 16.993 & $-$1.158 & 14.656 & $-$11.428 & 8.726  & 39.920 \\
VEU  & 2.116  & 17.130 & $-$1.263 & 16.119 & $-$12.046 & 7.854  & 39.107 \\
AAXJ & 1.776  & 19.861 & $-$0.486 & 7.041  & $-$12.073 & 8.711  & 45.935 \\
ACWX & 2.109  & 17.360 & $-$1.241 & 15.588 & $-$11.794 & 8.051  & 38.965 \\
SPEM & 1.988  & 19.102 & $-$0.805 & 8.728  & $-$12.064 & 7.293  & 39.317 \\
EEMV & 0.105  & 14.281 & $-$1.222 & 14.895 & $-$10.009 & 5.371  & 34.947 \\
DGS  & 1.157  & 17.711 & $-$1.552 & 16.799 & $-$13.976 & 5.312  & 48.341 \\
FEM  & 0.230  & 21.522 & $-$1.278 & 13.583 & $-$15.102 & 8.256  & 49.788 \\
EWX  & 2.691  & 17.655 & $-$1.740 & 18.525 & $-$13.688 & 6.523  & 46.235 \\
IXN  & 16.211 & 22.872 & $-$0.565 & 9.785  & $-$15.371 & 10.504 & 36.530 \\
DBEM & 1.651  & 16.675 & $-$0.545 & 5.769  & $-$9.286  & 7.021  & 35.409 \\
DEM  & $-$0.382 & 19.127 & $-$1.220 & 11.468 & $-$12.514 & 7.159  & 42.645 \\
ECON & $-$1.695  & 20.843 & $-$0.317 & 6.489  & $-$10.312 & 10.984 & 45.959 \\
CWI  & 2.377  & 17.046 & $-$1.095 & 13.134 & $-$11.333 & 9.955  & 38.336 \\
EEMA & 2.217  & 20.609 & $-$0.559 & 7.610  & $-$12.347 & 8.411  & 45.616 \\
IEMG & 1.269  & 19.997 & $-$0.893 & 10.625 & $-$13.552 & 7.222  & 42.224 \\
AIA  & 4.195  & 22.241 & $-$0.155 & 4.740  & $-$10.344 & 11.043 & 55.619 \\
SCHE & 1.227  & 19.712 & $-$0.788 & 9.209  & $-$12.528 & 7.625  & 39.897 \\
VWO  & 1.028  & 19.614 & $-$0.851 & 9.807  & $-$12.885 & 7.494  & 40.251 \\
GMF  & 3.449  & 19.365 & $-$0.601 & 7.462  & $-$10.827 & 8.907  & 42.176 \\
EDIV & 0.540  & 18.534 & $-$1.245 & 13.143 & $-$12.362 & 7.668  & 45.499 \\
PXH  & 1.409  & 20.994 & $-$0.919 & 9.322  & $-$13.710 & 7.639  & 44.427 \\
EEM  & 0.984  & 20.433 & $-$0.744 & 9.217  & $-$13.329 & 7.745  & 41.460 \\
ACWI & 6.836  & 16.844 & $-$1.053 & 15.275 & $-$11.896 & 7.822  & 33.525 \\
SMH  & 21.317 & 30.286 & $-$0.396 & 4.998  & $-$15.562 & 9.825  & 45.303 \\
VXUS & 2.049  & 17.133 & $-$1.260 & 15.799 & $-$11.802 & 8.048  & 39.922 \\
PIE  & 0.452  & 20.716 & $-$1.212 & 12.085 & $-$13.595 & 7.941  & 42.739 \\
SOXX & 19.064 & 30.716 & $-$0.395 & 4.747  & $-$16.520 & 10.270 & 46.245 \\
EELV & $-$0.567 & 14.551 & $-$1.343 & 12.734 & $-$9.581  & 5.615  & 42.178 \\
EWP  & 4.811  & 18.778 & $-$0.899 & 10.339 & $-$12.368 & 7.153  & 34.973 \\
\hline
DJIA & 8.597  & 17.242 & $-$0.968 & 23.887 & $-$13.842 & 10.764 & 37.086 \\
S\&P 500 & 10.280 & 17.535 & $-$0.828 & 16.543 & $-$12.765 & 8.968 & 33.925 \\

\hline
\end{tabular}
}
\vspace{0.2em}

\parbox{\textwidth}{\footnotesize
\textit{Note:} Mean and standard deviation are annualized (daily log returns multiplied by 252 and $\sqrt{252}$, respectively) and expressed as percentages. Skewness and excess kurtosis are computed from daily log returns. Min and Max denote the single largest daily log loss and gain over the sample period, expressed as percentages. Max DD denotes the maximum drawdown, defined as the largest peak-to-trough cumulative decline computed from the price series. Sample period: February 20, 2015 to February 20, 2025.
}
\end{threeparttable}
\end{table}

The diversity in portfolio composition is important because differences in market concentration and asset exposure may influence return asymmetry, volatility persistence, and tail-risk characteristics, thereby motivating the use of both variance-based and tail-sensitive portfolio methodologies in the empirical analysis.\footnote{Table~\ref{tab:etf_summary} (Appendix~\ref{app:etf_table}) summarizes the ETF sample, including ticker symbols, ETF categories, and inception dates, highlighting the substantial heterogeneity in geographic and sectoral exposures across the sample.}

Table~\ref{tab:summary_stats} reports the descriptive statistics for the full sample. All return distributions exhibit negative skewness and substantial excess kurtosis, indicating clear departures from normality across the ETF sample. Semiconductor-focused ETFs stand out by exhibiting the highest annualized mean returns and return volatility. Specifically, SMH and SOXX recorded annualized mean returns of approximately 21.3\% and 19.1\%, respectively, with annualized standard deviations of approximately 30.3\% and 30.7\%. In contrast, the low-volatility ETFs EELV and EEMV exhibited the lowest annualized standard deviations (14.6\% and 14.3\%) and generated near-zero or negative average annualized returns over the sample period. These cross-sectional differences in return and risk characteristics provide the empirical motivation for the tail-risk diagnostics and portfolio optimization analyses presented in the subsequent sections.

Within the sample, EWT provided direct exposure to Taiwan equities and served as the primary Taiwan-focused benchmark, while AAXJ, EEMA, GMF, VWO, and SCHE provided indirect Taiwan exposure through broader Asia--Pacific and emerging-market allocations. SMH, SOXX, and IXN provided concentrated exposure to the global semiconductor and technology sectors. Because Taiwan-based firms occupied dominant positions in global semiconductor supply chains throughout the sample period, these ETFs also provided indirect but economically significant exposure to Taiwan. In contrast, ACWX, VEU, IXUS, and VXUS provided broadly diversified international exposure with comparatively low sector concentration, serving as useful benchmarks for evaluating the effects of concentration and sector specialization on return asymmetry and tail risk. This heterogeneity, encompassing concentrated and diversified portfolios, semiconductor-focused and broad-market exposures, and emerging- and developed-market investments, motivated the tail-risk, volatility, portfolio optimization, and performance-evaluation analyses presented in the following sections.

\subsection{Performance Dynamics and the Benchmark Portfolio}

This section examines the cumulative investment performance of the ETF sample over the period February 20, 2015, to February 20, 2025. All cumulative return series are normalized to an initial investment value of \$100 to facilitate direct comparison across assets and portfolio structures. Throughout the paper, the equally weighted portfolio (EWP) serves as the primary benchmark. Unlike optimized portfolios, the EWP assigns equal weights to all constituent ETFs and therefore avoids reliance on estimated covariance matrices or expected returns. This benchmark is widely used in empirical portfolio studies because of its robustness to estimation error and its competitive out-of-sample performance \citep{demiguel2009optimal,jaffri2025}.

\begin{figure}[H]
    \centering
    \includegraphics[width=0.49\textwidth]{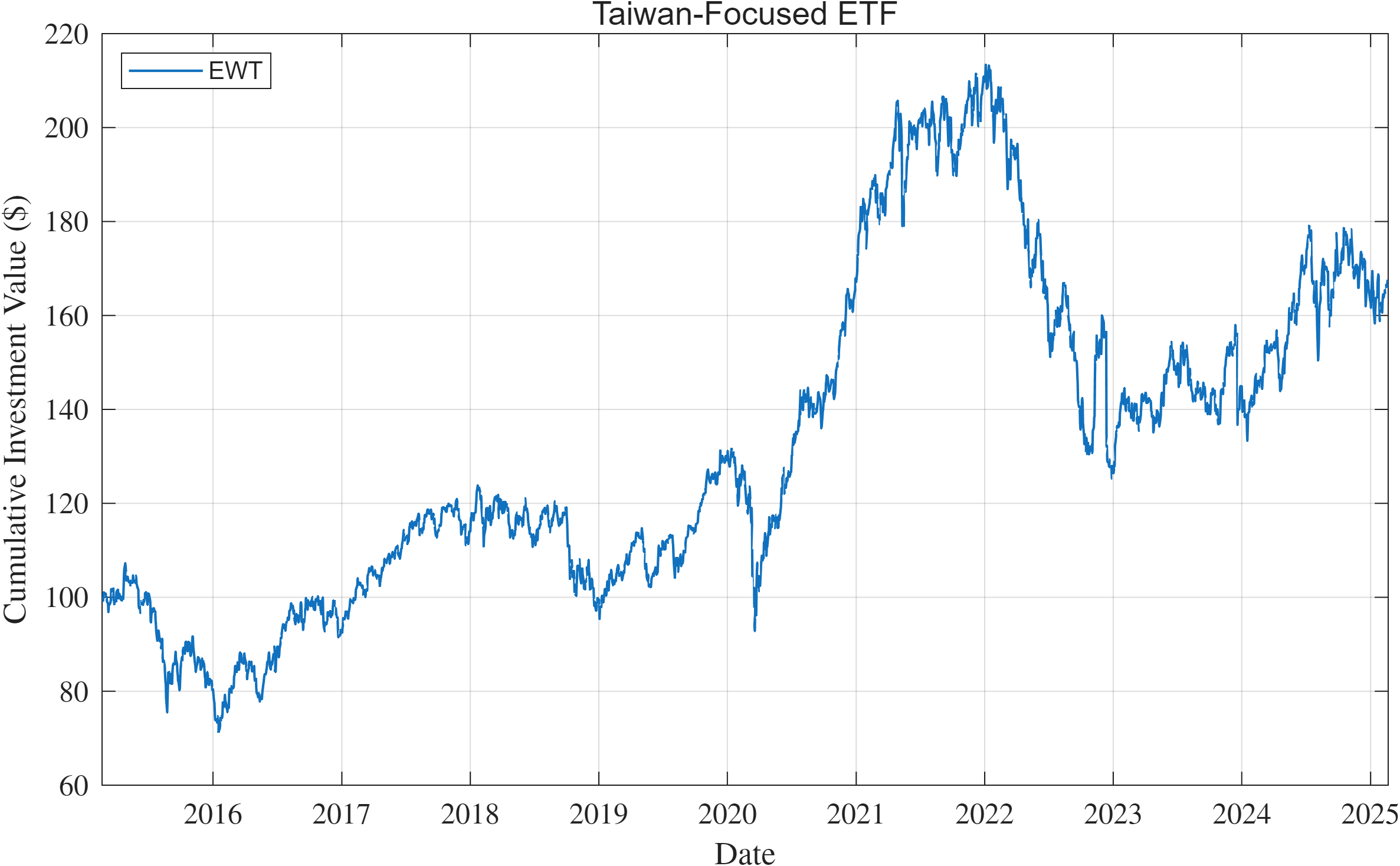}
    \includegraphics[width=0.49\textwidth]{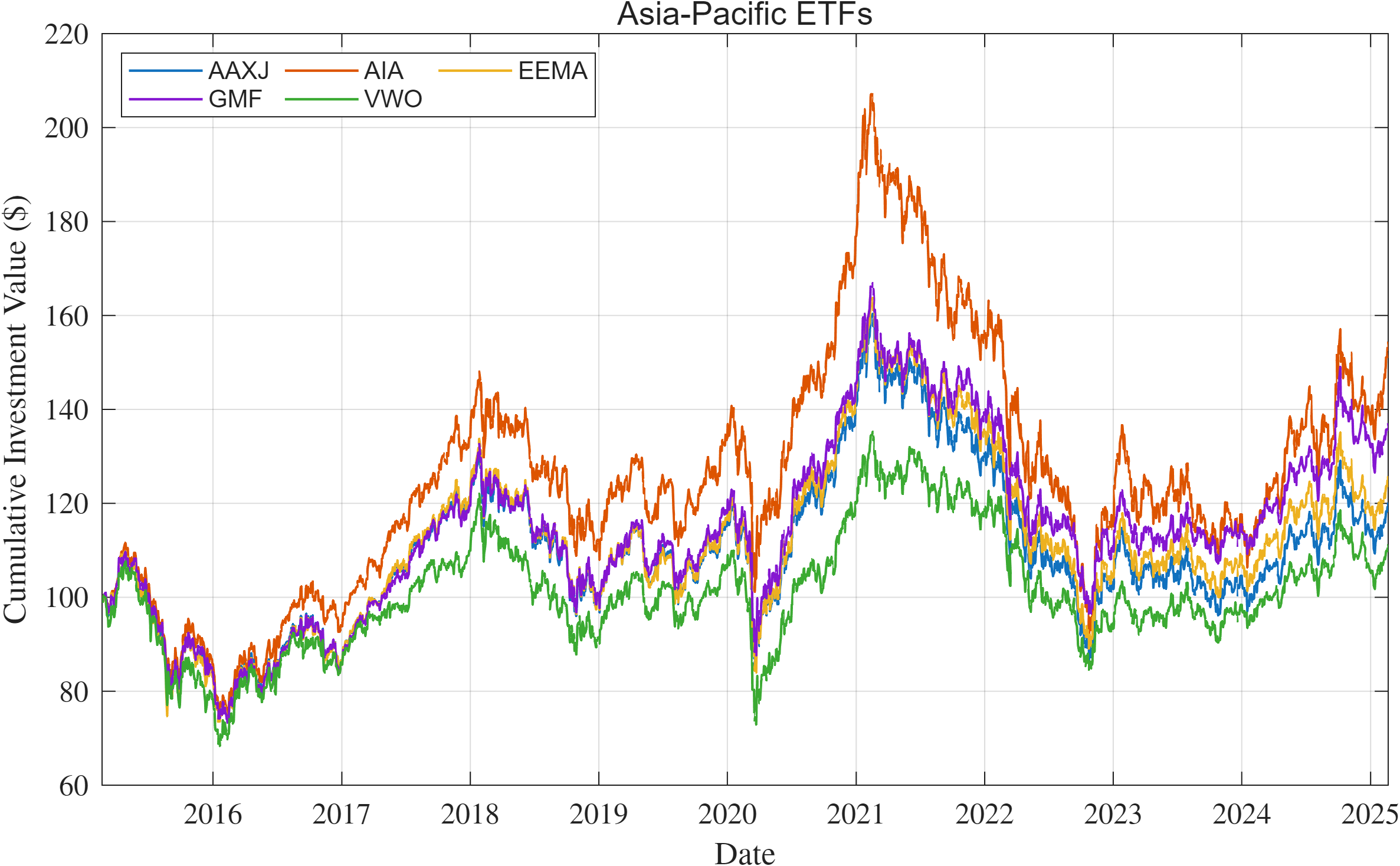}
    \includegraphics[width=0.49\textwidth]{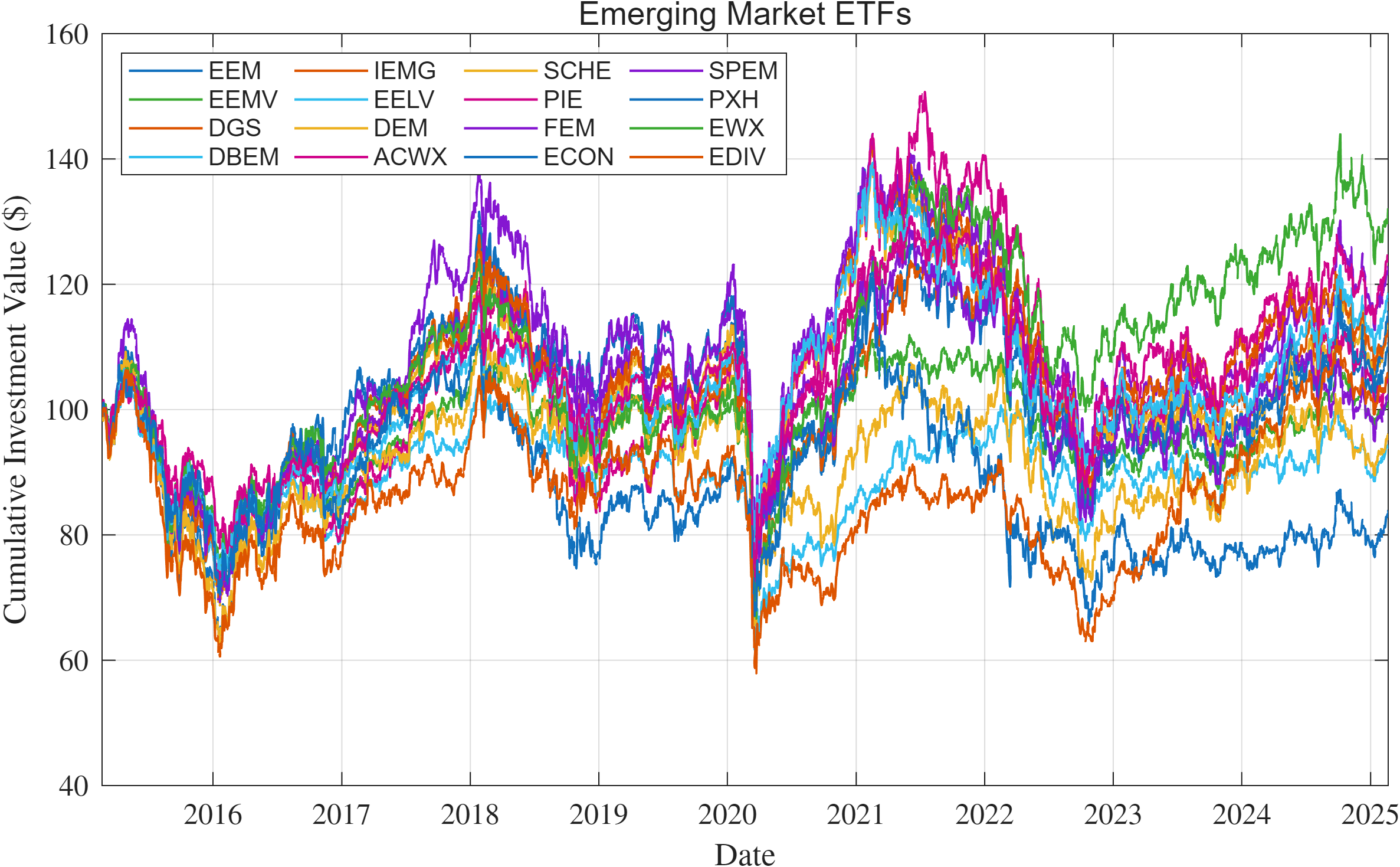}
    \includegraphics[width=0.49\textwidth]{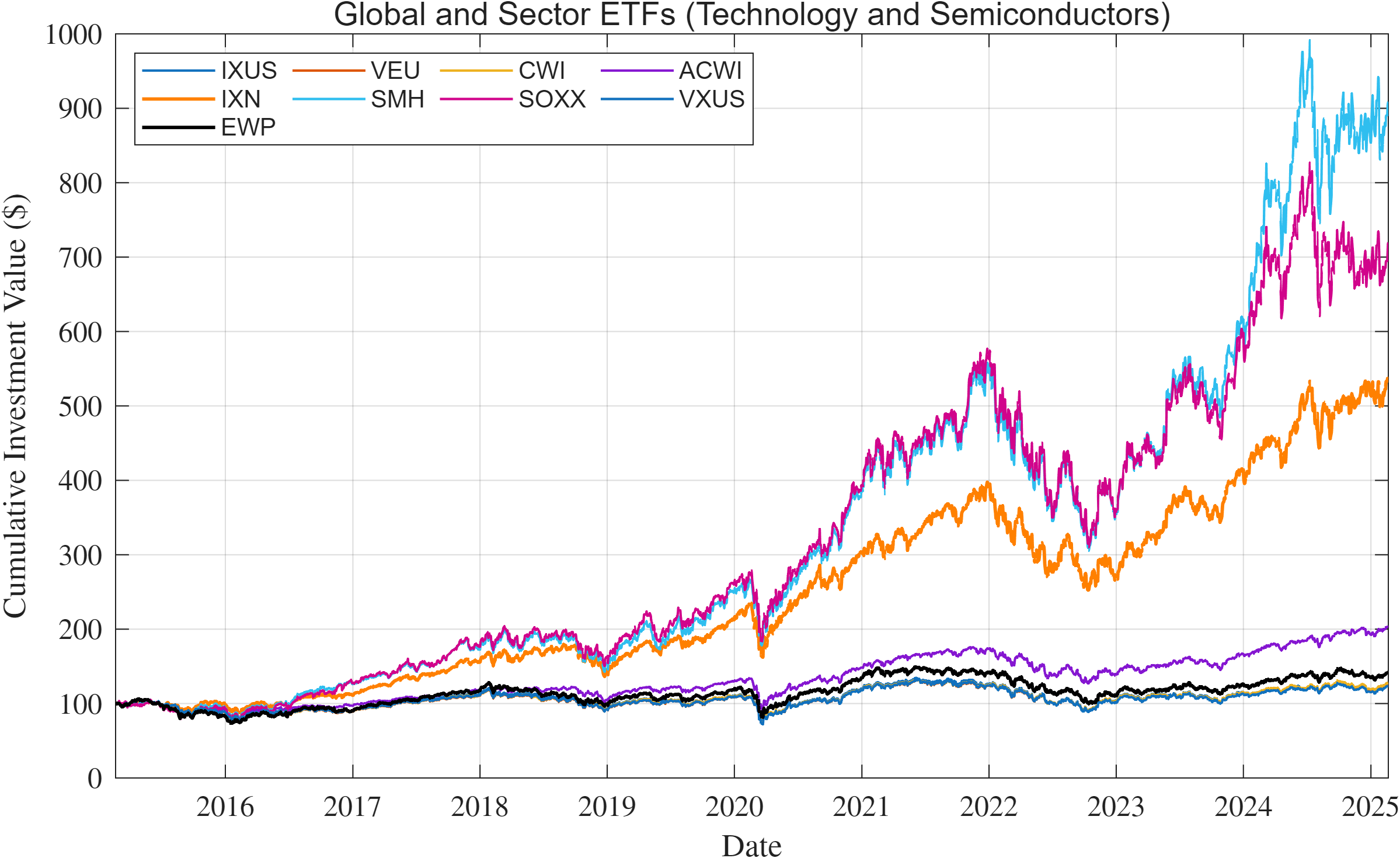}
    \caption{Cumulative investment values of ETFs with exposure to Taiwan and the equally weighted portfolio (EWP) benchmark from February 20, 2015, to February 20, 2025.}
    \label{fig:cum_price}
\end{figure}

Figure~\ref{fig:cum_price} presents the cumulative investment values of the ETF sample and the EWP benchmark over the full sample period. Several patterns are evident. Semiconductor-focused ETFs, particularly SOXX and SMH, substantially outperformed the remaining ETFs after 2020, with cumulative investment values approaching or exceeding \$1,000 by late 2024. This period coincided with rapid growth in artificial intelligence applications and advanced semiconductor manufacturing, during which TSMC reported strong demand for advanced-node semiconductor technologies \citep{TSMC2025}.

Semiconductor ETFs also exhibited substantially larger fluctuations than diversified global ETFs, including sharper drawdowns during periods of market stress and stronger rebounds during subsequent recoveries. In contrast, the EWP displayed smoother cumulative performance with lower sensitivity to extreme movements in individual assets, illustrating the diversification effect emphasized in portfolio theory \citep{elton2009modern,markowitz1952}.

The COVID-19 market collapse in early 2020 was evident across all ETF series, although recovery dynamics differed substantially across sectors. Semiconductor ETFs recovered rapidly and subsequently entered a prolonged expansion phase, whereas diversified emerging-market ETFs experienced slower and more moderate recoveries. From an economic perspective, this divergence reflected Taiwan's central role in global semiconductor supply chains and the sensitivity of semiconductor-related ETFs to common systematic factors, including AI-driven computing demand, supply-chain disruptions, geopolitical tensions, and fluctuations in global electronics production.

Overall, the cumulative performance patterns suggest that greater sector concentration was associated with higher cumulative returns, but also with greater volatility and exposure to tail risk. These observations motivate the formal analyses of tail risk, extreme-value behavior, volatility dynamics, and portfolio optimization presented in the subsequent sections.

%%%%%%%%%%%%%%%%%%%%%%%%%%%%%%%%%%%%%%%%%%%%%%%%%%
%%%
\section{Risk Measures and Tail Behavior}\label{sec:risk_tail_behavior}
Taiwan-exposed ETFs span a broad spectrum of concentration risk, ranging from globally diversified portfolios to semiconductor-focused funds whose underlying holdings are highly correlated through common exposure to technology demand, semiconductor investment cycles, and global supply-chain conditions. This heterogeneity raises a fundamental question: do conventional volatility-based performance measures adequately characterize downside risk across these investment vehicles, or do tail-sensitive measures reveal economically important differences that variance conceals? The question is particularly relevant in semiconductor-concentrated markets, where export restrictions, supply-chain disruptions, and abrupt shifts in technology demand can generate sharp, correlated drawdowns that are not fully reflected in annualized volatility measures.

This section addresses this question by examining ETF-level performance and downside risk using two complementary classes of measures. Conventional volatility-based measures—the Sharpe ratio, Sortino ratio, and maximum drawdown—provide a benchmark, whereas Value-at-Risk (VaR) and Conditional Value-at-Risk (CVaR) quantify extreme losses beyond what variance alone can capture. Comparisons with the DJIA and the S\&P~500 provide benchmark evidence on whether the risk characteristics of Taiwan-exposed ETFs differ materially from those of mature, broadly diversified developed-market portfolios.

The results indicate that variance-based measures conceal economically meaningful differences in downside risk between concentrated semiconductor ETFs and diversified global funds, motivating the tail-sensitive portfolio and volatility analyses presented in the subsequent sections.

\subsection{ETF-Level Risk-Adjusted Performance and Drawdown}

A natural starting point for evaluating Taiwan-exposed ETFs is the class of volatility-based reward-to-risk measures widely used in empirical financial econometrics to summarize portfolio performance relative to return variability \citep{campbell1997econometrics}. Although these measures provide a useful benchmark for comparing portfolio performance, they do not distinguish between upside and downside deviations and therefore provide only limited information about differences in downside risk across ETFs. We evaluate performance using the Sharpe ratio, Sortino ratio, and maximum drawdown, which belong to the broader class of reward-to-risk measures commonly used in portfolio evaluation \citep{cheridito2013reward}. The Sharpe and Sortino ratios summarize investment performance relative to total and downside risk, respectively, whereas maximum drawdown measures the largest peak-to-trough loss experienced over the sample period \citep{sharpe1998,sortino1994}.

\begin{figure}[H]
\centering
\includegraphics[width=\textwidth]{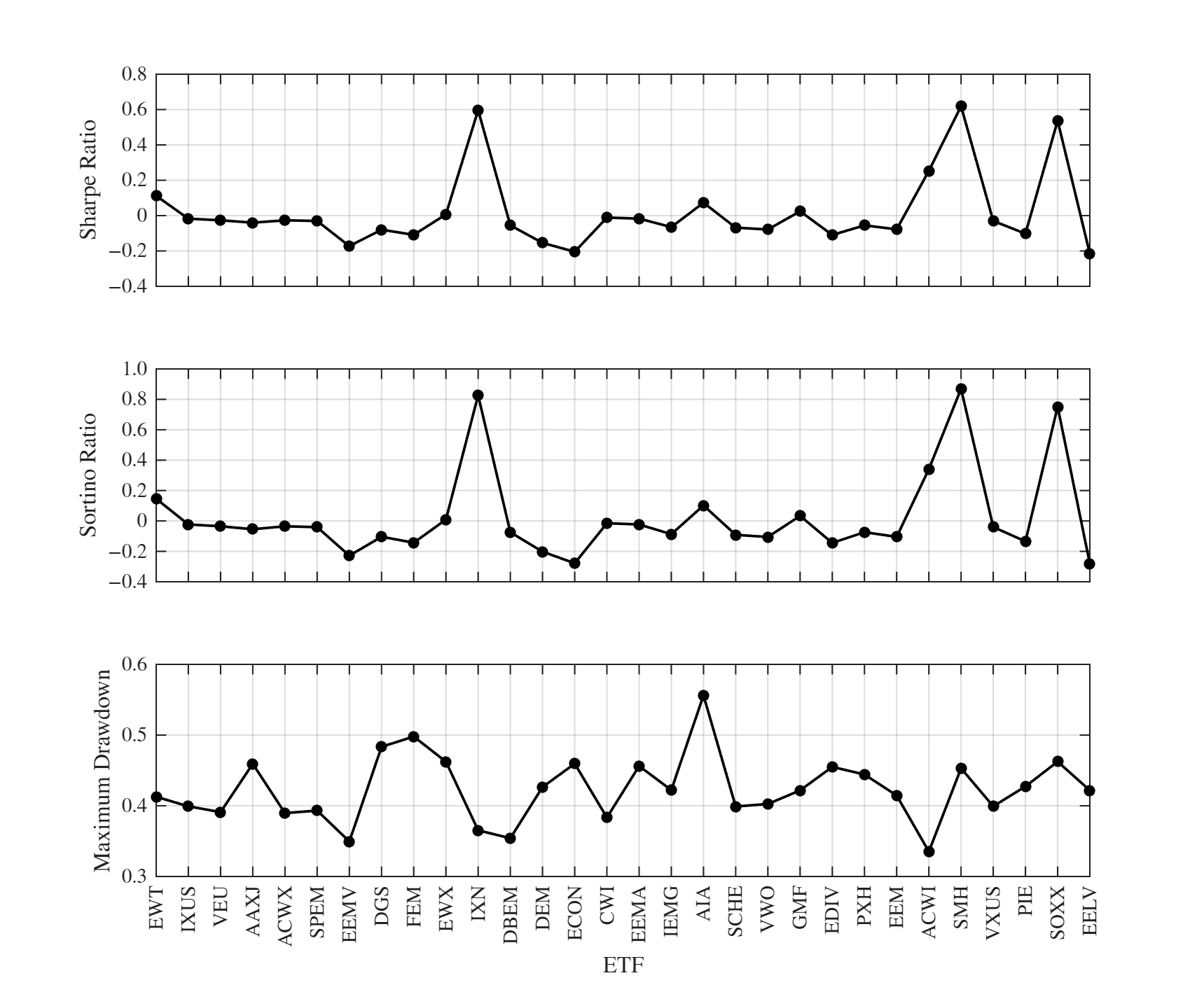}
\caption{Sharpe ratio, Sortino ratio, and maximum drawdown across ETFs.}
\label{fig:sharpe_sortino_drawdown}
\end{figure}

Figure~\ref{fig:sharpe_sortino_drawdown} shows the Sharpe ratio, Sortino ratio, and maximum drawdown across ETFs. Although the Sharpe and Sortino ratios display similar ranking patterns, neither aligns closely with maximum drawdown, indicating that these measures reflected different aspects of risk-adjusted performance. Most ETFs exhibited Sharpe ratios close to zero, generally ranging from approximately $-0.20$ to $0.10$. Three ETFs stood out from the rest of the sample. IXN, SMH, and SOXX recorded Sharpe ratios of approximately $0.60$, $0.62$, and $0.53$, respectively. At the lower end, EELV and ECON recorded Sharpe ratios of approximately $-0.22$ and $-0.20$, respectively. Several ETFs exhibited negative Sharpe and Sortino ratios, indicating that their average excess returns fell below the risk-free rate over the sample period. These ETFs therefore underperformed the risk-free rate on a risk-adjusted basis.

The Sortino ratio exhibited a closely related cross-sectional pattern. IXN, SMH, and SOXX again attained the highest values, approximately $0.83$, $0.87$, and $0.75$, respectively, while most of the remaining ETFs were concentrated between approximately $-0.20$ and $0.10$. The close correspondence between the Sharpe and Sortino ratios suggested that incorporating downside volatility did not materially change the relative ranking of the ETFs. Consequently, the Sortino ratio provided only limited additional information beyond the Sharpe ratio for this sample.

Maximum drawdown, however, varied substantially across the sample and did not follow the same pattern as the Sharpe and Sortino ratios. AIA exhibited the largest maximum drawdown, approximately $58\%$, followed by FEM at approximately $54\%$, while DGS and EDIV both experienced drawdowns of approximately $51\%$. ACWI recorded the smallest maximum drawdown, approximately $35\%$, followed by EEMV and DBEM at approximately $37\%$. The relatively small drawdown of ACWI was consistent with the benefits of broad geographic and sectoral diversification.

IXN combined one of the highest Sharpe and Sortino ratios in the sample with one of the smallest maximum drawdowns, approximately $39\%$. This result indicated that its strong risk-adjusted performance was not accompanied by unusually severe peak-to-trough losses over the sample period. In contrast, SMH and SOXX combined high Sharpe and Sortino ratios with above-average maximum drawdowns of approximately $49\%$ and $50\%$, respectively. These findings suggested that the two semiconductor-focused ETFs generated strong risk-adjusted returns while remaining exposed to substantial cumulative losses during adverse market periods.

The contrast between IXN and the SMH--SOXX pair illustrated that technology-oriented ETFs were not homogeneous with respect to downside risk. Funds with similarly high volatility-adjusted performance nevertheless differed materially in their exposure to extreme drawdowns. This pattern may have reflected differences in portfolio concentration and sensitivity to systematic market shocks, including semiconductor demand contractions, export restrictions, supply-chain disruptions, geopolitical tensions, and fluctuations in technology investment cycles. Narrowly focused semiconductor ETFs may have been more vulnerable to these shocks than broader technology funds.

Diversified global ETFs generally avoided the most severe drawdowns observed among the more concentrated sector ETFs. In particular, ACWI recorded the smallest maximum drawdown in the sample, while VEU and VXUS avoided the larger drawdowns observed for the semiconductor-focused ETFs. Overall, the results indicated that the Sharpe ratio, Sortino ratio, and maximum drawdown measured different aspects of portfolio risk and should not be interpreted as interchangeable measures of performance. Whereas the Sharpe and Sortino ratios summarized average excess returns relative to total and downside volatility, respectively, maximum drawdown measured the magnitude of cumulative peak-to-trough losses. These findings motivated the subsequent analysis using Value-at-Risk and Conditional Value-at-Risk, which examined the left tail of the return distribution more directly.

\subsection{ETF-Level VaR and CVaR Analysis}\label{sec:var_cvar}

The Sharpe and Sortino ratio analysis in the preceding subsection revealed limited discrimination across ETFs despite meaningful differences in sector concentration and drawdown behavior. To examine whether tail-sensitive measures revealed differences that variance-based measures concealed, downside tail risk was analyzed using Value-at-Risk (VaR) and Conditional Value-at-Risk (CVaR), both estimated from the historical return distribution to allow flexibility in the presence of non-Gaussian returns \citep{jorion2007var,lindquist2022advanced}.

For a loss variable $L$ and confidence level $\alpha \in (0,1)$, VaR is defined as the loss threshold not exceeded with probability $\alpha$,

\begin{equation*}
\mathrm{VaR}_{\alpha}(L)
=
\inf\{\ell \in \mathbb{R}: \mathbb{P}(L \leq \ell) \geq \alpha\},
\end{equation*}

while CVaR, also known as Expected Shortfall, measures the expected loss conditional on losses exceeding that threshold,

\begin{equation*}
\mathrm{CVaR}_{\alpha}(L)
=
\mathbb{E}\left[L \mid L \geq \mathrm{VaR}_{\alpha}(L)\right].
\end{equation*}

Unlike VaR, CVaR incorporates the entire tail of the loss distribution beyond the VaR threshold and satisfies the axioms of coherent risk measures \citep{artzner1999coherent}, providing a more complete representation of extreme downside risk.

\begin{figure}[H]
\centering
\includegraphics[width=\textwidth]{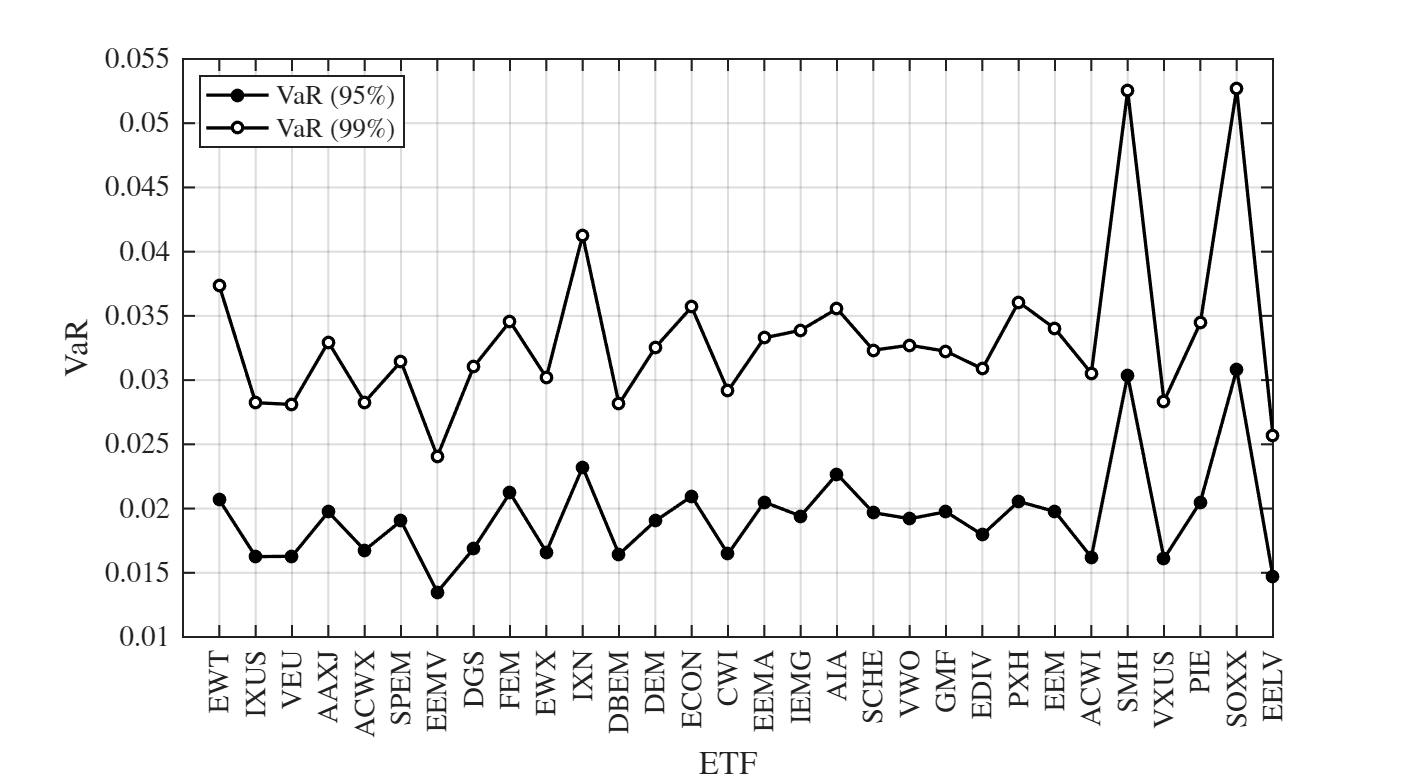}
\caption{Value at Risk (VaR) at 95\% and 99\% confidence levels.}
\label{fig:VaR}
\end{figure}

Figure~\ref{fig:VaR} reports VaR estimates across ETFs at the 95\% and 99\% confidence levels. At the 95\% confidence level, SMH and SOXX exhibit the largest VaR estimates, at approximately 3.0\% and 3.1\%, respectively. Diversified international ETFs, including ACWX, VEU, and VXUS, exhibit lower VaR estimates, generally between 1.6\% and 1.7\%. Taiwan-related ETFs, including EWT, lie between these two groups. Overall, the 95\% VaR estimates indicate substantial cross-sectional differences in downside risk across the ETF sample.

At the 99\% confidence level, VaR estimates are uniformly larger than their 95\% counterparts because of the higher confidence level. The cross-sectional pattern remains similar to that observed at the 95\% level. SMH and SOXX again exhibit the largest VaR estimates, at approximately 5.3\%, whereas diversified international ETFs remain between approximately 2.8\% and 3.0\%. The rankings are largely unchanged across the two confidence levels, with semiconductor-focused ETFs exhibiting larger downside tail risk than diversified international ETFs. These results motivate the subsequent examination of CVaR, which measures the expected loss beyond the VaR threshold.

\begin{figure}[H]
\centering
\includegraphics[width=\textwidth]{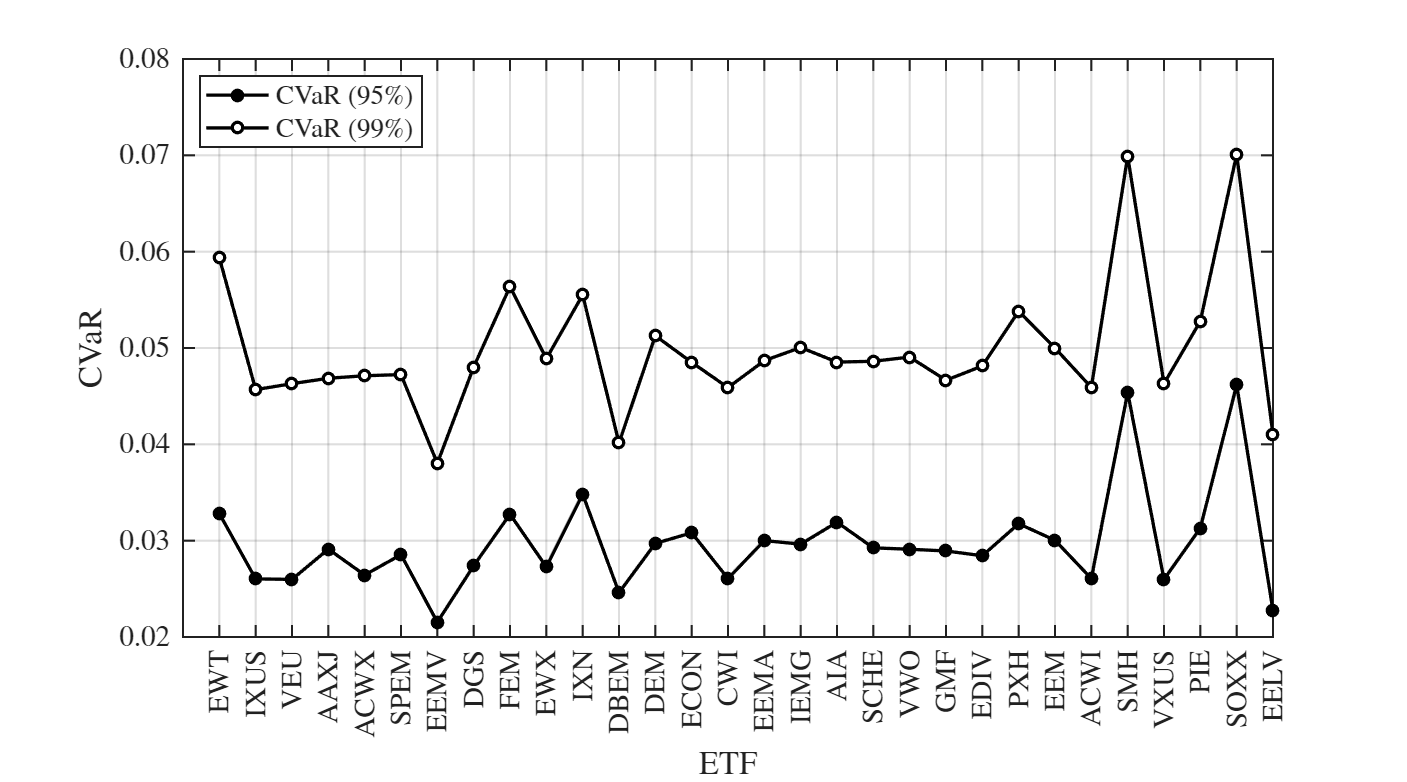}
\caption{Conditional Value at Risk (CVaR) at 95\% and 99\% confidence levels.}
\label{fig:CVaR}
\end{figure}

Figure~\ref{fig:CVaR} reports the corresponding CVaR estimates. Cross-sectional variation is larger than that observed for VaR. At the 95\% confidence level, SMH and SOXX exhibit the largest CVaR estimates, at approximately 4.5\% and 4.6\%, respectively, whereas diversified international ETFs generally range between 2.5\% and 3.0\%. Taiwan-related ETFs remain between these two groups. At the 99\% confidence level, the ranking is similar, with SMH and SOXX reaching approximately 7.0\%, while diversified international ETFs remain below 5\%.

Comparing VaR with CVaR highlights the magnitude of losses beyond the VaR threshold. For SMH, the 95\% CVaR is approximately 4.5\%, compared with a VaR of approximately 3.0\%. At the 99\% confidence level, the CVaR increases to approximately 7.0\%, compared with a VaR of approximately 5.3\%. The difference between VaR and CVaR is generally larger for semiconductor-focused ETFs than for diversified international ETFs, indicating heavier downside tails for the more concentrated funds.

Overall, the VaR and CVaR results show that downside risk differs across the ETF sample at both confidence levels. Semiconductor-focused ETFs consistently exhibit larger tail-risk measures than diversified international ETFs, whereas Taiwan-related ETFs remain between these two groups. These differences are larger than those observed for the Sharpe and Sortino ratios reported in the previous subsection, indicating that variance-based performance measures do not fully capture differences in downside tail risk. These results motivate the broader investigation of tail behavior and volatility dynamics presented in Section~\ref{sec:tailrisk_volatility}.

%%%%%%%%%%%%%%%%%%%%%%%%%%%%%%%%%%%%%%%%%%%%%%%%%%%
%%%%
\section{Extreme Tail Risk and Volatility Analysis across Benchmark Portfolios}\label{sec:tailrisk_volatility}

While the previous analysis examined downside risk using VaR and CVaR, quantile-based measures alone do not fully capture the extreme and dynamic features of financial returns. In particular, Taiwan-related and semiconductor-focused ETFs may exhibit heavy tails, volatility clustering, and asymmetric responses to market shocks.

To further investigate these characteristics, this section examines extreme-value behavior and volatility dynamics through benchmark portfolio comparisons involving the equally weighted portfolio (EWP), the Taiwan ETF (EWT), the Dow Jones Industrial Average (DJIA), and the S\&P~500 Index. The analysis combines Hill tail-index estimation, long-memory diagnostics, and asymmetric volatility modeling to provide a broader assessment of risk characteristics. Particular attention is given to the role of portfolio construction, sector concentration, and technology exposure in shaping tail behavior and volatility persistence. The inclusion of mature U.S. benchmarks such as the DJIA and S\&P~500 allows comparison between Taiwan-related ETFs and diversified developed-market indices.

\subsection{Hill Estimator and Tail--Risk Comparison} \label{subsec:hill}
Quantile-based risk measures such as VaR and CVaR summarize downside losses at specific probability levels but do not directly characterize the asymptotic behavior of the tail distribution. To examine return behavior beyond quantile-based measures, we used the Hill estimator, one of the most widely used semi-parametric methods for estimating the tail index of heavy-tailed distributions. The Hill estimator is based on extreme value theory and uses information from upper order statistics, making it particularly useful for analyzing financial returns because it focuses directly on tail observations without requiring a full parametric specification \citep{hill1975simple}. This flexibility is especially important in heavy-tailed settings, where Gaussian approximations may not adequately capture extreme outcomes \citep{embrechts1997,mcneil2015quantitative}.

Define the loss series as

\begin{equation*}
L_t=-R_t,
\end{equation*}

where \(R_t\) denotes portfolio returns. This transformation converts large negative returns into large positive values, allowing the standard Hill estimator to be applied. Following the standard framework of regular variation in extreme value theory \citep{resnick2007}, we assumed that the loss distribution followed

\[
P(L>x)=x^{-\alpha}\ell(x),\qquad x\rightarrow\infty,
\]

where \(\alpha>0\) is the tail index and \(\ell(x)\) is a slowly varying function. This regular variation assumption characterizes the asymptotic behavior of the extreme downside tail and provides the theoretical foundation for the Hill estimator.

Let

\[
L_{(1)}
\ge
L_{(2)}
\ge
\cdots
\ge
L_{(m)}
\]

denote the descending order statistics of the transformed loss series, where \(L_{(1)}\) represents the largest observed loss. Using the largest \(k\) order statistics, the Hill estimator is defined as

\begin{equation*}
\widehat{\alpha}_k
=
\left\{
\frac{1}{k}
\sum_{i=1}^{k}
\left(
\ln L_{(i)}
-
\ln L_{(k+1)}
\right)
\right\}^{-1},
\end{equation*}

where \(L_{(k+1)}\) serves as the threshold separating tail observations from the remaining observations in the sample \citep{hill1975simple}. Under standard regularity conditions, where \(k\to\infty\) and \(k/m\to0\), the Hill estimator is consistent and asymptotically normal. The tail index \(\alpha\) determines the rate at which the downside tail decays, where smaller values of \(\alpha\) indicate heavier tails and a greater probability of extreme losses.

A practical difficulty in applying the Hill estimator is selecting the threshold parameter \(k\). Small values of \(k\) produce high estimation variance because only a limited number of extreme observations are used, whereas large values introduce bias through the inclusion of observations that do not belong to the tail region. This bias--variance trade-off has long been recognized as one of the principal practical challenges in Hill estimation \citep{danielsson2001using}. To guide the choice of \(k\), we used Hill plots, which display the estimated tail index \(\widehat{\alpha}_k\) as a function of \(k\). The interpretation focused on an intermediate region where the estimates exhibited relatively stable behavior while avoiding the highly variable region at small values of \(k\) and the potential bias associated with larger values of \(k\).

%end blf

\medskip

\begin{figure}[h!]
    \centering
    \includegraphics[width=\textwidth]{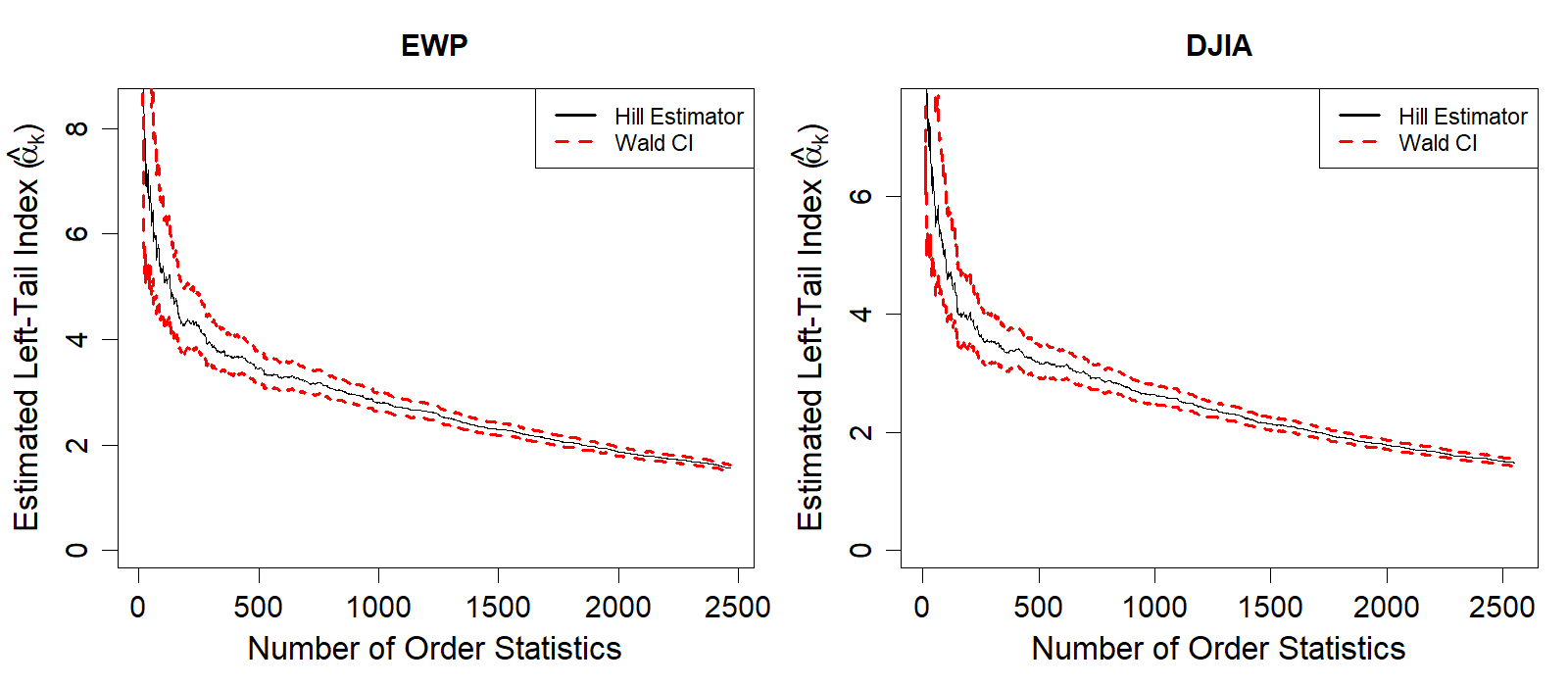}
    \includegraphics[width=\textwidth]{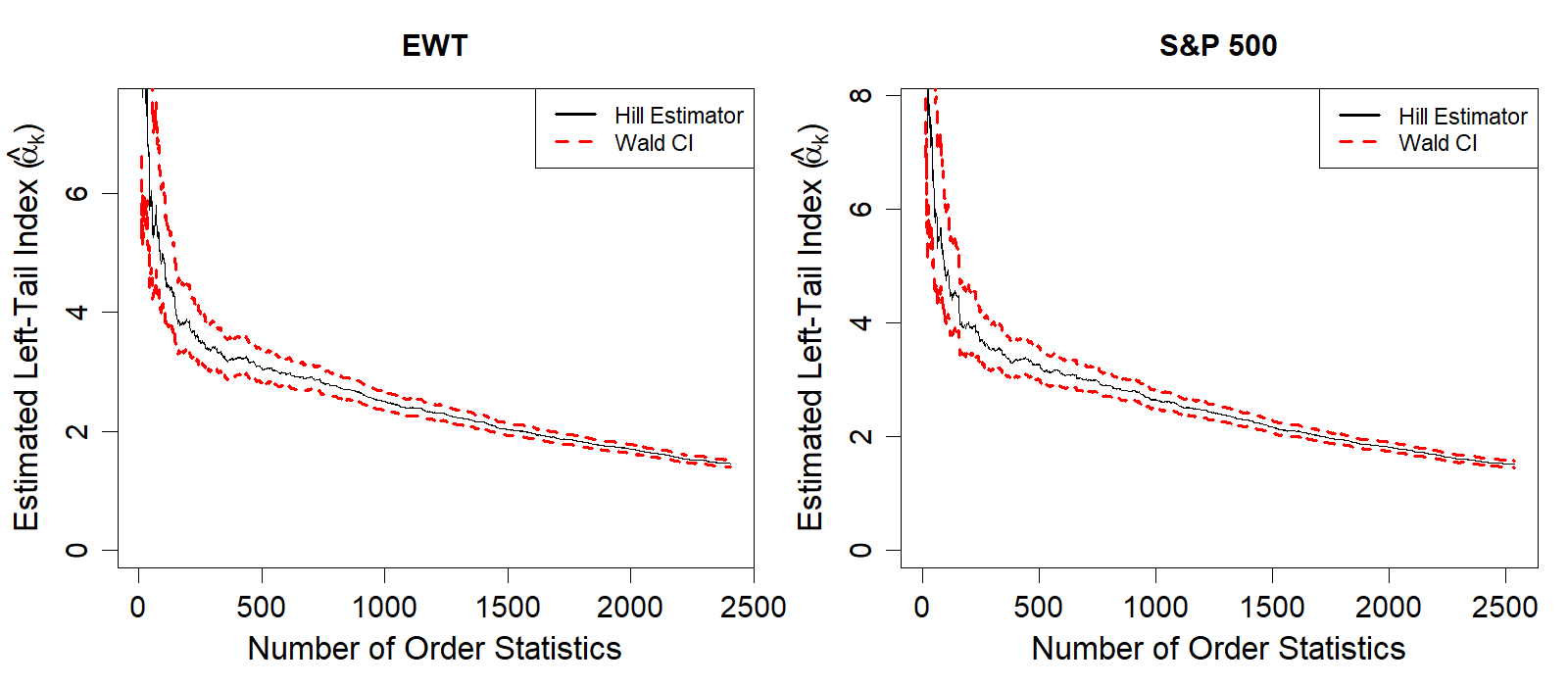}
    \caption{Left-tail Hill estimates ($\widehat{\alpha}_k$) with 95\% Wald confidence bands for (a) EWP, (b) DJIA, (c) EWT, and (d) S\&P~500.}
   % ,based on the transformed loss series $L_t=-R_t$.}
    \label{fig:left-tail}
\end{figure}

\paragraph{Benchmark comparisons:}

Figure~\ref{fig:left-tail} presents the empirical Hill estimates of the left tail (see for right tail in appendix \ref{sec:right-tail-hill}) for the benchmark comparisons considered in this study. The upper panels compare the equally weighted portfolio (EWP) with the DJIA, while the lower panels compare the Taiwan ETF (EWT) with the S\&P~500 Index. The performance of EWP and DJIA were compared since both of them represent diversified equity portfolios where EWT and S\&P~500 provide a comparison between concentrated Taiwan-focused investment vehicle and broad U.S. market benchmark. The comparisons therefore allow us to assess whether the tail-risk characteristics observed in the ETF-based portfolios differ materially from those of widely followed benchmark indices.

All four series exhibit substantial variation for small values of \(k\), followed by smoother behavior as additional order statistics are included. The interpretation focused on an intermediate range of \(k\), approximately \(k\in[500,1500]\), where \(\widehat{\alpha}_k\) exhibited relatively stable behavior. Within this region, the estimated tail indices for EWP, DJIA, EWT, and the S\&P~500 generally fell within the range \(\widehat{\alpha}_k\approx2.5\text{--}3.5\). These estimates indicated heavy-tailed behavior in the left tails of all four return series. Since the estimated tail indices remained above $2$, the results suggested that the return distributions retained finite variance while still deviating from Gaussian tail behavior. The estimated tail indices also imply substantially heavier tails than those associated with normal distributions, indicating that extreme losses occur more frequently than would be predicted under normality assumptions. This finding supports the use of tail-sensitive risk measures in the subsequent portfolio analysis.

\paragraph{EWP versus DJIA:}
Figure~\ref{fig:left-tail} (upper panels) compares the left-tail Hill estimates for EWP and DJIA. Both series produced similar tail-index estimates over the stable region, although EWP occasionally exhibited slightly lower values than DJIA. The EWP also exhibited somewhat wider confidence bands for smaller values of \(k\), indicating greater estimation uncertainty in the most extreme observations. Since smaller values of the tail index correspond to heavier left tails, the results suggested broadly similar downside tail behavior for the two portfolios.
The equal-weighted structure of EWP places relatively greater weight on smaller and potentially more volatile holdings, which may contribute to the wider confidence bands observed in the extreme region.

\paragraph{EWT versus S\&P~500:}

Figure~\ref{fig:left-tail} (lower panels) compares the left-tail Hill estimates for EWT and the S\&P~500 Index. The estimated tail indices again remained within a similar range, although EWT occasionally produced slightly lower estimates and somewhat wider confidence bands than the S\&P~500.
The results suggested broadly similar downside tail behavior between EWT and the S\&P~500. Compared with the broader and more diversified S\&P~500 benchmark, EWT concentrates exposure within Taiwan-related equities, which may contribute to the somewhat greater variability observed in the tail-index estimates.

\medskip

Taken together, the Hill estimates suggested that the downside tails of EWP and EWT were broadly comparable to those of the DJIA and S\&P 500. Although EWP and EWT occasionally exhibited slightly lower tail-index estimates and wider confidence intervals, the confidence bands largely overlapped with those of the corresponding benchmarks, suggesting no substantial differences in asymptotic tail-decay behavior. This result is noteworthy because it suggests that the elevated downside risk observed in the ETF-level VaR and CVaR analysis did not necessarily translate into substantially heavier asymptotic tails at the benchmark portfolio level.

The benchmark comparisons in Figure~\ref{fig:left-tail} suggest broadly similar asymptotic tail-decay behavior across the ETF universe, yet the VaR and CVaR estimates reported in Section~\ref{sec:var_cvar} reveal substantial cross-sectional differences in extreme downside risk. These two findings are not contradictory. The tail index governs the rate at which the tail of the return distribution decays asymptotically, whereas quantile-based risk measures such as VaR and CVaR depend jointly on the tail index and the scale of the return distribution. Two series can share a similar tail index while exhibiting materially different CVaR estimates if their return volatilities differ substantially. Semiconductor-focused ETFs such as SMH and SOXX have considerably higher return volatility than diversified international ETFs, which shifts the entire loss distribution to the right and produces larger quantile-based risk estimates at any given confidence level, even when the asymptotic tail-decay rate is similar. The cross-sectional variation in CVaR documented in Section~\ref{sec:var_cvar} therefore reflects primarily differences in return scale rather than differences in tail thickness, a distinction with direct implications for portfolio construction and risk budgeting.

\subsection{Long Memory and Asymmetric Volatility Modeling}\label{subsec:long_memory}

While the Hill estimator provides evidence on the extremal behavior of return distributions, tail behavior alone does not fully characterize financial risk. Financial return series often exhibit volatility clustering, persistent dependence in volatility measures, and asymmetric responses to market shocks. These features are particularly important when analyzing Taiwan-related assets and semiconductor-related sectors, where common market factors may influence return behavior over extended periods.

It is well established that raw financial returns typically exhibit limited serial dependence, whereas transformations such as squared and absolute returns frequently display strong and persistent autocorrelation patterns \citep{ding1993long,cont2001empirical}. This phenomenon is commonly referred to as volatility clustering and is often interpreted as evidence of long-range dependence in volatility measures.

An important econometric issue is distinguishing genuine long-memory behavior from persistence generated by conditional heteroskedasticity or other volatility structures. Long memory is characterized by slowly decaying dependence over long horizons and is often associated with hyperbolic rather than exponential decay in autocorrelations. However, highly persistent volatility dynamics may resemble long-memory behavior in finite samples even when true fractional integration is absent. Apparent long memory may arise from mechanisms other than genuine fractional dependence, including aggregation effects, structural changes, and multi-component volatility dynamics, all of which can generate persistence patterns resembling long-memory behavior \citep{granger1980,diebold2001long,corsi2009}.

Standard GARCH models capture volatility persistence through short-memory dynamics, where the effects of shocks decay at an exponential rate \citep{bollerslev1986generalized}. To accommodate more persistent dependence structures, fractionally integrated models such as FIGARCH were developed to allow for long-range dependence in conditional variance dynamics \citep{baillie1996fractionally}.

Empirical evidence has also consistently demonstrated asymmetric volatility responses, where negative shocks tend to have larger effects on future volatility than positive shocks of similar magnitude. This phenomenon, commonly referred to as the leverage effect, was incorporated into the GJR--GARCH framework of \citet{glosten1993relation}, which allows positive and negative innovations to have different effects on conditional variance.

To investigate these issues, we examined volatility persistence and asymmetry for both the equally weighted portfolio (EWP) and the Taiwan ETF (EWT), allowing a comparison between diversified portfolio exposure and concentrated Taiwan-related market exposure.

To investigate long-memory volatility persistence in the portfolios considered, three complementary approaches were employed: the Geweke--Porter--Hudak (GPH) estimator \citep{geweke1983estimation}, the Local Whittle estimator \citep{robinson1995gaussian}, and the Hurst exponent.

The GPH estimator provides a semi-parametric estimate of long-range dependence based on the low-frequency behavior of the periodogram, whereas the Local Whittle estimator provides an alternative frequency-domain approach with desirable asymptotic properties under relatively weak assumptions. Both estimators were implemented using the low-frequency ordinates of the periodogram, as long-memory behavior is primarily reflected in the low-frequency structure of the series. The bandwidth parameter was selected according to

\[
m=n^{0.5},
\]

where \(n\) denotes the sample size and \(m\) represents the number of low-frequency ordinates retained in the estimation. For the sample considered in this study, this choice yielded approximately \(m=40\). This rule follows common practice in semi-parametric long-memory estimation and balances the trade-off between estimation variance and bias. Under standard regularity conditions, both estimators are consistent and asymptotically normal.

\begin{table}[ht]
\centering
\caption{Long-memory estimates for EWP and EWT.}
\begin{tabular}{lccc}
\hline
Benchmark & GPH Estimate ($d$) & Local Whittle ($d$) & Hurst Exponent ($H$) \\
\hline
EWP & $0.187$ ($p<0.001$) & $0.235$ & $0.650$ \\
EWT & $0.163$ ($p=0.093$) & $0.073$ & $0.544$ \\
\hline
\end{tabular}
\label{tab:longmemory}
\end{table}

The initial results from above methods provided evidence of long-memory-type behavior in the squared returns of the portfolios, particularly for EWP. Table~\ref{tab:longmemory} reports the GPH, Local Whittle, and Hurst estimates for the selected portfolio return series. For EWP, all three measures provide evidence of stationary long-memory and persistent dependence. In particular, the estimated fractional differencing parameters lie within the stationary long-memory region $(0<d<0.5)$, while the Hurst exponent exceeds $0.5$, indicating persistence. Similar but weaker evidence is observed for EWT, where the estimated degree of dependence is smaller across all three measures. Overall, the results suggest stronger long-memory characteristics for EWP than for EWT.

Given the evidence of long-memory obtained from the GPH, Local Whittle, and Hurst exponent analyses, we further examine the robustness of these findings using a FIGARCH(1,$d$,1) model estimated over a four-year rolling window for both portfolios. 

\begin{figure}[H]
\centering
\begin{subfigure}[b]{0.49\textwidth}
\centering
\includegraphics[width=\textwidth]{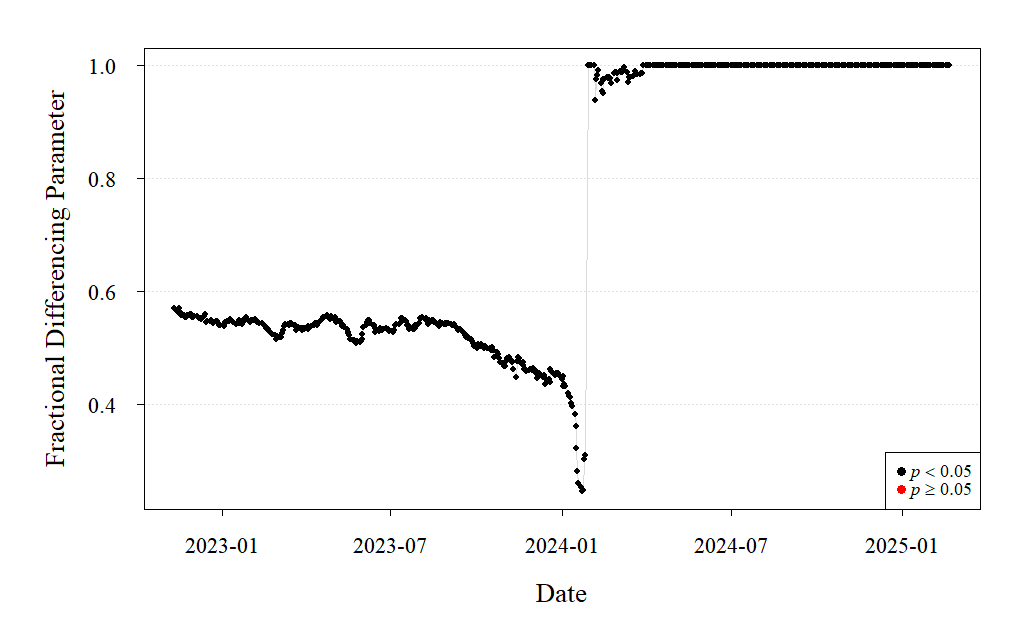}
\caption{EWP}
\label{fig:rollingfigarch_EWP}
\end{subfigure}
\hfill
\begin{subfigure}[b]{0.49\textwidth}
\centering
\includegraphics[width=\textwidth]{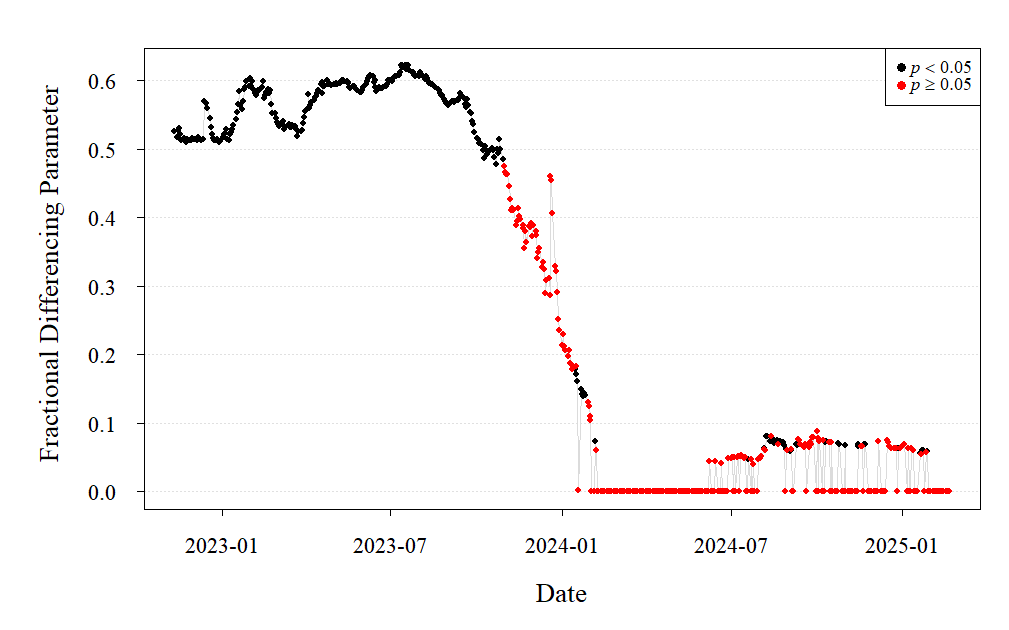}
\caption{EWT}
\label{fig:rollingfigarch_EWT}
\end{subfigure}
\caption{Rolling estimates of the FIGARCH(1,d,1) fractional differencing parameter under a four-year rolling window for (a) EWP and (b) EWT.
Red points denote estimates that are not statistically significant at the 5\% level.}
\label{fig:rollingfigarch}
\end{figure}

Figure~\ref{fig:rollingfigarch} presents the time-varying estimates of the fractional differencing parameter $d$ for the portfolio returns. For EWP, the estimated values of $d$ remain close to one for a significant portion of the sample period. Such values indicate extremely persistent volatility dynamics, placing the process near the boundary between long-memory and nonstationary behavior, where the effects of volatility shocks dissipate very slowly. Consequently, these estimates should be interpreted with caution, as values near unity may also reflect model instability, structural changes, or finite-sample estimation effects rather than genuine long-memory dependence. In contrast, for EWT, a considerable number of the estimated $d$ values are statistically insignificant, providing weaker evidence of long-memory behavior. Overall, while the earlier GPH, Local Whittle, and Hurst estimates suggest the presence of long-memory, the rolling FIGARCH results provide only mixed support and do not conclusively establish stable long-memory dependence across the entire sample period.

Next to observe more intuitively, we filter the return series using a GJR--GARCH$(1,1)$ model with Student-$t$ innovations, which captures volatility clustering, asymmetric responses, and heavy-tailed return distributions. The resulting GJR--GARCH specification can be written as

\begin{equation*}
\sigma_t^2
=
\omega
+
\alpha \epsilon_{t-1}^2
+
\gamma I_{\{\epsilon_{t-1}<0\}}\epsilon_{t-1}^2
+
\beta \sigma_{t-1}^2,
\end{equation*}

where $I_{\{\epsilon_{t-1}<0\}}$ is an indicator function equal to one when the shock is negative, allowing negative shocks to have a different impact on volatility. For the GJR--GARCH$(1,1)$ model, volatility persistence is quantified by
\[
\alpha+\beta+\gamma P(\epsilon_{t-1}<0),
\]
which simplifies to
\[
\alpha+\beta+\frac{\gamma}{2},
\]
under symmetric innovations. Values closer to one indicate stronger persistence and slower dissipation of volatility shocks. After filtering the return series with a GJR--GARCH(1,1) model, we re-estimate the fractional differencing parameter using the GPH, Local Whittle, and Hurst exponent methods. The post-filtering results provide additional insight into the source of the apparent long-memory observed in the raw data. 

\begin{table}[H]
\centering
\caption{Long-range dependence measures for squared standardized residuals after GJR--GARCH filtering.}
\label{tab:lrd}
\begin{tabular}{lccc}
\hline
Benchmark & GPH Estimate ($d$) & Local Whittle ($d$) & Hurst Exponent ($H$) \\
\hline
EWP & $-0.019$ $(p=0.828)$ & $-0.001$ & $0.557$ \\
EWT & $0.070$ $(p=0.548)$ & $-0.009$ & $0.501$ \\
\hline
\end{tabular}
\end{table}
As reported in Table~\ref{tab:lrd}, the estimated values of the fractional differencing parameter are close to zero and are no longer statistically significant for both EWP and EWT. Moreover, the Hurst exponents are close to 0.5, indicating little evidence of persistent long-range dependence. These findings suggest that the long-memory detected in the unfiltered series is largely attributable to volatility clustering and conditional heteroskedasticity, which are effectively captured by the GJR--GARCH(1,1) model.

\begin{figure}[H]
\centering
\includegraphics[width=0.82\textwidth]{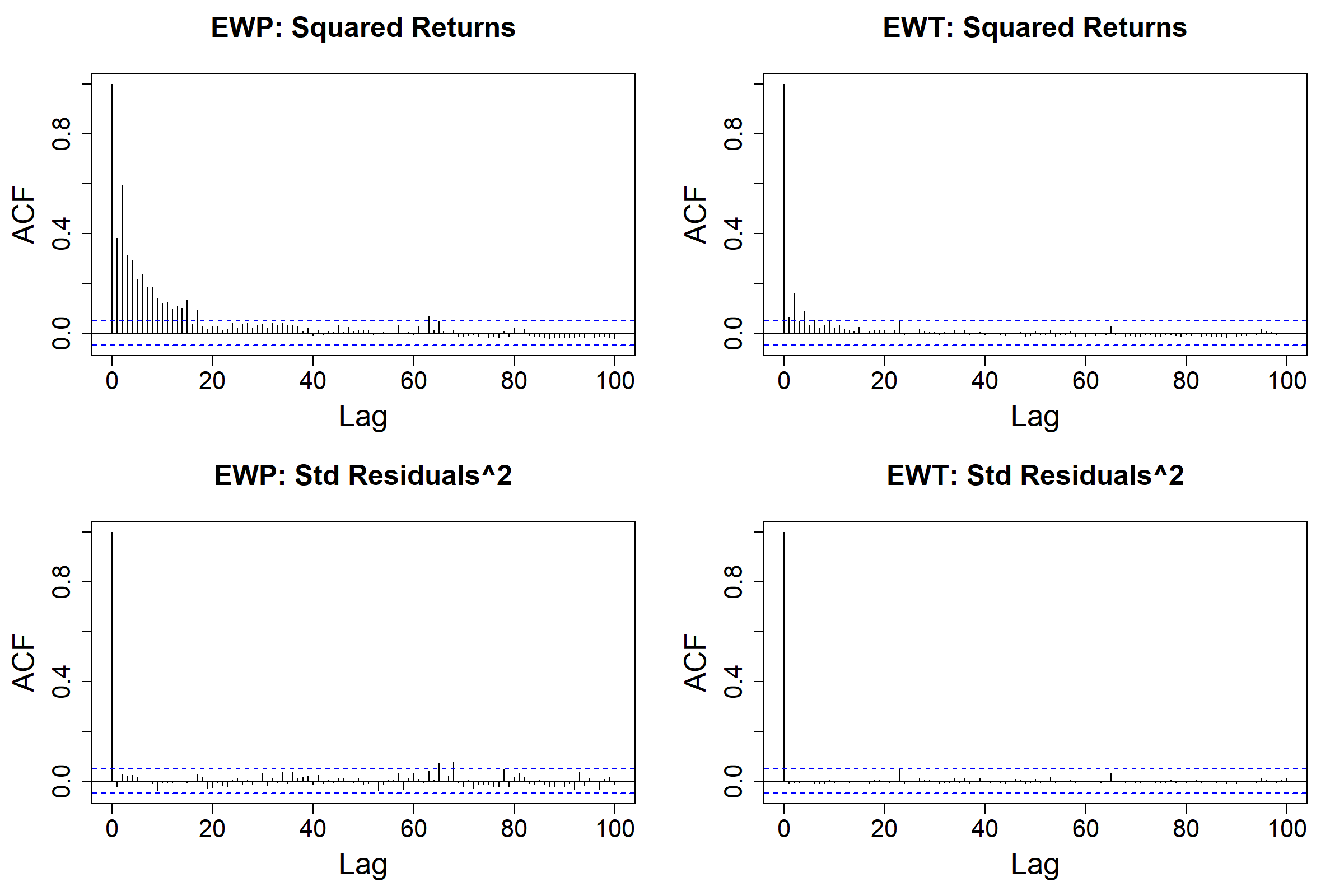}
\caption{Autocorrelation functions of squared returns (top panels) and squared standardized residuals (bottom panels) for (EWP, left) and the Taiwan ETF (EWT, right).}
\label{fig:acf_compare}
\end{figure}

Figure~\ref{fig:acf_compare} provides visual support for these findings. The autocorrelation functions of squared returns for both EWP and EWT decay slowly, indicating persistent volatility clustering, with the effect appearing more pronounced for EWP. In contrast, the autocorrelations of squared standardized residuals are largely insignificant across lags, with most values falling within the confidence bounds after GJR--GARCH filtering.
Consistent with the results in Table~\ref{tab:lrd}, the estimated long-memory measures for the filtered series are close to zero, while the Hurst exponent approaches 0.5, indicating no evidence of long-range dependence.

\begin{table}[H]
\centering
\caption{Estimated GJR--GARCH$(1,1)$-$t$ Parameters and Diagnostic Statistics}
\label{tab:gjr_results}

\begin{tabular}{lcccc}
\hline
Statistic & EWP & p-value & EWT & p-value\\
\hline

$\mu$
&0.00039
&0.070
&0.00101
&0.0002\\

$\omega$
&0.000004
&0.001
&0.000017
&0.000006\\

$\alpha$
&0.0157
&0.086
&0.0810
&0.030\\

$\beta$
&0.8857
&$<0.001$
&0.7837
&$<0.001$\\

$\gamma$
&0.1175
&$<0.001$
&0.0746
&0.098\\

$\nu$
&6.994
&$<0.001$
&4.738
&$<0.001$\\

Persistence 
$\left(\alpha+\beta+\gamma/2\right)$
&0.960
&--
&0.902
&--\\

\hline
\end{tabular}
\end{table}

Table~\ref{tab:gjr_results} reports the estimated GJR--GARCH$(1,1)$-$t$ parameters and diagnostic statistics for EWP and EWT. Both series exhibited substantial volatility persistence, with estimated persistence measures of approximately \(0.96\) for EWP and \(0.90\) for EWT, indicating that volatility shocks dissipated slowly over time. The estimated asymmetry parameters were positive for both series, suggesting the presence of asymmetric volatility responses. The asymmetry effect was stronger and statistically significant for EWP (\(\gamma\approx0.118\)), whereas the estimated asymmetry parameter for EWT (\(\gamma\approx0.075\)) was smaller and not statistically significant at conventional levels. The estimated Student-$t$ degrees-of-freedom parameter was lower for EWT (\(\nu\approx4.74\)) than for EWP (\(\nu\approx6.99\)), indicating heavier-tailed return behavior for the Taiwan ETF. \footnote{Appendix Figure~\ref{fig:rolling_gjr_EWP} reports rolling estimates of the asymmetry parameter in the GJR--GARCH$(1,1)$-$t$ model. The estimated asymmetry parameter remained positive throughout the sample but exhibited noticeable variation over time, suggesting that asymmetric volatility effects were not constant across periods.}

Taken together, these results suggested that the apparent long-memory behavior observed in squared returns was largely attributable to volatility clustering and asymmetric volatility dynamics rather than genuine fractional integration. From a practical perspective, this finding implies that standard asymmetric GARCH-type models may adequately capture the persistence present in ETF volatility without requiring more complex fractionally integrated specifications. Consequently, volatility forecasting and risk management for Taiwan-related ETFs may benefit more from modeling asymmetric volatility responses than from incorporating long-memory structures.

The present analysis focused primarily on univariate volatility dynamics. Dynamic correlations and tail dependence among ETFs were not examined and may provide additional information regarding common risk exposures and co-movement patterns. Extending the analysis to multivariate frameworks such as Dynamic Conditional Correlation GARCH (DCC--GARCH) models or tail-dependence measures may provide further insight into the dependence structure of Taiwan-related assets.

%%%%%%%%%%%%%%%%%%%%%%%%%%%%%%%%%%%%%%%%%%%%%%
%%%%%  

\section{Historical Portfolio Optimization}
\label{sec:historical_optimization}

The previous sections demonstrated that Taiwan-related ETFs exhibit asymmetric returns, heavy tails, volatility clustering, and persistent conditional heteroskedasticity. These characteristics suggest that traditional variance-based portfolio optimization may provide only a partial representation of investment risk. Accordingly, this section examines portfolio allocation under both variance-based and tail-sensitive optimization frameworks. Specifically, we compare the classical mean--variance framework of \citet{markowitz1952} with Conditional Value-at-Risk (CVaR) portfolio optimization using rolling-window estimation. The analysis considers both long-only and long--short investment constraints and evaluates how diversification, downside risk, concentration, and estimation uncertainty influence portfolio allocations over time. Particular attention is given to the role of semiconductor concentration and Taiwan exposure in shaping optimal portfolios and efficient frontiers.

Section~\ref{sec:optimization_methodology} presents the estimation methodology. Sections~\ref{sec:mv_optimization} and \ref{sec:cvar_optimization} compare the mean--variance and CVaR efficient frontiers, respectively. Section~\ref{sec:price_return} evaluates the realized performance of the optimized portfolios under long-only and long--short constraints, while Appendix~\ref{app:robust} presents a subsample robustness analysis.

\subsection{Overview and Methodology}\label{sec:optimization_methodology}

This section describes the portfolio optimization frameworks and estimation procedures employed in the empirical analysis. Portfolio weights are estimated dynamically under both the classical mean--variance framework of \citet{markowitz1952} and a CVaR-based optimization framework using a four-year (1,008 trading day) rolling window of historical returns. The estimation window advances one trading day at a time, and portfolio weights are re-optimized and rebalanced daily using the trailing 1,008-day return history so that $\mu_t$ and $\Sigma_t$ are updated continuously rather than at fixed calendar intervals. This rolling-window approach allows model parameters to vary over time and facilitates out-of-sample portfolio evaluation under changing market conditions \citep{lindquist2022advanced}. Similar rolling-window procedures are commonly employed in empirical portfolio optimization and ETF studies \citep{divelgama2026}.

The four-year estimation window follows the approach of \citet{lindquist2022advanced} and \citet{divelgama2026} and is consistent with common practice in empirical portfolio studies. This window length provides approximately 1,000 daily observations per estimation period, which is sufficient for estimating the covariance matrix of the 30-ETF universe with reasonable precision. Expected return estimates remain sensitive to sample variation as in any finite-sample setting. The window length also maintains responsiveness to structural changes in market conditions, including the COVID-19 market disruption, semiconductor supply-chain shocks, the subsequent AI-driven technology expansion, and periods of elevated market volatility.

Let $\mu_t$ denote the vector of sample mean returns and let $\Sigma_t$ denote the corresponding covariance matrix estimated from the rolling window ending at time $t$. The benchmark variance-based allocation is defined as the global minimum-variance portfolio

\[
\min_{w}\; w^\top \Sigma_t w
\qquad
\text{subject to}
\qquad
w^\top \mathbf{1}=1.
\]

Additional constraints depend on the investment strategy. For long-only portfolios, the constraints require

\[
w_i \geq 0,
\qquad i=1,\ldots,N.
\]

For long--short portfolios, portfolio weights satisfy

\[
-0.3 \leq w_i \leq 1.3,
\qquad i=1,\ldots,N.
\]

These bounds follow \citet{lindquist2022advanced} and reflect realistic institutional constraints that permit limited short exposure while preventing extreme leveraged positions that would be impractical for most investors.

In addition to mean--variance optimization, CVaR-based portfolios are evaluated at the 95\% and 99\% confidence levels using the scenario-based framework of \citet{rockafellar2000optimization}. Two specifications are considered. The first is a minimum-CVaR portfolio, formulated following \citet{rockafellar2000optimization,rockafellar2002conditional} as a tractable linear programming problem. Introducing an auxiliary variable $\gamma$ interpreted as VaR, the discrete CVaR minimization problem is written as

\[
\min_{\mathbf{w},\gamma}
\left\{
\gamma
+
\frac{1}{\alpha T}
\sum_{t=1}^{T}
\max\bigl(0,-r_t^\top\mathbf{w}-\gamma\bigr)
\right\},
\]

where $\{r_t\}_{t=1}^{T}$ denotes historical return scenarios, $\mathbf{w}$ denotes portfolio weights, and $\alpha$ denotes the tail probability. This formulation incorporates the empirical return distribution directly and avoids imposing normality assumptions.

The second specification is a CVaR-efficient tangent portfolio that maximizes the STARR ratio \citep{rachev2008advanced},

\[
\mathrm{STARR}_{\alpha}
=
\frac{E[R_p]-R_f}
{\mathrm{CVaR}_{\alpha}(R_p)},
\]

where $R_p$ denotes the portfolio return, $R_f$ denotes the risk-free rate, and $\mathrm{CVaR}_{\alpha}(R_p)$ denotes the Conditional Value-at-Risk of the portfolio return at confidence level $\alpha$. Unlike variance-based approaches, CVaR optimization places greater emphasis on downside risk and is therefore particularly relevant for portfolios exhibiting asymmetric or heavy-tailed return distributions.

The rolling estimation procedure generates a sequence of time-varying portfolio weights $\{w_t\}$, allowing portfolio allocations to adjust as market conditions evolve. This feature is particularly relevant for Taiwan-related ETFs, whose return dynamics may be influenced by changing semiconductor-sector conditions, macroeconomic developments, and geopolitical events.

Throughout the analysis, the equally weighted portfolio (EWP) serves as the benchmark portfolio. Unlike optimized portfolios, the EWP assigns an equal weight of $1/N$ to each of the $N$ assets and does not rely on estimated parameters. Consequently, it is less sensitive to estimation error arising from expected-return and covariance estimates and provides a simple and robust benchmark. \citet{demiguel2009optimal} showed that equally weighted strategies often perform competitively relative to optimized portfolios in finite samples, a finding that motivates the use of the EWP as the primary benchmark throughout the analysis.

Additional efficient-frontier comparisons are presented in Sections~\ref{sec:mv_optimization} and \ref{sec:cvar_optimization}, where SMH and SOXX are plotted alongside the optimized portfolios to assess the role of concentrated semiconductor exposure under both mean--variance and CVaR criteria.

The following subsections present the mean--variance and CVaR efficient frontiers and evaluate the realized performance of the resulting portfolios under rolling-window estimation.

\subsubsection{Mean--Variance Optimization}
\label{sec:mv_optimization}

The Markowitz mean--variance framework minimizes portfolio variance subject to a target expected return and a full-investment constraint, generating a set of efficient portfolios whose risk--return combinations form the efficient frontier in standard deviation--return space \citep{markowitz1952,lindquist2022advanced}. When a risk-free asset is introduced, the efficient set becomes the capital market line (CML), tangent to the frontier at the portfolio that maximizes the Sharpe ratio and represents the optimal risky allocation under the mean--variance criterion.

\begin{figure}[h!]
\centering
\includegraphics[width=0.8\textwidth]{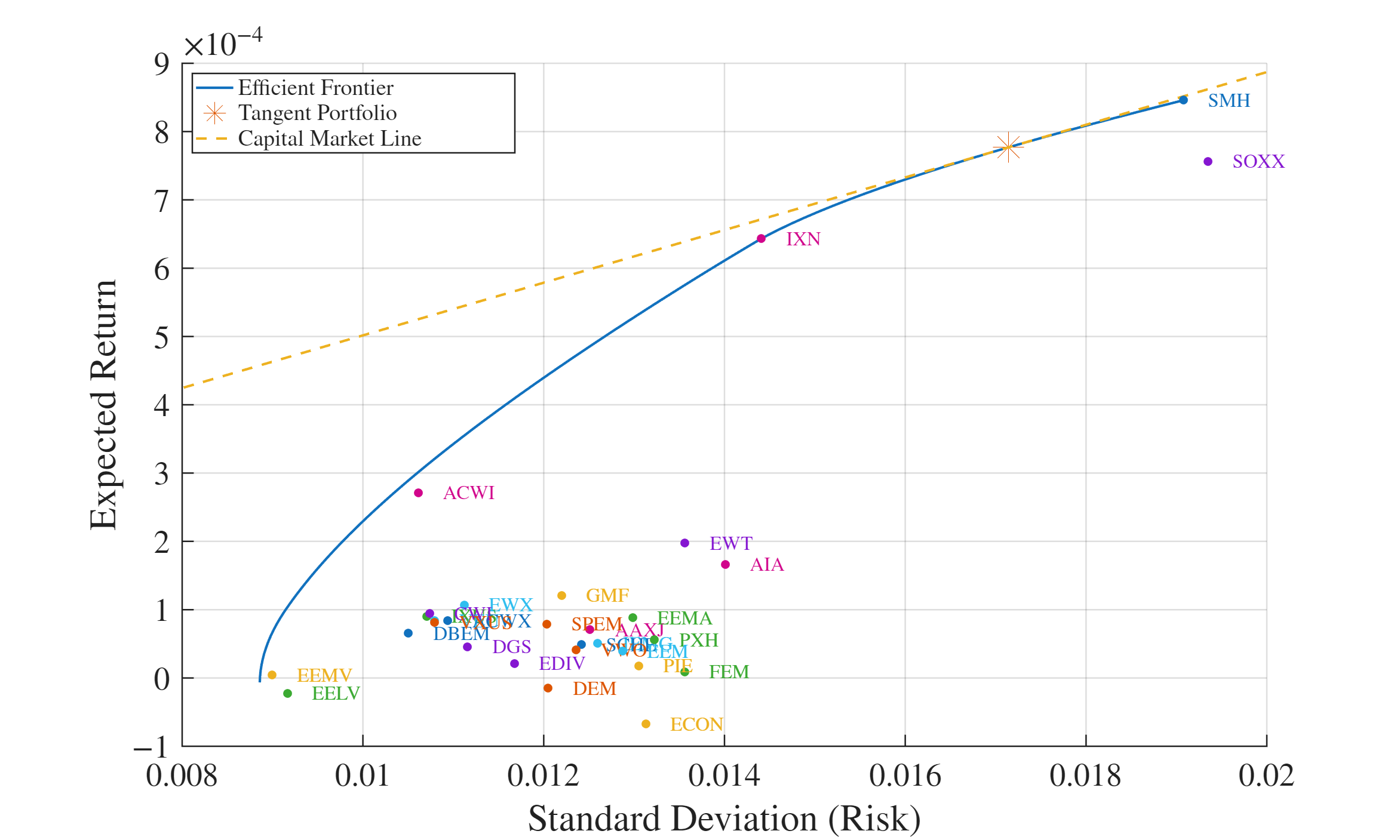}
\caption{Markowitz efficient frontier using the 3-month U.S. Treasury yield as the risk-free rate.}
\label{fig:MV_US}
\end{figure}

Figure~\ref{fig:MV_US} presents the mean--variance efficient frontier for the 30-ETF universe. A prominent feature of the figure is the wide dispersion of ETFs in risk--return space, reflecting substantial heterogeneity in sector concentration and geographic exposure across the sample. Broadly diversified ETFs such as ACWI, VEU, VXUS, and ACWX clustered in the lower-left region, exhibiting relatively low volatility and moderate expected returns consistent with broad geographic diversification. SMH occupied the upper-right region with the highest expected return in the sample and, relative to SOXX, lay closer to the efficient frontier, indicating a more favorable risk--return trade-off under the mean--variance criterion. SOXX itself lay below the frontier, exhibiting high volatility relative to its expected return. Technology-oriented ETFs such as IXN occupied an intermediate position, reflecting the strong performance of the global technology sector and its partial overlap with semiconductor-sector dynamics.

Despite the strong standalone performance of semiconductor ETFs, the tangent portfolio was not concentrated in a single asset. The dashed capital market line, constructed using the 3-month U.S.\ Treasury yield as the risk-free rate, was tangent to the frontier at a point well below SMH in risk--return space, indicating that the mean--variance optimal allocation diversified across multiple ETFs rather than concentrating in the highest-return asset. This finding was consistent with \citet{markowitz1952}, who argued that combining imperfectly correlated assets reduced portfolio variance below that of individual holdings. This diversified outcome contrasted sharply with the CVaR framework examined next, where tail-risk-based optimization produced a substantially more concentrated portfolio, underscoring how the choice of risk measure could materially alter portfolio composition even when the investment universe remained unchanged.

\subsubsection{CVaR Optimization}
\label{sec:cvar_optimization}

As defined in Section~\ref{sec:var_cvar}, CVaR measures the expected loss conditional on losses exceeding the VaR threshold and satisfies the coherence axioms of \citet{artzner1999coherent}. The CVaR minimization problem and STARR ratio are formally presented in Section~\ref{sec:optimization_methodology}. The analysis below examines the resulting CVaR efficient frontiers and portfolio allocations.

\begin{figure}[h!]
\centering
\includegraphics[width=0.495\textwidth]{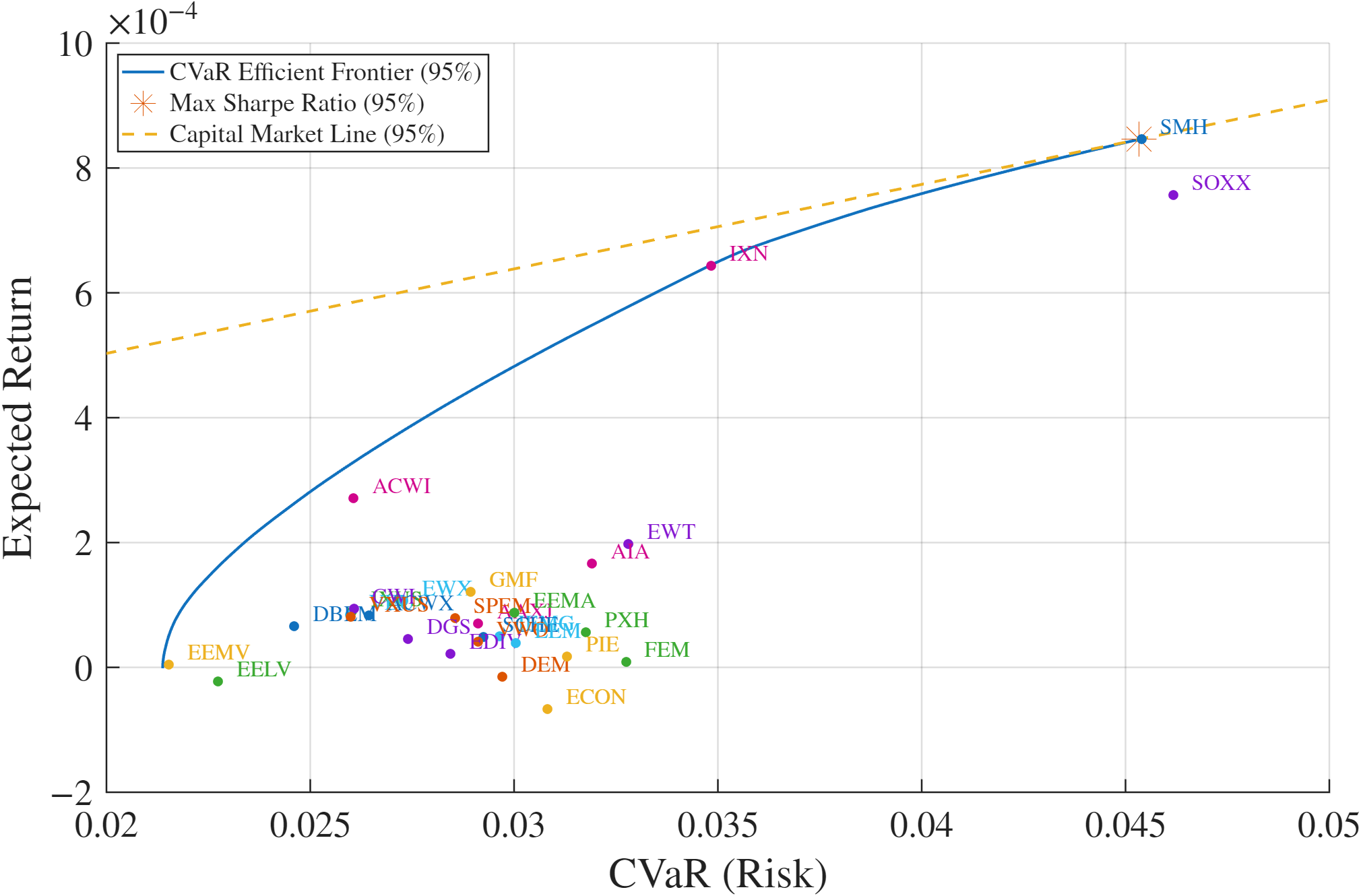}
\includegraphics[width=0.495\textwidth]{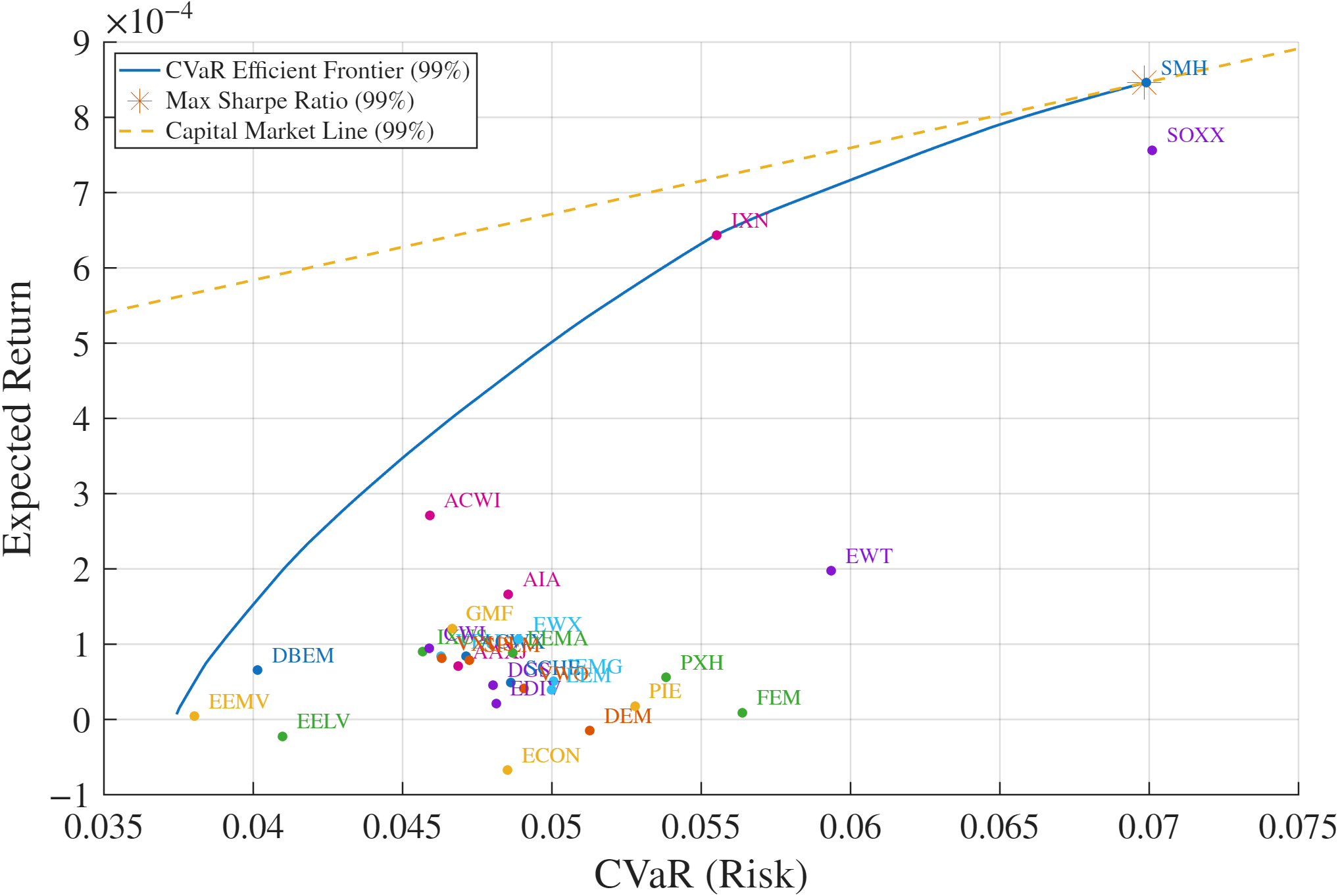}
\caption{CVaR-efficient frontiers at the 95\% and 99\% confidence levels using 3-month U.S.\ Treasury yields as risk-free rates. The tangent portfolio maximizes the STARR ratio, defined as the ratio of excess expected return to CVaR.}
\label{fig:CVaR_US}
\end{figure}

Figure~\ref{fig:CVaR_US} reports the CVaR efficient frontiers at the 95\% and 99\% confidence levels together with the corresponding capital market lines. Individual ETFs are plotted in mean--CVaR space using their sample mean returns and asset-level CVaR estimates computed at the same confidence level. The risk-free rate is proxied by the 3-month U.S.\ Treasury yield and converted to a daily rate to match the frequency of the return data \citep{lindquist2022advanced}.

The 99\% frontier lies to the right of the 95\% frontier, reflecting the larger tail losses captured at the higher confidence level. Despite this shift, the overall shape of the frontier remains similar across the two confidence levels. Semiconductor-focused ETFs occupy the upper-right region of both panels. SMH exhibits the highest expected return in the sample, and the capital market line passes very close to SMH at both confidence levels, indicating that SMH exhibits a more favorable trade-off between expected return and tail risk than most other ETFs in the sample. In contrast, SOXX exhibits comparable tail risk but lower expected return and lies below the efficient frontier. Technology-oriented ETFs such as IXN also appear near the upper portion of the frontier, reflecting their exposure to the broader technology sector and partial overlap with semiconductor-sector dynamics.

The tangent portfolio lies very close to SMH at both confidence levels. Relative to the other ETFs in the sample, SMH combines unusually high expected returns with a level of tail risk that remains favorable under the STARR criterion. Consequently, the CVaR-optimal allocation places substantially greater weight on SMH than does the corresponding mean--variance allocation. This outcome differs markedly from the diversified tangent portfolio observed in Figure~\ref{fig:MV_US} and indicates that the choice of risk measure can lead to materially different portfolio compositions even when the efficient frontiers appear broadly similar in shape.

This result should be interpreted within the context of the sample period. The post-COVID sample included a period of unusually strong semiconductor-sector performance associated with rapid growth in AI-related investment and demand, and the prominence of SMH in the CVaR optimization may partly reflect sample-specific conditions rather than a persistent structural feature of semiconductor ETFs. The subsample analysis reported in Appendix~\ref{app:robust} supports this interpretation. In the post-COVID subsample, the CVaR tangent portfolio lies substantially closer to SMH than in the pre-COVID period, indicating a greater influence of semiconductor-sector performance on the CVaR-efficient set. This pattern is consistent with increased concentration in semiconductor exposure during the AI-driven expansion.

Overall, the CVaR results differ substantially from the mean--variance results. Whereas mean--variance optimization produces a diversified tangent portfolio, the CVaR framework places much greater emphasis on assets with favorable expected-return-to-tail-risk characteristics. In the present sample, this leads to substantially greater concentration in semiconductor exposure, with SMH emerging as the dominant holding in the CVaR-optimal portfolio at both the 95\% and 99\% confidence levels. The contrast with the mean--variance results demonstrates that alternative measures of risk can lead to substantially different portfolio allocations within the same investment universe.
At the same time, the two frameworks' efficient frontiers remain broadly similar in shape, so the divergence documented above is a difference in portfolio \emph{composition} rather than in the underlying risk--return structure of the ETF universe---a distinction that motivates the performance evaluation in Section~\ref{sec:portfolio_performance}.

\subsection{Basic Portfolio Strategies: Price and Return Performance}\label{sec:price_return}

To evaluate the practical implications of the portfolio optimization frameworks developed in the previous sections, we compared the realized performance of three portfolio construction approaches based on Taiwan-exposed ETFs: the equally weighted portfolio (EWP), the minimum-variance portfolio (MVP), and portfolios constructed under conditional value-at-risk (CVaR) constraints. These approaches differed primarily in their definition of risk and therefore generated different portfolio allocations and return characteristics.

The MVP followed the mean--variance framework of \citet{markowitz1952}, whereas the CVaR approach focused on downside tail risk \citep{rockafellar2000optimization}. The equally weighted portfolio served as a simple diversification benchmark and has been shown to perform competitively in the presence of estimation error \citep{demiguel2009optimal}.

Portfolio performance was evaluated under both long-only and long--short constraints using rolling-window estimation. Particular attention was given to the interaction between diversification, concentration risk, semiconductor exposure, and downside protection.
The empirical results presented below illustrate how alternative definitions of risk can lead to materially different portfolio compositions and realized performance outcomes.

\subsubsection{Long--Only Portfolios}

We first examine portfolio strategies under the long-only constraint described in Section~\ref{sec:optimization_methodology}. This specification prohibits short selling and serves as the baseline portfolio construction framework.

\begin{figure}[htb]
	\centering
	\includegraphics[width=0.49\textwidth]{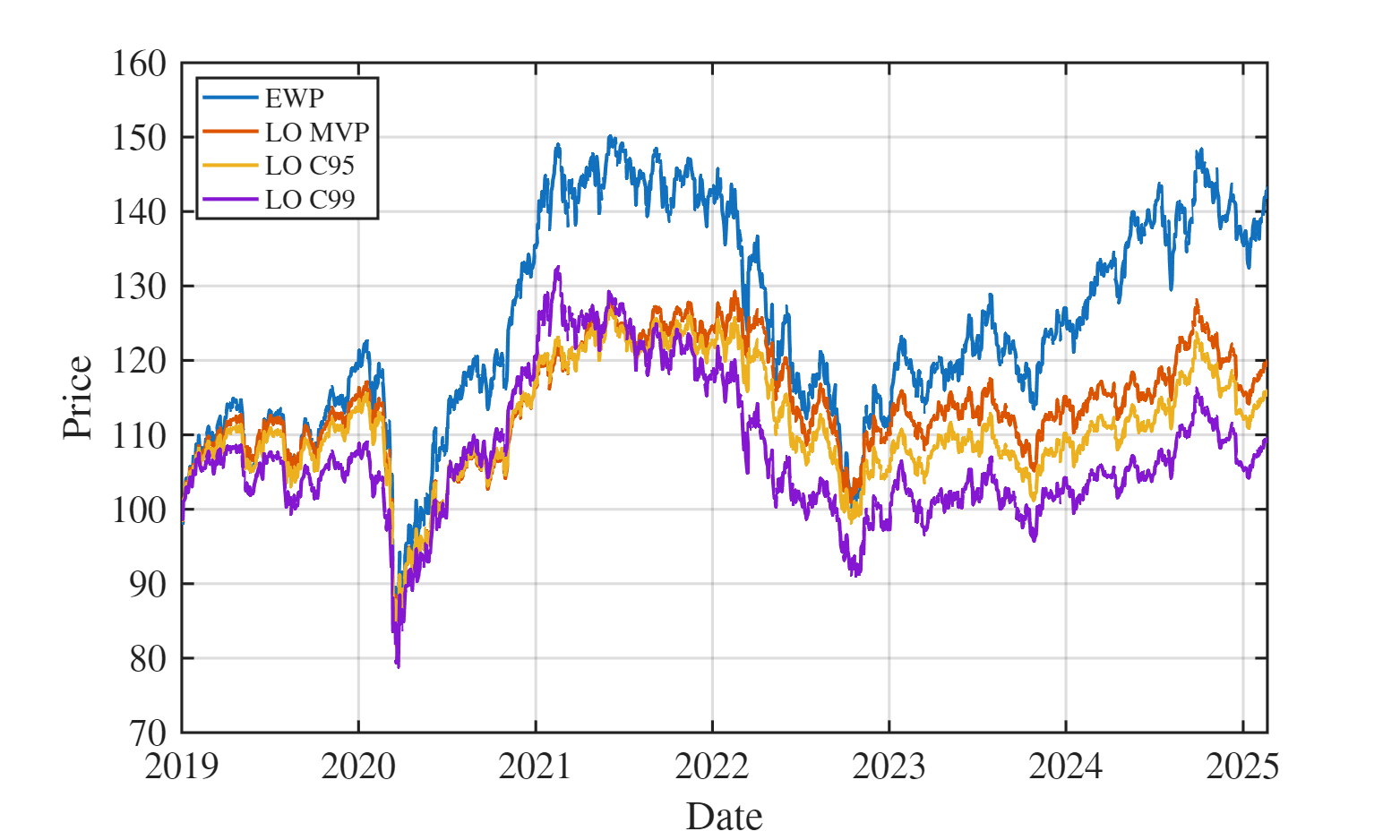}
	\includegraphics[width=0.49\textwidth]{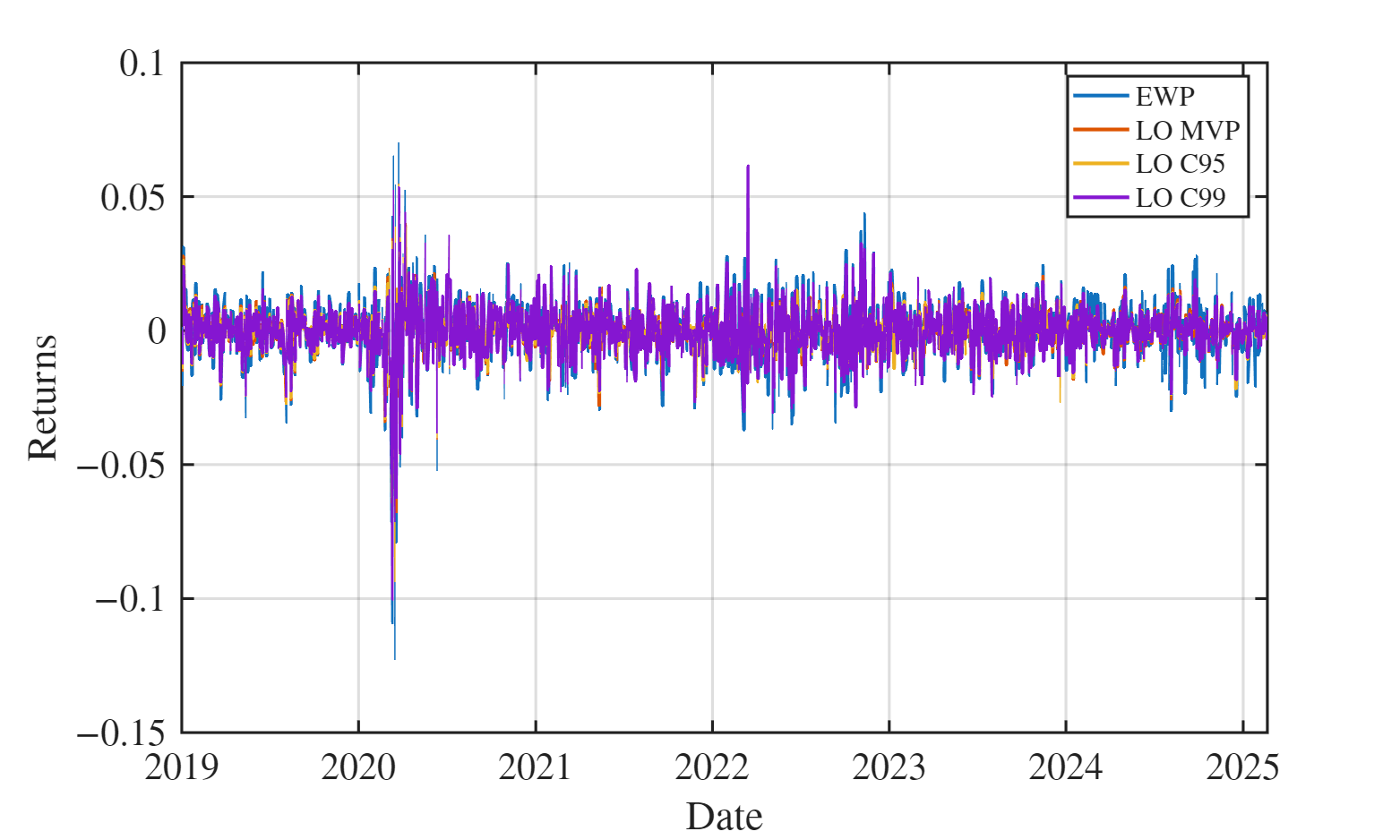}
	\caption{Long--only portfolio performance over the 2019--2025 evaluation period. Portfolio weights are re-estimated and portfolios are rebalanced daily using a trailing four-year (1,008 trading day) rolling window.}
	\label{fig:long-only}
\end{figure}

Figure~\ref{fig:long-only} shows the performance of the long-only portfolios over the 2019--2025 period, with performance evaluated assuming an initial investment of \$100. The price paths in the left panel reveal clear differences across portfolio strategies. The equally weighted portfolio (EWP) achieved the highest cumulative value over the sample period and remained the strongest-performing strategy following the market disruption in early 2020. Because the EWP maintained broad exposure across the ETF universe, it participated more fully in the appreciation of semiconductor- and technology-related assets during the latter part of the sample period.

In contrast, the long-only minimum-variance portfolio (LO MVP) followed a substantially smoother trajectory with lower cumulative growth. This behavior reflected the objective of variance minimization, which systematically reduced exposure to more volatile assets, including semiconductor-focused ETFs. The CVaR portfolios (LO C95 and LO C99) occupied an intermediate position. Although they exhibited lower volatility and smaller drawdowns than the EWP, their more conservative allocations limited participation in strong upward market movements. Among these portfolios, LO C99 displayed the slowest growth because the 99\% CVaR criterion placed greater emphasis on extreme downside losses.

The return series shown in the right panel reinforce these differences. The EWP exhibited the largest return fluctuations, including both deeper drawdowns and stronger rebounds, consistent with its broader exposure to volatile technology and semiconductor assets. By comparison, the LO MVP exhibited the smallest return fluctuations, reflecting the smoothing effect of variance minimization. The CVaR portfolios showed intermediate behavior, with return fluctuations generally smaller than those of the EWP while remaining comparable to those of the MVP. All strategies experienced a sharp decline in early 2020, visible in both panels of Figure~\ref{fig:long-only}, consistent with the broad market disruption associated with the COVID-19 pandemic.

The strong out-of-sample performance of the EWP was consistent with \citet{demiguel2009optimal}, who showed that estimation error in expected returns and covariances could offset the theoretical benefits of optimization in finite samples. Because the EWP required no parameter estimates, it was immune to this source of performance degradation. This robustness was compounded by the extraordinary appreciation of semiconductor-related assets during the AI-driven technology expansion. Because MVP and CVaR optimization systematically underweighted high-volatility holdings, both approaches limited participation in the sector-specific gains that dominated returns during the latter part of the sample period.

Overall, the long-only results indicated that strategies designed to reduce volatility or tail risk achieved smoother return paths and smaller drawdowns, but at the cost of weaker participation in the sustained appreciation of semiconductor- and technology-related assets that the broadly diversified EWP captured more fully.

\subsubsection{Long--Short Portfolios}

We also examine long--short portfolios under the constraints described in Section~\ref{sec:optimization_methodology}. Portfolio weights are permitted to take negative values, subject to the bounds
$-0.3 \leq w_i \leq 1.3$, allowing limited short positions and moderate leverage. This specification provides a useful comparison with the long-only setting and follows the portfolio construction framework of \citet{lindquist2022advanced}.

\begin{figure}[htb]
	\centering
	\includegraphics[width=0.49\textwidth]{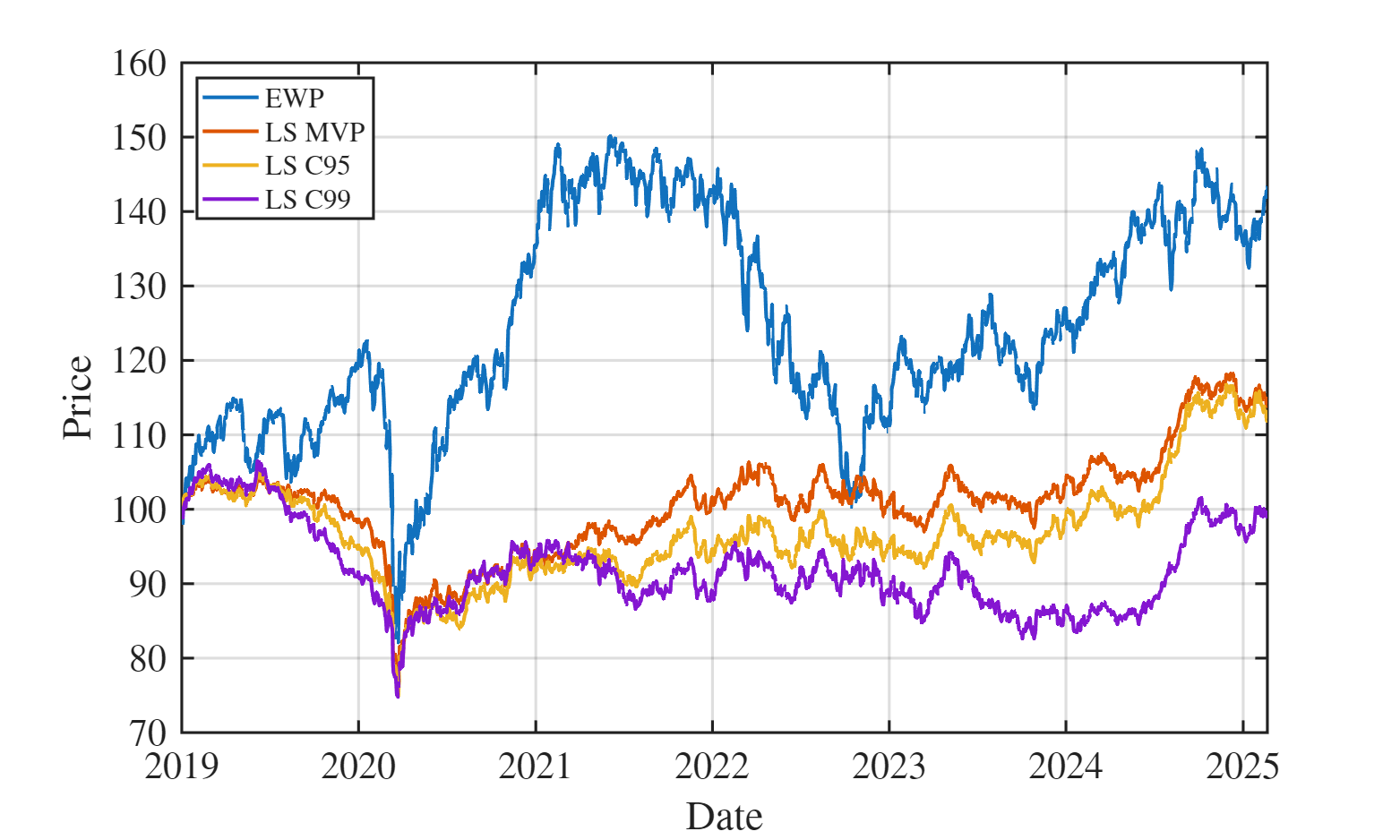}
	\includegraphics[width=0.49\textwidth]{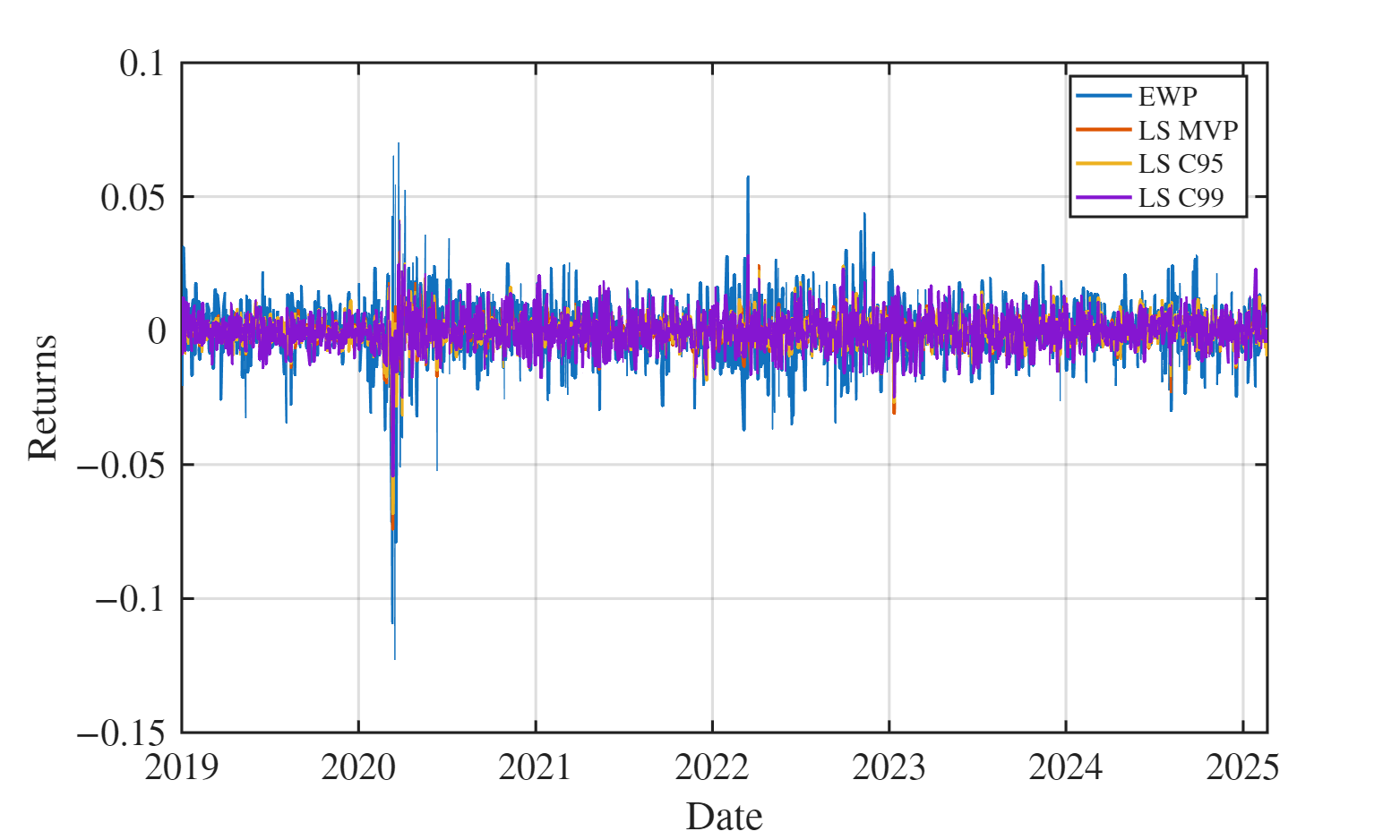}
	\caption{Long--short portfolio performance over the 2019--2025 evaluation period. Portfolio weights are re-estimated and portfolios are rebalanced daily using a trailing four-year (1,008 trading day) rolling window.}
	\label{fig:long-short}
\end{figure}

Figure~\ref{fig:long-short} shows the performance of the long--short portfolios over the 2019--2025 period, with performance evaluated assuming an initial investment of \$100. The price paths in the left panel reveal clear differences across portfolio strategies. The equally weighted portfolio (EWP) achieved the highest cumulative value by the end of the sample period, although its trajectory remained more volatile than those of the optimized portfolios. The long--short minimum-variance portfolio (LS MVP) followed a smoother trajectory with lower cumulative growth. The CVaR portfolios (LS C95 and LS C99) occupied an intermediate position, with LS C99 exhibiting the lowest cumulative value among all strategies. All portfolios experienced a sharp decline in early 2020, visible in both panels of Figure~\ref{fig:long-short}, followed by a recovery during the remainder of the sample period.

The return series shown in the right panel reinforce these observations. The optimized portfolios exhibited smaller return fluctuations than the EWP, with the LS MVP displaying the least variation. Volatility increased during periods of market stress, particularly in early 2020, although the optimized portfolios generally remained less volatile than the EWP throughout the sample period.

The weaker cumulative performance of the optimized portfolios was consistent with the findings of \citet{demiguel2009optimal}, who showed that estimation error in expected returns and covariances could offset the theoretical benefits of portfolio optimization in finite samples. The introduction of short positions further increased sensitivity to estimation error because negative portfolio weights amplified the effect of errors in estimated correlations and covariances. Consequently, even a small estimation error could generate a poorly performing short position, reducing realized out-of-sample performance relative to the EWP.

LS C99 exhibited the lowest cumulative value among all strategies because the 99\% CVaR criterion imposed the most restrictive constraint on tail losses. As a result, the portfolio systematically avoided assets with both high tail risk and high return potential. During a sample period characterized by exceptionally strong performance in semiconductor- and technology-related assets, this conservative allocation limited participation in the gains that drove cumulative returns during the latter part of the sample.

Overall, the long--short results indicated that allowing limited short positions reduced return variability but did not improve cumulative performance during the sample period. The optimized portfolios achieved smoother return paths at the cost of lower participation in the semiconductor- and technology-driven appreciation that the EWP, through its broad and unconstrained exposure, captured more fully. This weaker realized performance reflected both the estimation-error sensitivity introduced by short positions and the exceptional strength of semiconductor-related assets over the sample period.

%%%%%%%%%%%%%%%%%%%%%%%%%%%%%%%%%%%%%%%%%%%%%%
%%%%%%

\section{Portfolio-Level Risk-Adjusted Performance}
\label{sec:portfolio_performance}

Having established that mean--variance and CVaR optimization produced materially different portfolio allocations (Sections~\ref{sec:mv_optimization}--\ref{sec:cvar_optimization}), and that long-only portfolios captured more of the semiconductor-driven upside while long--short portfolios traded that upside for smoother return paths (Section~\ref{sec:price_return}), the next question was whether these allocation differences translated into meaningfully different realized performance. This section evaluated portfolio performance using three complementary reward-to-risk measures: the Sharpe ratio, the Rachev ratio, and the STARR ratio.

Together, these measures provide different perspectives on portfolio performance. The Sharpe ratio evaluates return relative to total volatility, the Rachev ratio compares extreme gains with extreme losses, and the STARR ratio evaluates excess return relative to conditional value-at-risk. No single measure fully characterizes portfolio quality in non-normal return environments, motivating the use of all three. As shown in the following analysis, portfolio rankings varied across performance measures. Variance-based measures generally favored the equally weighted portfolio, whereas tail-sensitive measures tended to favor CVaR-based allocations.

Performance distributions were summarized using rolling-window estimates and boxplots, facilitating comparisons of central tendency, dispersion, and stability across different market conditions. The analysis considered equally weighted portfolios, minimum-variance portfolios, and CVaR-based portfolios under both long-only and long--short investment constraints. This approach allowed both average portfolio performance and the consistency of performance over time to be evaluated, providing a basis for comparing variance-based and tail-sensitive strategies in the ETF universe examined throughout the paper.

\subsection{Sharpe Ratio}

The Sharpe ratio \citep{sharpe1998} measures excess return per unit of return volatility and serves as a standard benchmark for risk-adjusted performance evaluation:

\[
SR(T)
=
\frac{\mathbb{E}[r_p(t)-r_f(t)]_{[0,T]}}
{\sqrt{\operatorname{Var}[r_p(t)-r_f(t)]_{[0,T]}}},
\]

where $r_p(t)$ denotes the portfolio return at time $t$, $r_f(t)$ denotes the risk-free rate, and the expectation and variance are computed over the evaluation period $[0,T]$.

\begin{figure}[h!]
\centering
\includegraphics[width=0.495\textwidth]{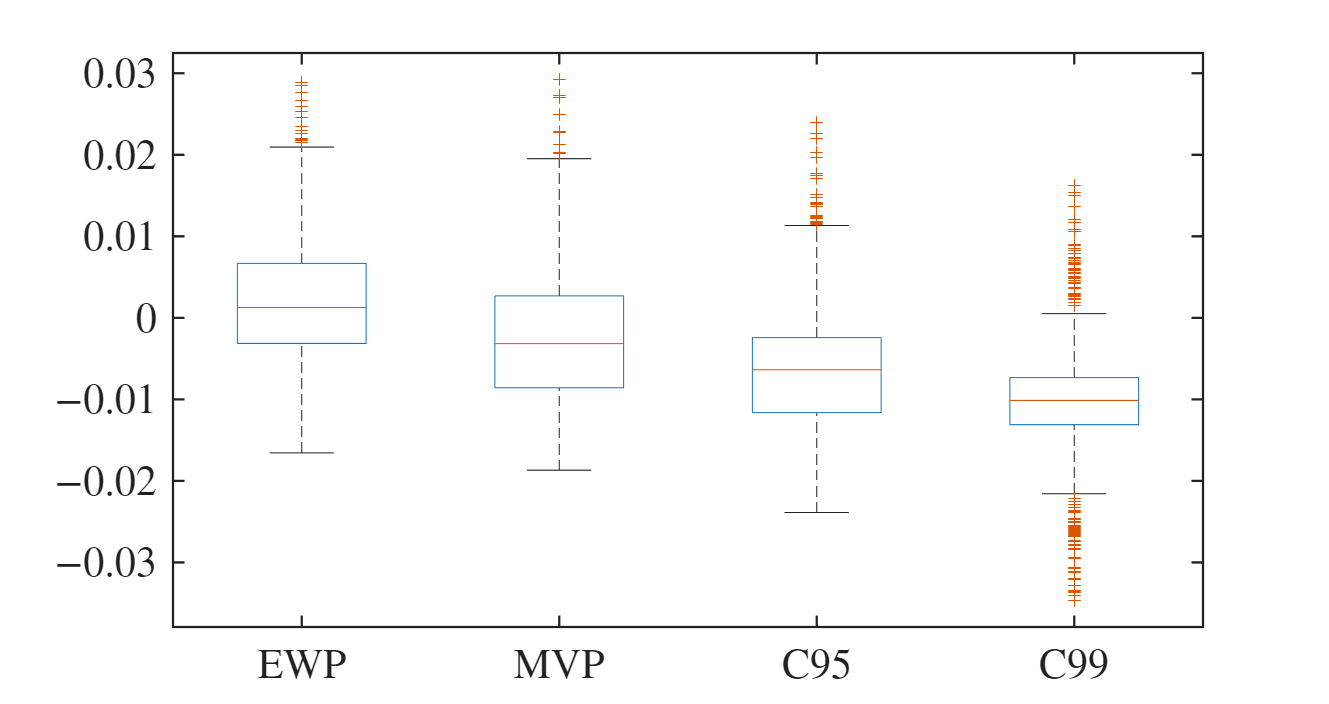}
\includegraphics[width=0.495\textwidth]{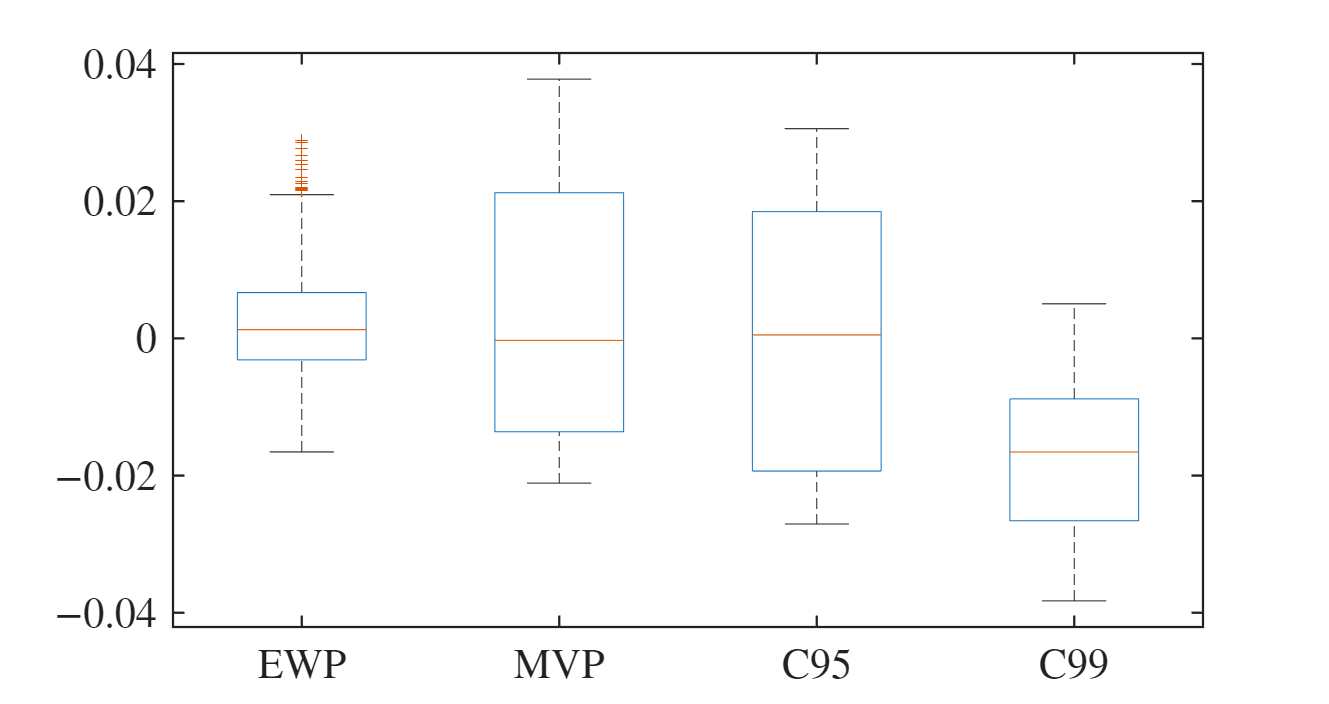}
\caption{Boxplots of Sharpe ratios for the equally weighted portfolio (EWP) and optimized portfolio strategies under long-only (LO) and long-short (LS) constraints.}
\label{fig:sharpe_ratio}
\end{figure}

Figure~\ref{fig:sharpe_ratio} presents boxplots of Sharpe-ratio values for the equally weighted portfolio (EWP), the minimum-variance portfolio (MVP), and the minimum-CVaR portfolios at the 95\% and 99\% confidence levels (C95 and C99). The left panel corresponds to the long-only case, whereas the right panel corresponds to the long-short case.

Under long-only constraints, the EWP exhibits the highest median Sharpe ratio together with relatively low dispersion, indicating stronger volatility-adjusted performance than the alternative strategies. The MVP exhibits a slightly negative median Sharpe ratio, suggesting that variance minimization alone did not necessarily improve excess-return generation. The minimum-CVaR portfolios display progressively lower median Sharpe ratios as the confidence level increases, with the C99 portfolio producing the weakest performance. This pattern is consistent with the cost of imposing increasingly restrictive downside-risk constraints.

The relatively strong Sharpe-ratio performance of the EWP is consistent with the findings of \citet{demiguel2009optimal}, who showed that estimation error in expected returns and covariance matrices can offset the theoretical advantages of optimized portfolios in finite samples. Because the EWP does not rely on parameter estimation, it is less affected by this source of performance deterioration. In addition, its broad exposure to the ETF universe, including semiconductor-related assets, allowed greater participation in the post-COVID technology expansion, contributing to stronger excess returns than portfolios that systematically underweighted higher-volatility semiconductor holdings.

Under long-short constraints, the dispersion increased across all portfolio strategies. This result suggests that the introduction of short positions increased sensitivity to estimation error and amplified portfolio variability. Because negative portfolio weights magnified the effects of errors in estimated returns and covariances, even relatively small estimation errors could produce larger changes in portfolio allocations. Although the median Sharpe ratios of the EWP and MVP remained relatively close, both minimum-CVaR portfolios, particularly C99, exhibited lower medians and wider interquartile ranges, indicating weaker and less stable volatility-adjusted performance when short selling was permitted.

Overall, the results indicate a trade-off between downside-risk protection and Sharpe-ratio performance. Portfolios that placed greater emphasis on tail-risk control generally achieved lower Sharpe ratios, whereas the equally weighted portfolio remained comparatively competitive, particularly under long-only constraints.

Although the Sharpe ratio provides a useful benchmark for risk-adjusted performance evaluation, it relies exclusively on variance and therefore does not distinguish between upside and downside fluctuations. In particular, it cannot distinguish between portfolios that differ primarily in their exposure to extreme losses rather than overall volatility, a distinction that the Rachev ratio is designed to capture. This limitation motivated the use of tail-sensitive performance measures, beginning with the Rachev ratio.

\subsection{Rachev Ratio}

The Rachev ratio \citep{rachev2008advanced,lindquist2022advanced} is defined as

\[
RR_{\beta,\gamma}(T)
=
\frac{\mathrm{CVaR}_{\beta}[r_f(t)-r_p(t)]_{[0,T]}}
{\mathrm{CVaR}_{\gamma}[r_p(t)-r_f(t)]_{[0,T]}},
\qquad \beta,\gamma \in (0,1).
\]

Here $\beta$ and $\gamma$ denote the tail probability levels used in the CVaR calculations. The numerator captures the expected gain in the upper tail of the excess-return distribution, measured as the CVaR of the negative excess return $r_f(t)-r_p(t)$, while the denominator captures the expected loss in the lower tail, measured as the CVaR of the excess return $r_p(t)-r_f(t)$. A higher Rachev ratio therefore indicates a more favorable balance between extreme gains and extreme losses. For notational convenience, the Rachev ratio is denoted by RR throughout this section.

In contrast to reward-to-variability measures such as the Sharpe ratio, the Rachev ratio belongs to the class of reward-to-risk measures because it evaluates extreme gains and losses using tail-based risk measures. This construction makes the Rachev ratio particularly suitable for return distributions that exhibit asymmetry and heavy tails. Throughout the analysis, tail probability levels of $\beta=\gamma=0.95$ and $\beta=\gamma=0.99$ are considered, representing moderate and more severe tail events, respectively.

\begin{figure}[h!]
\centering
\includegraphics[width=0.495\textwidth]{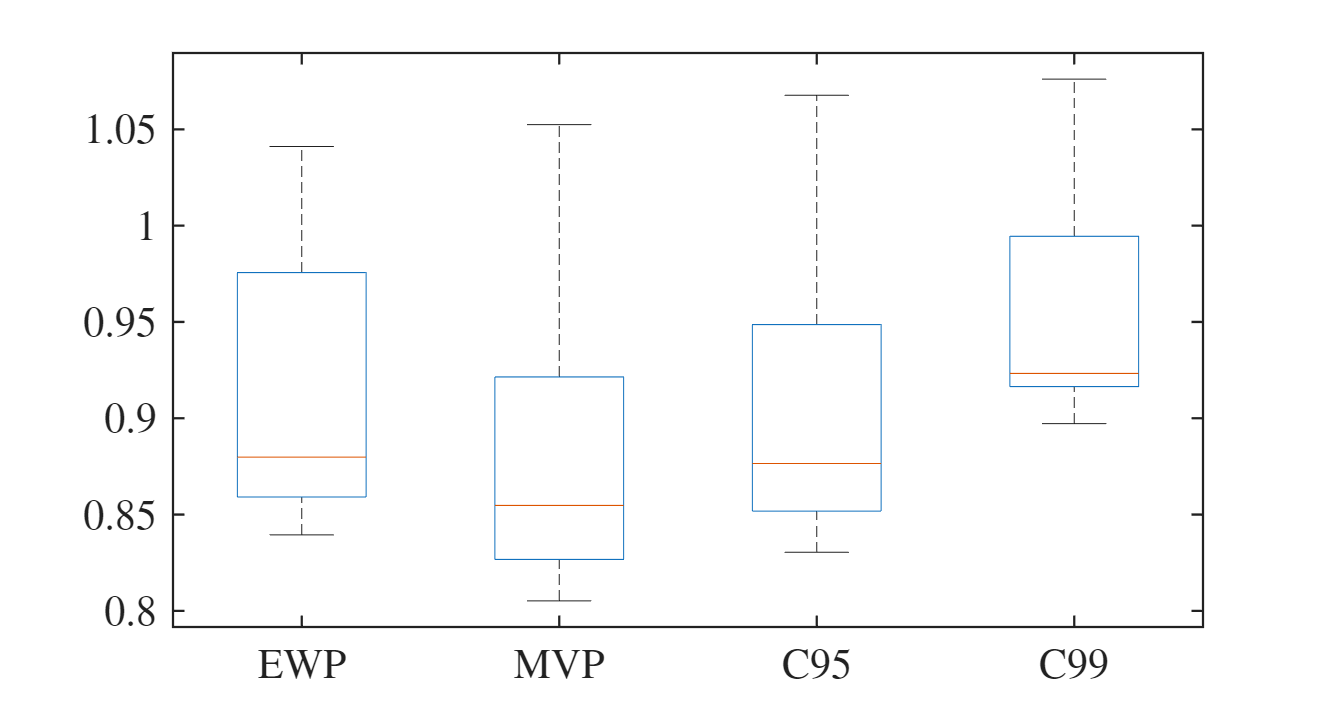}
\includegraphics[width=0.495\textwidth]{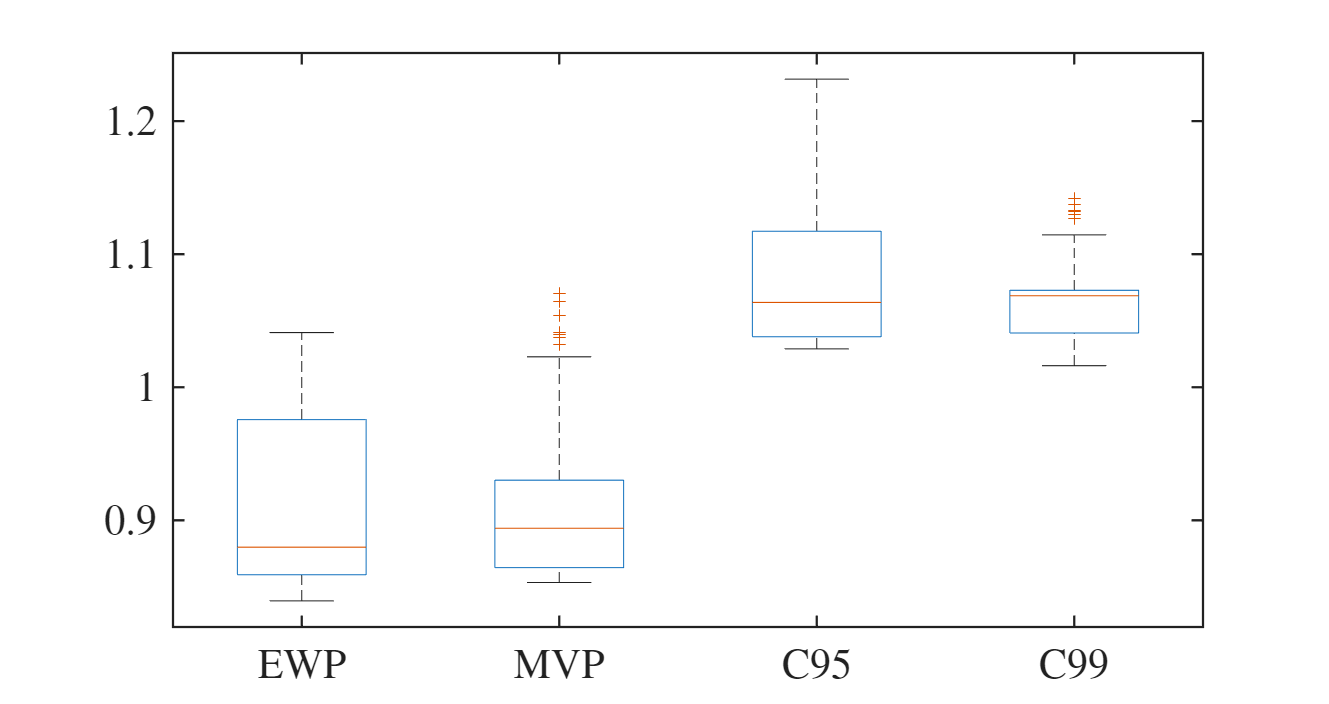}
\includegraphics[width=0.495\textwidth]{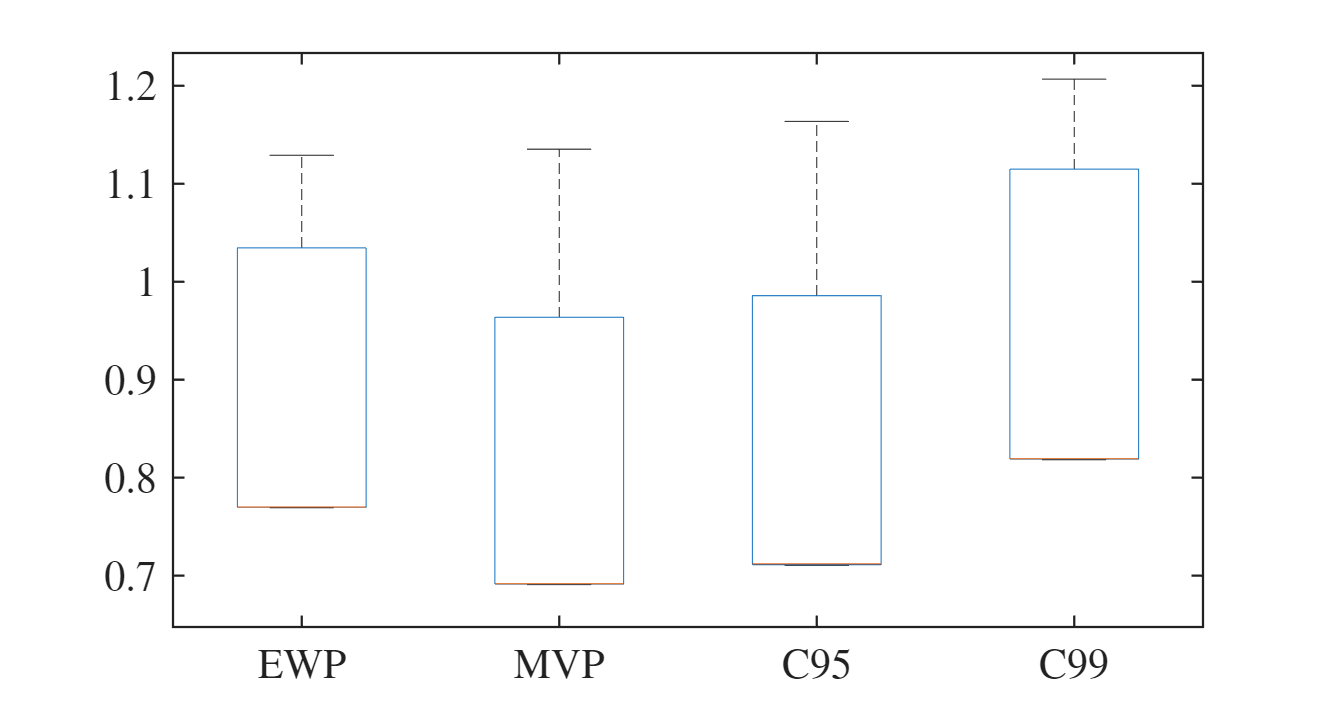}
\includegraphics[width=0.495\textwidth]{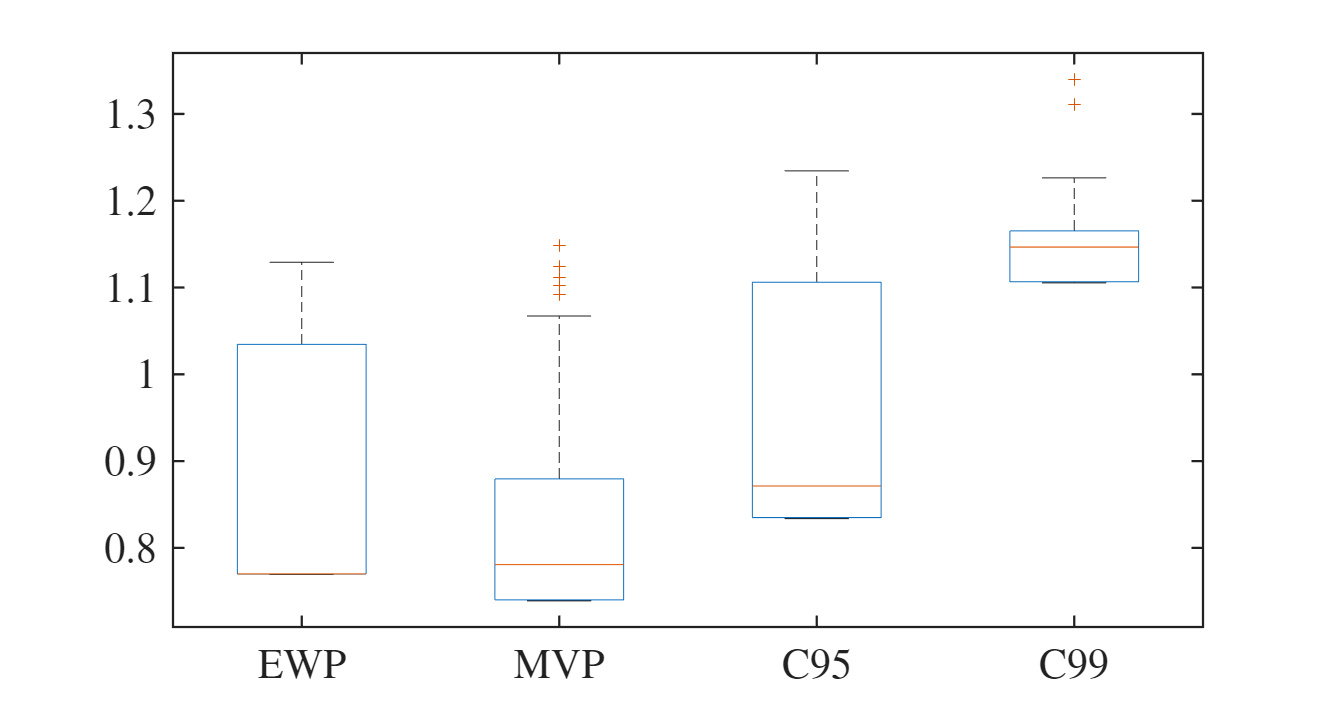}
\caption{Box-Whisker Plot of RR for EWP and portfolio strategies under LO (left) and LS (right) constraints, with the 95\% confidence level shown in the upper panels and the 99\% confidence level in the lower panels.}
\label{fig:rachev_ratio}
\end{figure}

Figure~\ref{fig:rachev_ratio} presents boxplots of RR values for the equally weighted portfolio (EWP), the minimum-variance portfolio (MVP), and the minimum-CVaR portfolios at the 95\% and 99\% confidence levels (C95 and C99). The upper panels correspond to the 95\% level and the lower panels correspond to the 99\% level, while the left and right panels present results under long-only and long-short constraints, respectively.

Under long-only constraints at the 95\% confidence level, C99 exhibits the highest median RR value, while C95 performs comparably to the EWP. The MVP exhibits the lowest median RR among all strategies, indicating that variance minimization produces a less favorable balance between extreme gains and extreme losses than either diversification or tail-risk optimization. The advantage of the CVaR portfolios becomes more pronounced at the 99\% confidence level, where C99 achieves a median RR substantially above those of the EWP and MVP. C95 also improves relative to the EWP at the 99\% level, consistent with the greater emphasis that higher confidence levels place on extreme tail events.

The stronger Rachev-ratio performance of the CVaR portfolios is consistent with their construction. By explicitly controlling extreme downside losses during portfolio optimization, CVaR-based portfolios reduce the magnitude of losses in the lower tail relative to overall portfolio performance. At the same time, these portfolios retain meaningful exposure to semiconductor-related ETFs, which benefit from the strong performance of the technology sector during the post-COVID period. As a result, the CVaR portfolios exhibit a more favorable balance between extreme gains and extreme losses, which is precisely the characteristic measured by the Rachev ratio.

When short selling is permitted, RR values increase further for the CVaR-based portfolios across both confidence levels. The improvement is most evident for the C99 portfolio, which exhibits the highest median RR in both long-short panels. The additional flexibility provided by short positions allows the optimization procedure to adjust portfolio exposures more aggressively in response to tail-risk characteristics, improving the balance between extreme gains and extreme losses. However, the dispersion of RR values also increases across all strategies, indicating greater sensitivity to extreme observations and additional estimation uncertainty associated with short positions.

Overall, the results indicate that CVaR-based portfolios provide superior tail reward-to-risk performance relative to both the equally weighted and minimum-variance portfolios, particularly at higher confidence levels and when short selling is permitted. These findings suggest that tail-sensitive portfolio construction performs better when evaluation is based on measures that explicitly account for tail asymmetry rather than overall return volatility.

Although the Rachev ratio incorporates both upside and downside tail behavior, it does not focus exclusively on downside risk. This motivates the use of the STARR ratio, which evaluates excess return relative to conditional value-at-risk and therefore provides a complementary perspective on portfolio performance.

\subsection{STARR Measure}

The STARR measure (Stable Tail Adjusted Return Ratio), introduced within the coherent risk measure framework of \citet{rachev2008advanced}, provides a downside-risk-adjusted alternative to the Sharpe ratio. For a confidence level $\beta \in (0,1)$, the STARR measure is defined as

\[
\mathrm{STARR}_{\beta}(T)
=
\frac{\mathbb{E}[r_p(t)-r_f(t)]_{[0,T]}}
{\mathrm{CVaR}_{\beta}[r_p(t)-r_f(t)]_{[0,T]}}.
\]

The STARR measure evaluates average excess return relative to downside tail risk measured by conditional value-at-risk. Unlike the Rachev ratio, which compares extreme gains with extreme losses, STARR focuses directly on downside-risk-adjusted performance.

Because CVaR is a coherent risk measure \citep{artzner1999coherent,rockafellar2002conditional}, the STARR framework is particularly relevant for return distributions exhibiting skewness, heavy tails, and asymmetric downside behavior. Throughout the analysis, confidence levels of 95\% and 99\% are considered, corresponding to moderate and more extreme downside tail events, respectively.

\begin{figure}[h!]
\centering
\includegraphics[width=0.495\textwidth]{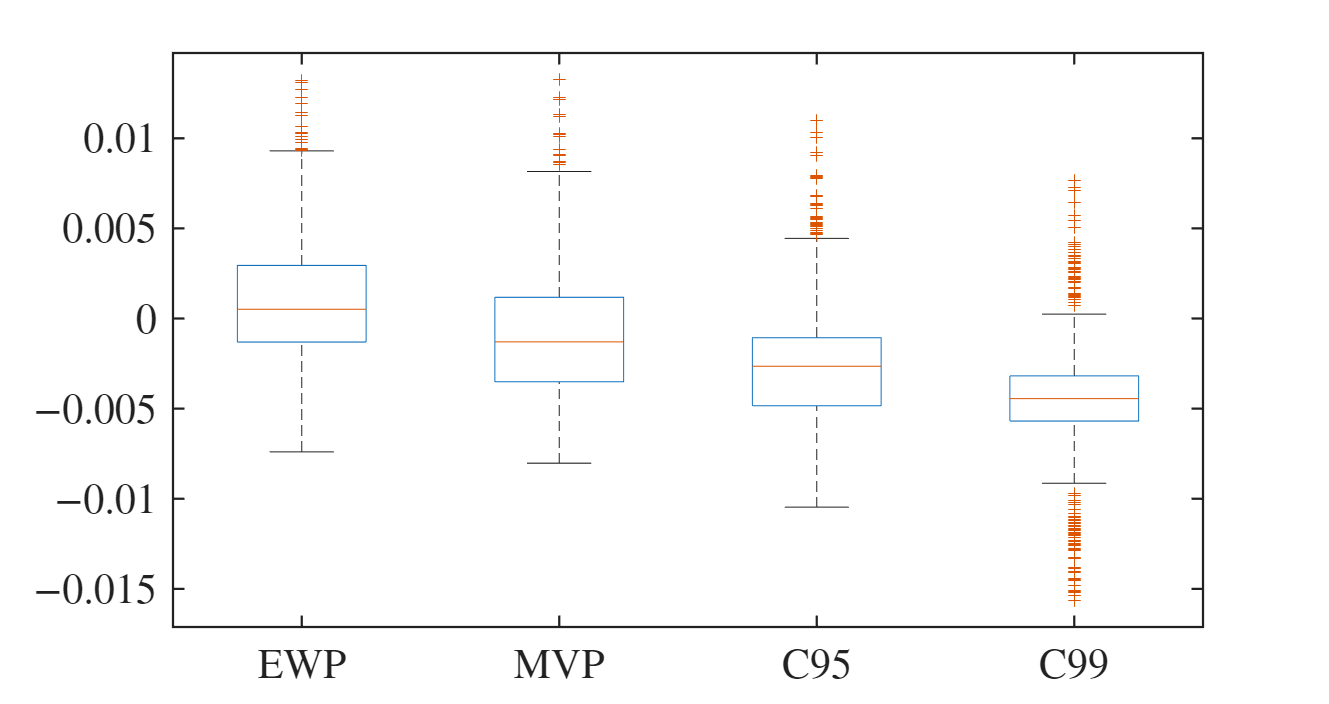}
\includegraphics[width=0.495\textwidth]{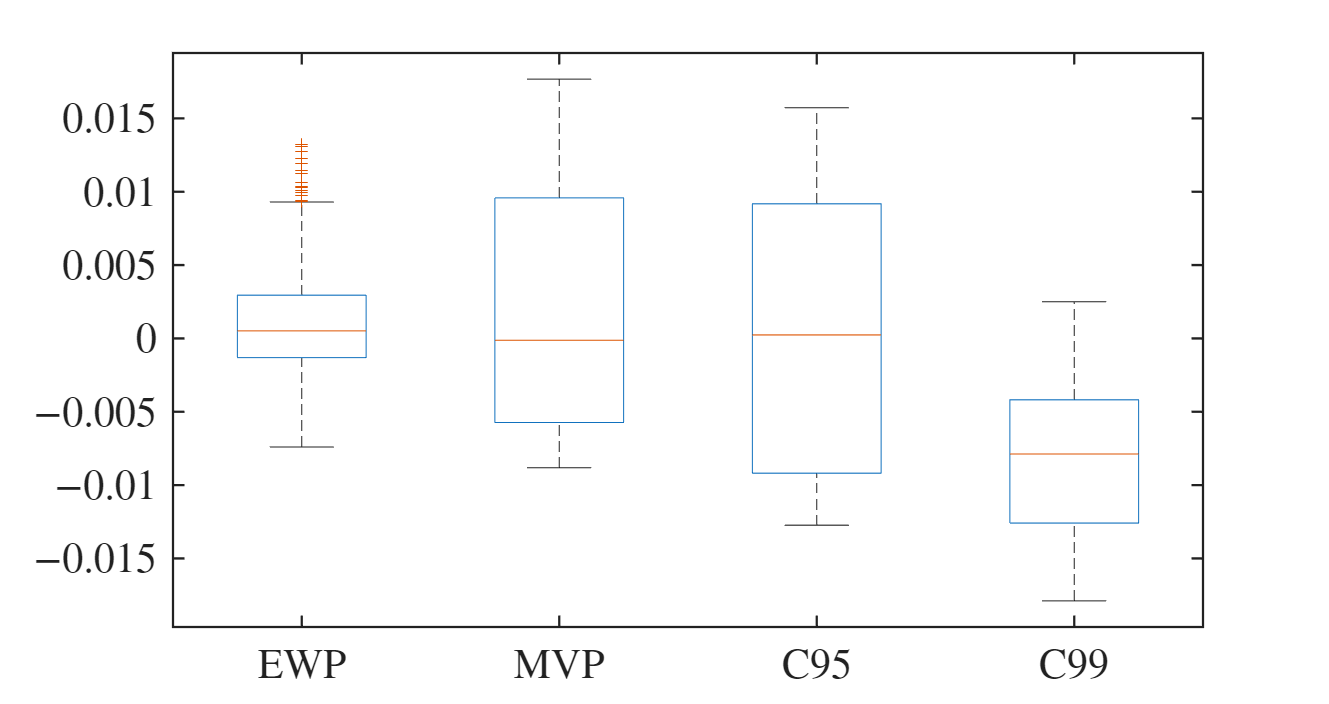}
\includegraphics[width=0.495\textwidth]{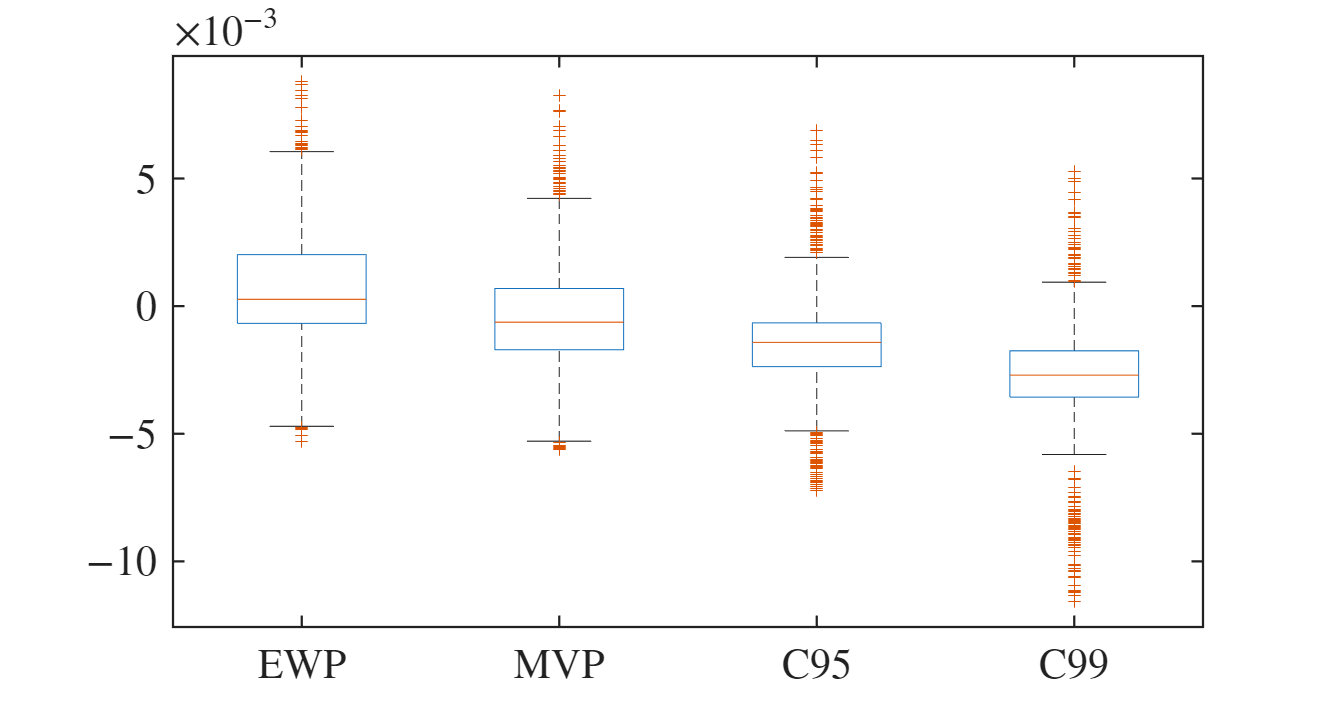}
\includegraphics[width=0.495\textwidth]{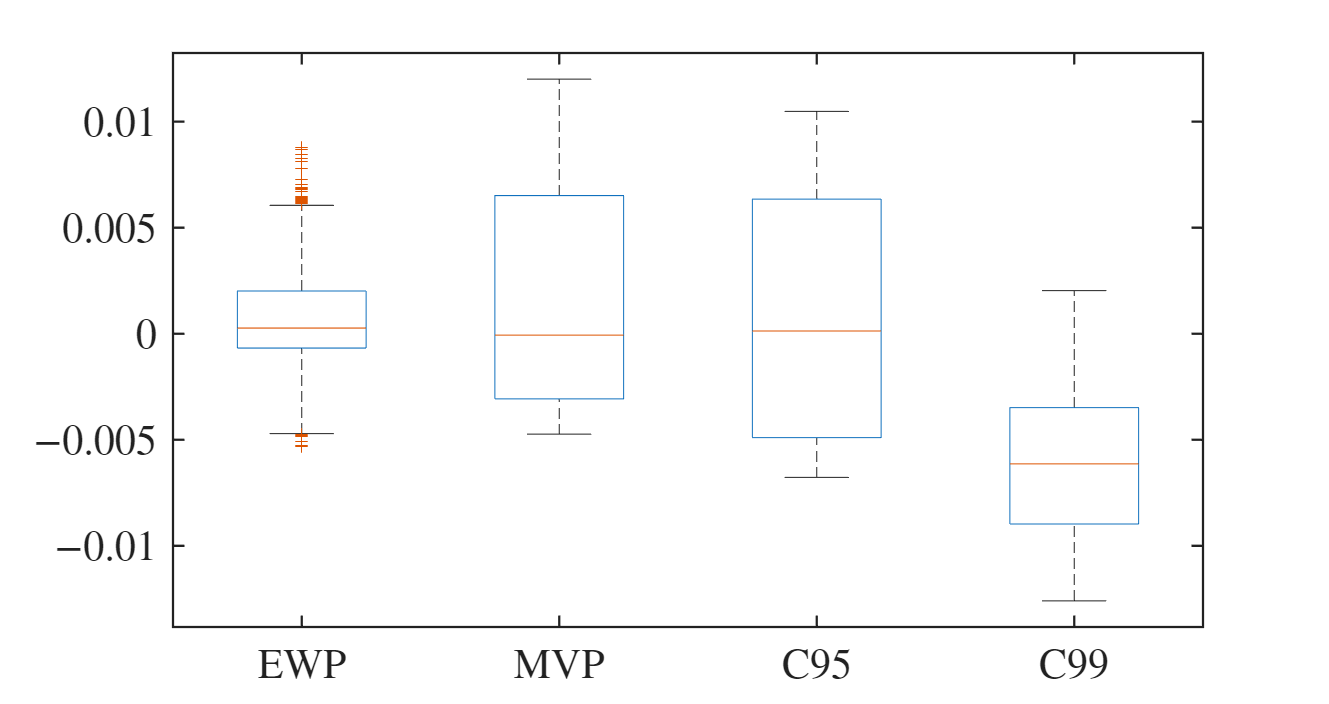}
\caption{Box-Whisker Plot of STARR for EWP and portfolio strategies under LO (left) and LS (right) constraints, with the 95\% confidence level shown in the upper panels and the 99\% confidence level in the lower panels.}
\label{fig:starr}
\end{figure}

Figure~\ref{fig:starr} presents boxplots of STARR values for the equally weighted portfolio (EWP), the minimum-variance portfolio (MVP), and the minimum-CVaR portfolios at the 95% and 99% confidence levels (C95 and C99). The upper panels correspond to the 95% confidence level and the lower panels to the 99% confidence level, while the left and right panels present results under long-only and long-short constraints, respectively.

Under long-only constraints, the EWP exhibits the highest median STARR values among all portfolio strategies at both confidence levels. This result indicates comparatively strong compensation for downside tail risk despite the absence of explicit tail-risk optimization. The strong STARR performance of the EWP reflects two reinforcing effects. First, broad diversification across the ETF universe reduces exposure to idiosyncratic tail losses, limiting the magnitude of the CVaR denominator. Second, the equal-weighted allocation maintains substantial exposure to semiconductor- and technology-related ETFs, allowing the portfolio to participate fully in the strong technology- and semiconductor-driven market expansion observed during the sample period. Together, these effects produce a favorable excess-return-to-CVaR ratio that the optimized portfolios did not consistently match during the sample period.

The MVP exhibits lower median STARR values than the EWP under long-only constraints. Variance minimization systematically reduces allocations to high-volatility assets, including semiconductor-related ETFs, thereby limiting participation in the strong positive returns generated by these assets during the sample period. Because STARR evaluates excess return relative to downside tail risk rather than total volatility, this conservative allocation reduces the excess-return numerator without proportionally reducing the CVaR denominator, resulting in weaker downside-risk-adjusted performance.

The minimum-CVaR portfolios exhibit negative median STARR values across most specifications, with the C99 portfolio generally producing the lowest median values across the four panels. STARR values are also noticeably smaller at the 99\% confidence level than at the 95\% level, as reflected in the scale of the lower-left panel. This result reflects the cost of aggressive downside-risk control. By reducing exposure to assets that contribute substantially to downside tail risk, the optimization also reduces exposure to positive tail realizations and return-generating opportunities, weakening the excess-return numerator without achieving a proportional reduction in CVaR.

When short selling is permitted, the dispersion of STARR values increases substantially across all portfolio strategies. This wider spread reflects the combined effects of leverage, estimation uncertainty, and increased sensitivity to extreme observations. Under long-short constraints, the C99 portfolio exhibits the most negative median STARR values and the widest dispersion among all strategies. These results suggest that aggressive downside-risk minimization combined with short positions can materially reduce downside-risk-adjusted return efficiency.

Overall, the STARR results reveal a trade-off between downside-risk protection and return generation. Portfolios optimized primarily to reduce tail losses do not necessarily achieve superior downside-risk-adjusted performance because reductions in downside exposure are often accompanied by weaker excess-return generation. In contrast, the equally weighted portfolio maintained comparatively stable performance throughout the sample period, highlighting the robustness of simple diversification strategies in heavy-tailed market environments.

Taken together with the Sharpe-ratio and Rachev-ratio results, the STARR analysis shows that portfolio rankings depended on the chosen performance criterion. Variance-based measures emphasize overall volatility, the Rachev ratio emphasizes tail asymmetry, and STARR emphasizes downside-risk-adjusted efficiency. This dependence is particularly relevant in the Taiwan-exposed ETF universe, where semiconductor concentration generates asymmetric return dynamics that affect variance-based and tail-sensitive measures differently. The results suggest that the equally weighted portfolio performed more favorably under volatility-based evaluation criteria, whereas CVaR-based portfolios performed more favorably under measures that explicitly account for tail asymmetry.

\subsection{Discussion of Portfolio-Level Performance}

\noindent\textbf{Summary.}
The Sharpe ratio and STARR measure favored EWP, whereas the Rachev ratio favored the CVaR-based portfolios. These results indicated that portfolio rankings depended on the dimension of risk captured by the performance measure, whether by overall volatility, downside risk relative to return, or the balance between extreme gains and extreme losses. The results also showed that different performance measures led to different conclusions regarding portfolio quality in the Taiwan-related ETF universe examined in this study.

\FloatBarrier

\section{Discussion and Conclusion}
\label{sec:conclusion}

Taiwan's dominant position in global advanced-node semiconductor manufacturing, combined with AI-driven demand cycles and geopolitical uncertainty, created an investment environment in which tail-sensitive risk measures provided insights that conventional variance-based frameworks did not capture. This study analyzed thirty U.S.-listed ETFs with Taiwan exposure over February 2015--February 2025, combining Hill tail-index estimation, GJR-GARCH and FIGARCH volatility modeling, CVaR-based portfolio optimization, and tail-sensitive performance evaluation. Four main findings emerge.

\subsection{Summary of Findings}

\noindent\textbf{Variance-based measures understated tail risk.} Sharpe and Sortino ratios clustered narrowly across the ETF universe, while VaR and CVaR revealed that semiconductor-focused ETFs, particularly SMH and SOXX, carried substantially greater downside risk than diversified benchmarks (Section~\ref{sec:risk_tail_behavior}). Because Hill tail-index estimates remained similar across ETFs, at approximately 2.5--3.5 in the stable estimation region (Section~\ref{subsec:hill}), this cross-sectional variation in CVaR reflected differences in return \emph{scale} rather than in asymptotic tail decay---a distinction that Sharpe-ratio comparisons alone cannot reveal.

\medskip

\noindent\textbf{Apparent long memory reflected conditional heteroskedasticity, not fractional integration.} GPH, Local Whittle, and Hurst estimates on raw squared returns suggested long-memory-type behavior, particularly for EWP, but all three diagnostics collapsed toward zero after GJR-GARCH(1,1)-\emph{t} filtering (Section~\ref{subsec:long_memory}, Tables~\ref{tab:longmemory}--\ref{tab:lrd}). Rolling FIGARCH estimates near unity for EWP were therefore more plausibly attributed to model instability under structural change than to genuine fractional integration, favoring asymmetric GARCH-type specifications over fractionally integrated models in this setting.

\medskip

\noindent\textbf{CVaR optimization produced qualitatively different allocations from mean--variance optimization.} The mean--variance tangent portfolio remained diversified across multiple ETFs in both the full sample (Figure~\ref{fig:MV_US}, Section~\ref{sec:mv_optimization}) and the pre- and post-COVID subsamples (Figure~\ref{fig:mv_robust}, Appendix~\ref{app:robust}). In contrast, the CVaR tangent portfolio lay very close to SMH at both the 95\% and 99\% confidence levels (Figure~\ref{fig:CVaR_US}, Section~\ref{sec:cvar_optimization}), with the concentration effect substantially stronger post-COVID than pre-COVID (Figure~\ref{fig:cvar_robust}, Appendix~\ref{app:robust}). This represented a difference in portfolio \emph{composition}, not merely frontier shape, and likely reflected SMH's realized return-to-tail-risk profile during the AI-driven expansion rather than a persistent structural property of semiconductor ETFs.

\medskip

\noindent\textbf{Portfolio rankings depended on the performance criterion.} EWP achieved the highest median Sharpe ratio and STARR value under long-only constraints, reflecting both its immunity to estimation error \citep{demiguel2009optimal} and its full participation in semiconductor-sector appreciation. The Rachev ratio produced the opposite ranking, favoring CVaR-based portfolios, particularly at the 99\% confidence level and under long--short constraints (Section~\ref{sec:portfolio_performance}). This reversal likely occurred because the assets contributing most to downside risk during the sample period were also the strongest performers, so tail-loss-minimizing portfolios reduced the STARR numerator proportionally more than they reduced the CVaR denominator. No single measure therefore fully characterized portfolio quality in the present study; rather, the appropriate evaluation criterion depended on whether an investor prioritized volatility-adjusted return, downside-risk-adjusted return, or the balance between extreme gains and extreme losses.

\subsection{Limitations}
Three limitations should be considered when interpreting these findings. First, the portfolio comparisons in Section~\ref{sec:portfolio_performance} relied on the distribution of rolling-window performance measures rather than formal statistical tests of ranking differences; Jobson--Korkie or Memmel-type tests, or bootstrap confidence intervals for the differences shown in Figures~\ref{fig:sharpe_ratio}--\ref{fig:starr}, would be needed to establish statistical significance. Second, the reported returns did not account for transaction costs or portfolio turnover. This omission was particularly relevant because the CVaR portfolios exhibited substantial shifts toward SMH during the post-COVID period, and realized net performance could therefore differ from the gross returns reported in this study. Third, the results reflected a single historical period encompassing the COVID-19 disruption and the subsequent AI-driven semiconductor expansion. Consequently, the concentration of the CVaR portfolios in SMH documented in Section~\ref{sec:cvar_optimization} should not be extrapolated to future periods without considering changes in AI investment cycles and Taiwan-related geopolitical risk.

\subsection{Future Research}
Several directions for future research remain open. Multivariate extensions incorporating dynamic copula models could capture time-varying tail dependence among ETFs during periods of market stress. Regime-switching volatility models that allow risk parameters to vary across expansion and stress regimes could improve both volatility forecasting and dynamic portfolio rebalancing. Incorporating a geopolitical risk index, such as that proposed by \citet{caldara2022}, directly into the CVaR optimization framework could provide a more forward-looking treatment of cross-strait tensions and export-restriction scenarios. Finally, incorporating transaction costs and examining the sensitivity of portfolio performance to alternative rebalancing frequencies would help determine whether the performance differences documented in this study remain robust after realistic implementation frictions are taken into account.

\normalem		% disables the ulem package

\bibliographystyle{apacite}
\bibliography{ref}

\appendix
\renewcommand{\thetable}{A\arabic{table}}
\setcounter{table}{0}
\renewcommand{\thefigure}{A\arabic{figure}}
\setcounter{figure}{0}

\section{ETF Sample Summary}\label{app:etf_table}
\begin{table}[H] 
\centering
\caption{Summary of U.S.-listed ETFs and their market exposure.}
\label{tab:etf_summary}
\begin{threeparttable}
\resizebox{\textwidth}{!}{
\begin{tabular}{lllc}
\hline

\textbf{Ticker} & \textbf{ETF Name} & \textbf{ETF Category} & \textbf{Inception Date}\\
\hline

EWT & iShares MSCI Taiwan ETF & Taiwan Equities & 20-Jun-2000 \\

SMH & VanEck Semiconductor ETF & Semiconductor Equities & 20-Dec-2011 \\
SOXX & iShares Semiconductor ETF & Semiconductor Equities & 10-Jul-2001 \\
IXN & iShares Global Tech ETF & Global Technology Sector Equities & 12-Nov-2001 \\

ACWI & iShares MSCI ACWI ETF & Global Equities & 26-Mar-2008 \\
ACWX & iShares MSCI ACWI ex U.S. ETF & Foreign All-Cap Equities (ex U.S.) & 26-Mar-2008 \\
CWI & SPDR MSCI ACWI ex-US ETF & Foreign All-Cap Equities (ex U.S.) & 10-Jan-2007 \\
IXUS & iShares Core MSCI Total International Stock ETF & Foreign All-Cap Equities (ex U.S.) & 18-Oct-2012 \\
VEU & Vanguard FTSE All-World ex-US ETF & Foreign All-Cap Equities (ex U.S.) & 02-Mar-2007 \\
VXUS & Vanguard Total International Stock ETF & Foreign All-Cap Equities (ex U.S.) & 26-Jan-2011 \\

AAXJ & iShares MSCI All Country Asia ex Japan ETF & Asia ex-Japan Equities & 13-Aug-2008 \\
AIA & iShares Asia 50 ETF & Asia Large-Cap Equities & 13-Nov-2007 \\
EEMA & iShares MSCI Emerging Markets Asia ETF & Asia Pacific Equities & 08-Feb-2012 \\
GMF & SPDR S\&P Emerging Asia Pacific ETF & Asia Pacific Equities & 19-Mar-2007 \\

EEM & iShares MSCI Emerging Markets ETF & Emerging Markets Equities & 07-Apr-2003 \\
ECON & Columbia Research Enhanced Emerging Economies ETF & Emerging Markets Equities & 14-Sep-2010 \\
IEMG & iShares Core MSCI Emerging Markets ETF & Emerging Markets Equities & 18-Oct-2012 \\
SCHE & Schwab Emerging Markets Equity ETF & Emerging Markets Equities & 14-Jan-2010 \\
SPEM & SPDR Portfolio Emerging Markets ETF & Emerging Markets Equities & 19-Mar-2007 \\
VWO & Vanguard FTSE Emerging Markets ETF & Emerging Markets Equities & 04-Mar-2005 \\

DBEM & Xtrackers MSCI Emerging Markets Hedged Equity ETF & Currency-Hedged Emerging Markets Equities & 09-Jun-2011 \\
DEM & WisdomTree Emerging Markets High Dividend Fund & Emerging Markets High Dividend & 13-Jul-2007 \\
DGS & WisdomTree Emerging Markets SmallCap Dividend Fund & Emerging Markets Small-Cap Dividend & 30-Oct-2007 \\
EDIV & SPDR S\&P Emerging Markets Dividend ETF & Emerging Markets Dividend & 23-Feb-2011 \\
EELV & Invesco S\&P Emerging Markets Low Volatility ETF & Emerging Markets Low Volatility & 13-Jan-2012 \\
EEMV & iShares Edge MSCI Min Vol Emerging Markets ETF & Emerging Markets Low Volatility & 18-Oct-2011 \\
EWX & SPDR S\&P Emerging Markets Small Cap ETF & Emerging Markets Small-Cap Equities & 12-May-2008 \\
FEM & First Trust Emerging Markets AlphaDEX Fund & Emerging Markets Smart Beta & 18-Apr-2011 \\
PIE & Invesco Dorsey Wright Emerging Markets Momentum ETF & Emerging Markets Momentum & 28-Dec-2007 \\
PXH & Invesco RAFI Emerging Markets ETF & Fundamental Emerging Markets Equities & 27-Sep-2007 \\

\hline
\end{tabular}
}

\begin{tablenotes}
\footnotesize
\item \textit{Note:} ETF classifications are based on each fund's primary geographic, sector, or investment-style exposure.
\end{tablenotes}
\end{threeparttable}
\end{table}

\section{Right-Tail Hill Estimation Results}\label{sec:right-tail-hill}

\begin{figure}[H]
    \centering
    \includegraphics[width=\textwidth]{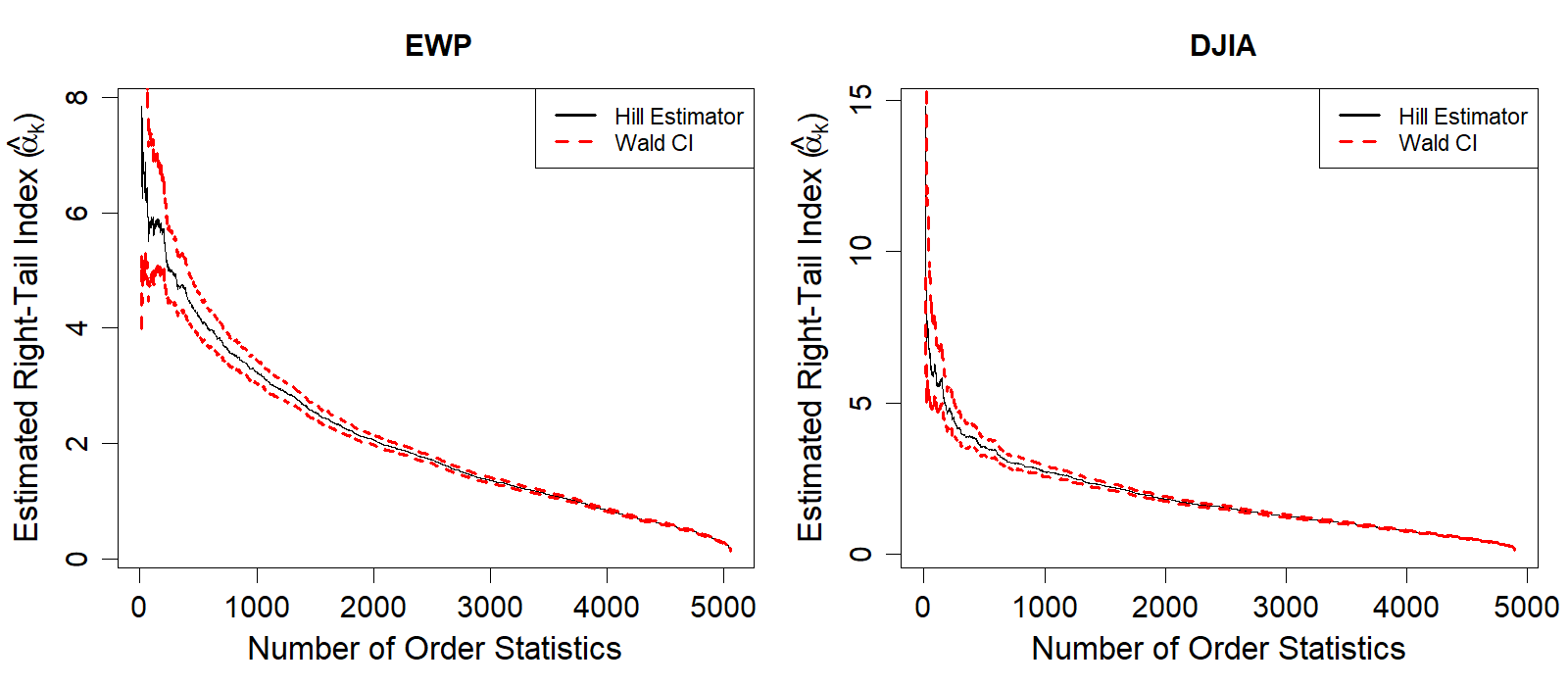}
    \includegraphics[width=\textwidth]{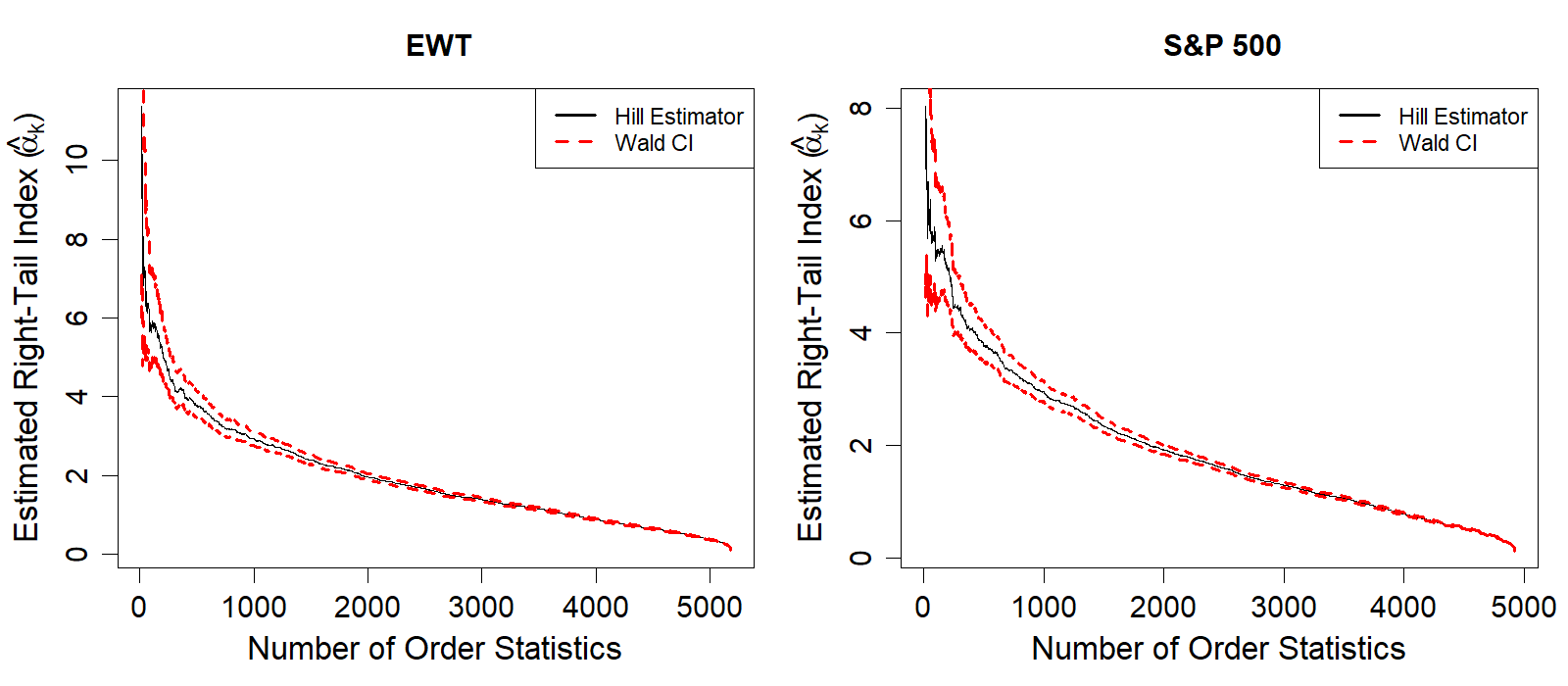}
    \caption{Right-tail Hill estimates ($\widehat{\alpha}_k$) with 95\% Wald confidence bands for (a) EWP, (b) DJIA, (c) EWT, and (d) S\&P~500, based on the transformed loss series $L_t=-R_t$.}
    \label{fig:right-tail}
\end{figure}

\section{Additional Rolling Estimation Results}

\begin{figure}[H]
\centering
\includegraphics[width=0.9\textwidth]{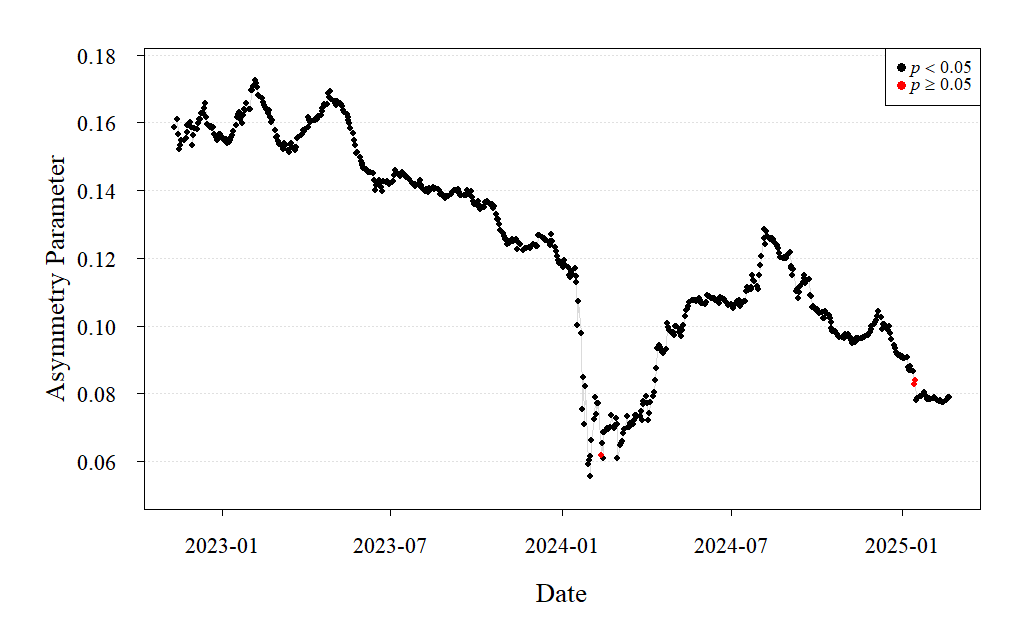}
\caption{Rolling estimates of the asymmetry parameter $\gamma$ in the GJR--GARCH$(1,1)$-$t$ model for EWP using a four-year rolling window. Black points indicate statistically significant estimates at the 5\% level.}
\label{fig:rolling_gjr_EWP}
\end{figure}

The estimated asymmetry parameter was relatively large during the earlier part of the sample, with values near \(0.18\text{--}0.20\), before declining around 2023. Although the parameter subsequently recovered, it remained below earlier levels and gradually decreased toward the end of the sample period. These results suggest that the impact of negative shocks on volatility was not constant across periods and may reflect changing market conditions over time.

\begin{figure}[H]
\centering
\includegraphics[width=0.9\textwidth]{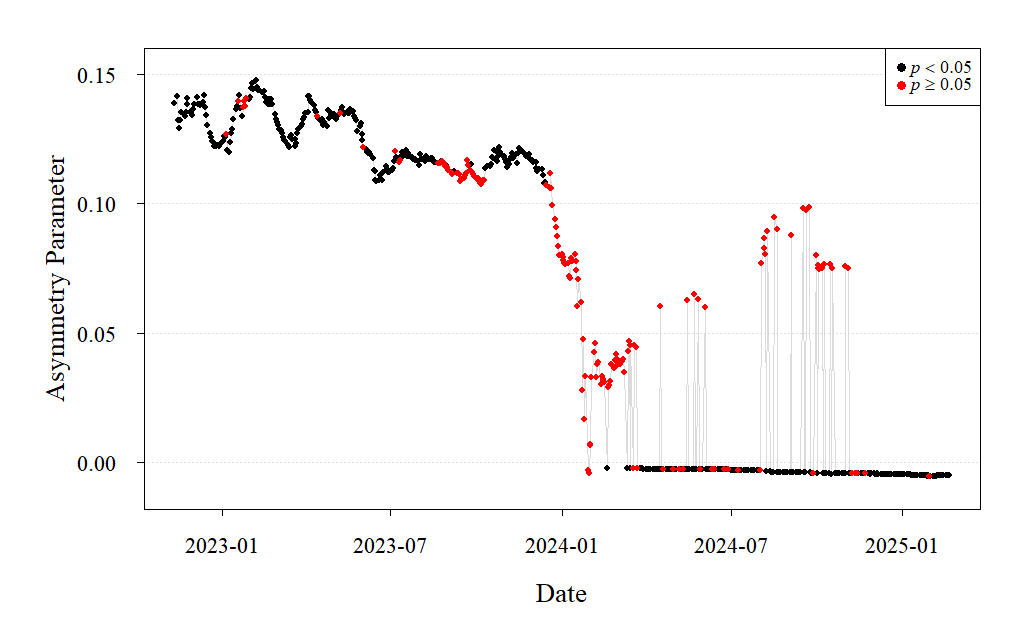}
\caption{Rolling estimates of the asymmetry parameter $\gamma$ in the GJR--GARCH$(1,1)$-$t$ model for EWT using a four-year rolling window. Black points indicate statistically significant estimates at the 5\% level.}
\label{fig:rolling_gjr_EWT}
\end{figure}

\noindent
Figure~\ref{fig:rolling_gjr_EWT} presents the rolling estimates of the GJR-GARCH(1,1)-$t$ asymmetry parameter for EWT. The estimated asymmetry parameter followed a pattern broadly similar to that of EWP through late 2023, with statistically significant estimates generally ranging between approximately 0.10 and 0.15. Beginning in early 2024, however, the rolling estimates declined sharply toward zero and became predominantly statistically insignificant, accompanied by substantially greater variability over time. This deterioration in estimation precision coincided with the instability observed in the rolling FIGARCH estimates of the fractional differencing parameter $d$ reported in Figure~\ref{fig:rollingfigarch_EWT}, where a similar increase in statistically insignificant estimates also occurred during the same period. Because this pattern appeared in both the GJR-GARCH and FIGARCH models, it was unlikely to have resulted solely from model specification and instead suggested a change in the volatility dynamics of EWT beginning in early 2024. The results indicated that the asymmetric and long-memory volatility characteristics of EWT were less stable after early 2024 than during the earlier sample period. Therefore, the post-2024 estimates should be interpreted with caution because of the increased estimation uncertainty observed during this period.

\section{Robustness Analysis}\label{app:robust}
\begin{figure}[H]
\centering
\includegraphics[width=0.9\textwidth]{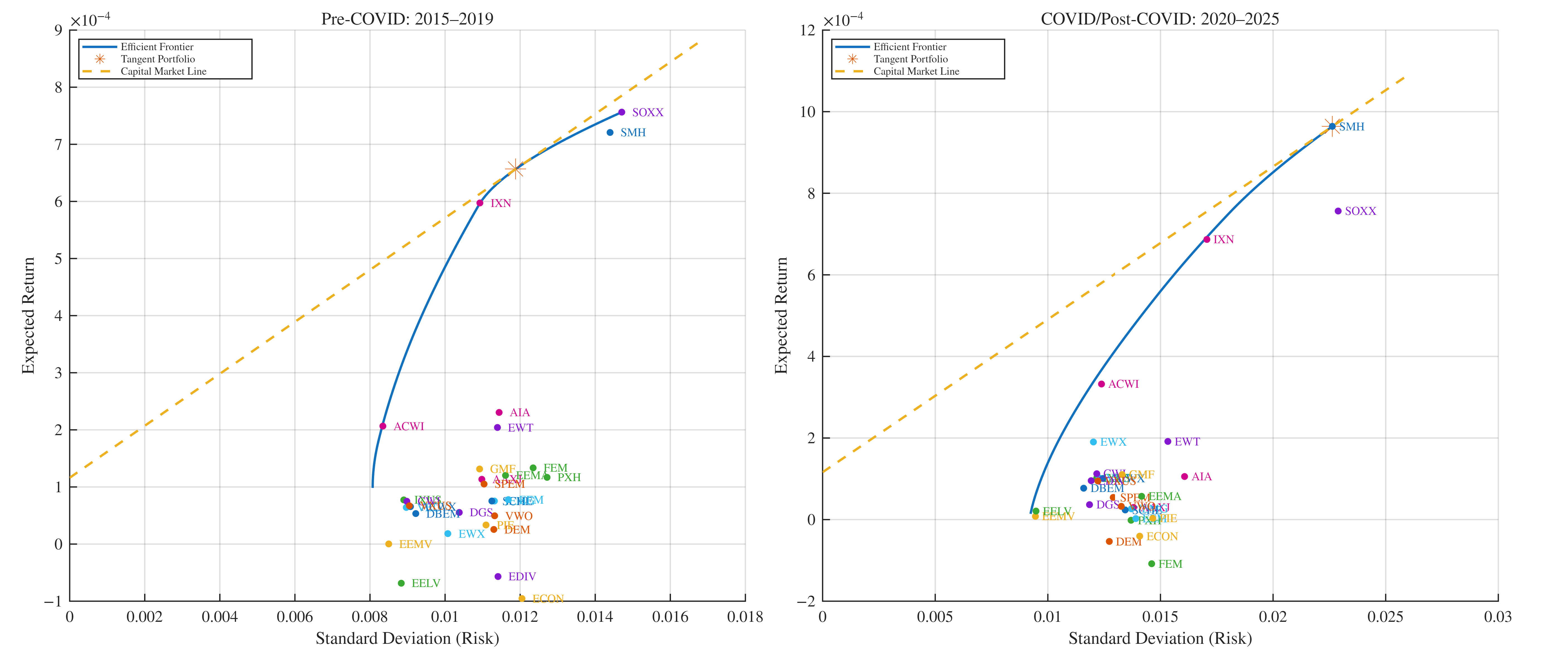}
\caption{Mean--variance efficient frontiers:(left) pre-COVID (2015--2019) and (right) post-COVID (2020--2025) subsamples for 3-months treasury yields.}
\label{fig:mv_robust}
\end{figure}

\begin{figure}[H]
\centering
\includegraphics[width=0.9\textwidth]{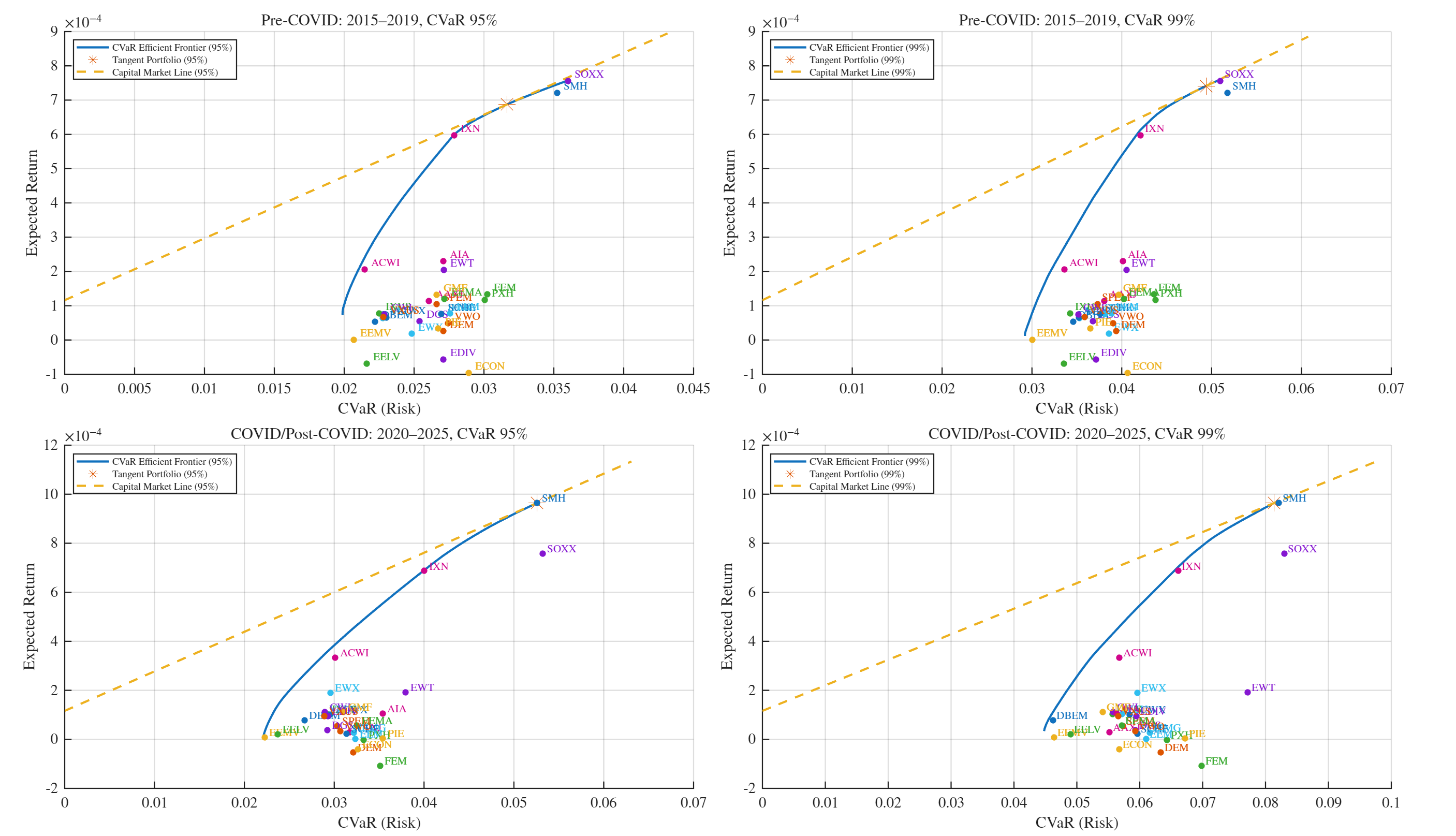}\caption{CVaR-efficient frontiers:(top row) pre-COVID (2015--2019) and (bottom row) post-COVID (2020--2025) subsamples at the 95\% and 99\% confidence levels for 3-months treasury yields.}
\label{fig:cvar_robust}
\end{figure} 

\end{document}